%% file: ms_1.tex
\documentclass[twocolumn]{aastex701}

\usepackage{xcolor} 
\usepackage{amsmath}
\usepackage{threeparttable}
\usepackage{multirow}
\usepackage{ltablex}
\usepackage{graphicx} 
\usepackage{array}
\usepackage{supertabular}
\usepackage{longtable}
\usepackage{url} 
\usepackage{booktabs}
\usepackage[caption=false]{subfig}
\usepackage[T1]{fontenc}
\usepackage{geometry}
\begin{document}
\begin{sloppypar}
    
\newcommand{\PMO}{Purple Mountain Observatory, Chinese Academy of Sciences, Nanjing 210008, China}
\newcommand{\USTC}{School of Astronomy and Space Sciences, University of Science and Technology of China, Hefei 230026, China}
\newcommand{\NAOC}{National Astronomical Observatories, Chinese Academy of Sciences, 20A Datun Road, Chaoyang District, Beijing 100101, China}
\newcommand{\Skynet}{Department of Physics and Astronomy, University of North Carolina at Chapel Hill, Chapel Hill, NC 27599, USA}
\newcommand{\IHEP}{State Key Laboratory of Particle Astrophysics, Institute of High Energy Physics, Chinese Academy of Sciences, Beijing 100049, China}
\newcommand{\IJCLab}{IJCLab, Univ Paris-Saclay, CNRS/IN2P3, Orsay, France}
\newcommand{\IAPP}{Institut d'Astrophysique de Paris, Paris, France}
\newcommand{\ERAU}{Department of Physical Sciences, Embry-Riddle Aeronautical University, 1 Aerospace Boulevard, Daytona Beach, Fl 32114, USA}
\newcommand{\CASS}{Center for Astrophysics and Space Science (CASS), New York University Abu Dhabi, Saadiyat Island, PO Box 129188, Abu Dhabi, UAE}
\newcommand{\NYUAD}{New York University Abu Dhabi, PO Box 129188, Saadiyat Island, Abu Dhabi, UAE}
\newcommand{\SKLRAT}{State Key Laboratory of Radio Astronomy and Technology, Purple Mountain Observatory, Chinese Academy of Sciences, 10 Yuanhua Road, Nanjing 210023, China}
\newcommand{\CEA}{Université Paris-Saclay, Université Paris Cité, CEA, CNRS, AIM, 91191, Gif-sur-Yvette, France}
\newcommand{\NARIT}{National Astronomical Research Institute of Thailand (NARIT), Chiang Mai 50180, Thailand.}
\newcommand{\NJU}{School of Astronomy and Space Science, Nanjing University, Nanjing 210023, China}

\title{GRB 250424A: A Case Study of Energy Injection with Multiwavelength Observations}

\author[orcid=0009-0005-0170-192X]{Yifang Liang}
\affiliation{\PMO}
\affiliation{\USTC}
\email[]{yfliang@pmo.ac.cn}

\author[orcid=0000-0002-8385-7848]{Yun Wang}
\affiliation{\PMO} 
\affiliation{Department of Astronomy, University of California, Berkeley, CA 94720-3411, USA}
\email[]{wangyun@pmo.ac.cn}

\author[orcid=0000-0002-2636-6508]{WeiKang~Zheng$^\dag$}
\affiliation{Department of Astronomy, University of California, Berkeley, CA 94720-3411, USA}
\email[show]{weikang@berkeley.edu}
\correspondingauthor{WeiKang Zheng, Di Xiao}

\author[orcid=0009-0006-0547-1030]{Priyadarshini Gokuldass}  
\affiliation{\ERAU}
\email[]{gokuldap@my.erau.edu}

\author[orcid=0000-0002-7457-4192]{Huali Li}  
\affiliation{\NAOC}
\email[]{lhl@nao.cas.cn}

\author[orcid=0009-0008-8053-2985]{Chenwei Wang}  
\affiliation{\IHEP}
\email[]{cwwang@ihep.ac.cn}

\author[orcid=0009-0000-0564-7733]{Riccardo Brivio}  
\affiliation{INAF-Osservatorio Astronomico di Brera, Via E. Bianchi 46,
23807 Merate, (LC), Italy}
\email[]{riccardo.brivio@inaf.it}

\author[orcid=0009-0009-2360-4396]{Donovan Schlekat}  
\affiliation{\Skynet}
\email[]{dschlekat@unc.edu}

\author[orcid=0000-0003-3460-0103]{Alexei~V.~Filippenko}  
\affiliation{Department of Astronomy, University of California, Berkeley, CA 94720-3411, USA}
\email[]{afilippenko@berkeley.edu}

\author[orcid=0000-0003-3224-2146]{Pillas Marion}  
\affiliation{\IAPP}
\email[]{pillas@iap.fr}

\author{Dalya Akl}   
\affiliation{\IJCLab}
\affiliation{\CASS}
\affiliation{\NYUAD} 
\email[]{dahliaaql01@gmail.com}

\author[orcid=0000-0002-7686-3334]{Sarah Antier}  
\affiliation{\IJCLab}
\email[]{sarah.antier@ijclab.in2p3.fr}

\author[]{Manasanun Tanasan}  
\affiliation{National Astronomical Research Institute of Thailand (NARIT), Chiang Mai 50180, Thailand.}
\email[]{}

\author[orcid=0000-0001-9109-8311]{Kanthanakorn Noysena}  
\affiliation{National Astronomical Research Institute of Thailand (NARIT), Chiang Mai 50180, Thailand.}
\email[]{kanthanakorn@narit.or.th}

\author[orcid=0000-0002-4304-2759]{Di Xiao$^{\ddag}$}
\affiliation{\PMO} 
\affiliation{\SKLRAT}
\email[show]{dxiao@pmo.ac.cn}


\author[orcid=0009-0000-5068-3434]{Jie An}
\affiliation{\NAOC}
\email[]{anjie@bao.ac.cn}

\author[orcid=0000-0001-5955-2502]{Thomas~G.~Brink}
\affiliation{Department of Astronomy, University of California, Berkeley, CA 94720-3411, USA}
\email[]{tgbrink@berkeley.edu}

\author[orcid=0000-0002-9650-4371]{Krittapas Chanchaiworawit}
\affiliation{National Astronomical Research Institute of Thailand, 260 Moo 4, Donkaew, Maerim, Chiang Mai 50180, Thailand}
\email[]{krittapas@narit.or.th}

\author[orcid=0000-0003-3144-7369]{Dylan A. Dutton}
\affiliation{\Skynet}
\email[]{astrodyl@live.unc.edu}

\author[]{Matteo Ferro}
\affiliation{INAF-Osservatorio Astronomico di Brera, Via E. Bianchi 46, 23807 Merate, (LC), Italy}
\email[]{matteo.brivio@ferro.it}

\author[orcid=0009-0005-4287-7198]{Michael Freeberg}
\affiliation{KNC, AAVSO, Hidden Valley Observatory(HVO), Colfax, WI.; iTelescope, UDRO, Beryl Junction, Utah.}
\email[]{}

\author[orcid=0000-0002-9037-8642]{Ren Jia}
\affiliation{\PMO}
\email[]{renjia@pmo.ac.cn}

\author[]{Alain Klotz}
\affiliation{IRAP, Université de Toulouse, CNRS, CNES, UPS, France}
\email[]{aklotz@irap.omp.eu}

\author[orcid=0000-0001-5931-2381]{Ye Li}
\affiliation{\PMO}
\email[]{yeli@pmo.ac.cn}

\author[orcid=0000-0002-4072-6899]{Xing Liu}
\affiliation{\NAOC}
\email[]{liuxing@nao.cas.cn}

\author[orcid=0000-0002-5060-3673]{Dan Reichart}
\affiliation{\Skynet}
\email[]{reichart@physics.unc.edu}

\author[0000-0003-4189-9668]{Antonio C. Rodriguez}
\affiliation{Center for Astrophysics $|$ Harvard \& Smithsonian, 60 Garden Street, Cambridge, MA, 02138, USA}
\affiliation{California Institute of Technology, Department of Astronomy, 
1200 E. California Blvd., Pasadena, CA, 91125, USA}
\email[]{acrodrig@astro.caltech.edu}

\author[]{Wenjun Tan}
\affiliation{\IHEP}
\email[]{tanwj@ihep.ac.cn}

\author[orcid=0000-0002-1481-4676]{Samaporn Tinyanont}
\affiliation{National Astronomical Research Institute of Thailand, 260 Moo 4, Donkaew, Maerim, Chiang Mai 50180, Thailand}
\email[]{samaporn@narit.or.th}

\author[]{Jing~Wang}
\affiliation{\NAOC}
\email[]{wj@bao.ac.cn}

\author[0009-0001-8025-3205]{Zi-Qi Wang}
\affiliation{Guangxi Key Laboratory for Relativistic Astrophysics, School of Physical Science and Technology, Guangxi University, Nanning 530004, China}
\email[]{ziqi.wang@st.gxu.edu.cn}

\author[]{Jianyan Wei}
\affiliation{\NAOC}
\email[]{}

\author[]{Samuel E. Whitebook}
\affiliation{Division of Physics, Mathematics and Astronomy, California Institute of Technology, Pasadena, CA 91125, USA}
\email[]{sewhitebook@astro.caltech.edu}

\author[orcid=0000-0002-6299-1263]{Xuefeng Wu}
\affiliation{\PMO}
\email[]{xfwu@pmo.ac.cn}

\author[orcid=0000-0003-3257-9435]{Dong Xu}
\affiliation{\NAOC}
\email[]{dxu@nao.cas.cn}

\author[orcid=0000-0002-6535-8500]{Yi Yang}
\affiliation{Department of Physics, Tsinghua University, Qinghua Yuan, Beijing 100084, China}
\email[]{yi_yang@mail.tsinghua.edu.cn}

\author[]{Jinpeng Zhang}
\affiliation{\IHEP}
\email[]{zhangjinpeng@ihep.ac.cn}

\author[orcid=0009-0008-6247-0645]{Wenlong Zhang}
\affiliation{\PMO}
\affiliation{\USTC}
\email[]{wlzhang@pmo.ac.cn}

\author[orcid=0000-0003-2915-7434]{Hao Zhou}
\affiliation{\PMO}
\email[]{haozhou@pmo.ac.cn}

\author[]{Valerio D'Elia}
\affiliation{Space Science Data Center (SSDC) - Agenzia Spaziale Italiana (ASI), 00133 Roma, Italy}
\email[]{valerio.delia@ssdc.asi.it} 

\author[]{Massimiliano De Pasquale}
\affiliation{University of Messina, Mathematics, Informatics, Physics and Earth Science Department, Via F.S. D’Alcontres 31, Polo Papardo, 98166 Messina, Italy}
\email[]{massimiliano.depasquale@unime.it}  

\author[]{Dino Fugazza}
\affiliation{INAF-Osservatorio Astronomico di Brera, Via E. Bianchi 46, 23807 Merate, (LC), Italy}
\email[]{dino.fugazza@inaf.it}  

\author[]{Luciano Nicastro}
\affiliation{INAF–Osservatorio di Astrofisica e Scienza dello Spazio di Bologna, Via Piero Gobetti 93/3, 40129 Bologna, Italy}
\email[]{luciano.nicastro@inaf.it}  

\author[orcid=0000-0003-1835-1522]{Damien Turpin}
\affiliation{\CEA} 
\email[]{damien.turpin@cea.fr}

\author[orcid=0009-0007-8085-6683]{Roger Hellot}
\affiliation{KNC, AITP, 23 rue sainte odile, 67560 Rosheim, France}
\email[]{}

\author[orcid=0000-0003-3358-4834]{Frederic Dux}
\affiliation{European Southern Observatory, Alonso de Córdova 3107, Vitacura, Santiago, Chile}
\affiliation{Institute of Physics, Laboratory of Astrophysics, Ecole Polytechnique F\'ed\'erale de Lausanne (EPFL), Observatoire de Sauverny, 1290 Versoix, Switzerland}
\email[]{frederic.dux@epfl.ch}

\author[]{Xiangyu Wang}
\affiliation{\NJU}
\affiliation{Key Laboratory of Modern Astronomy and Astrophysics (Nanjing University), Ministry of Education, China}
\email[]{xywang@nju.edu.cn}

\author[]{Frédéric Daigne}
\affiliation{Sorbonne Universite, CNRS, UMR 7095, Institut d’Astrophysique de Paris, 98 bis bd Arago, F-75014 Paris, France}
\email[]{daigne@iap.fr}

\author[orcid=0000-0001-7199-2906]{Yongfeng Huang}
\affiliation{\NJU}
\affiliation{Key Laboratory of Modern Astronomy and Astrophysics (Nanjing University), Ministry of Education, China}
\email[]{hyf@nju.edu.cn}


\author[]{HongBo Cai}
\affiliation{\NAOC}
\email[]{chb@nao.cas.cn}

\author[]{Alexis Coleiro}
\affiliation{Université Paris Cité, CNRS, Astroparticule et Cosmologie, F75013 Paris, France}
\email[]{coleiro@apc.in2p3.fr}

\author[]{Bertrand Cordier}
\affiliation{\CEA}
\email[]{bertrand.cordier@cea.fr}

\author[]{Stefano Crepaldi}
\affiliation{Centre National d’Etudes Spatiales, Centre spatial de Toulouse, 18 avenue Edouard Belin, 31401 Toulouse
Cedex 9, France}
\email[]{stefano.crepaldi@cnes.fr}

\author[]{YongWei Dong}
\affiliation{\IHEP}
\email[]{dongyw@ihep.ac.cn}

\author[]{Olivier Godet}
\affiliation{IRAP, Universite de Toulouse, CNRS, CNES, Toulouse, France}
\email[]{olivier.godet@irap.omp.eu}

\author[]{XuHui Han}
\affiliation{\NAOC}
\email[]{hxh@nao.cas.cn}

\author[]{Frédéric Piron}
\affiliation{Laboratoire Univers et Particules de Montpellier, Université Montpellier, CNRS/IN2P3, F-34095 Montpellier, France}
\email[]{piron@in2p3.fr}

\author[]{YuLei Qiu}
\affiliation{\NAOC}
\email[]{qiuyl@bao.ac.cn}

\author[]{Stéphane Schanne}
\affiliation{\CEA}
\email[]{stephane.schanne@cea.fr}

\author[]{Chao Wu}
\affiliation{\NAOC}
\email[]{cwu@nao.cas.cn}

\author[]{LiPing Xin}
\affiliation{\NAOC}
\email[]{xlp@nao.cas.cn}

\author[]{Yang Xu}
\affiliation{\NAOC}
\email[]{yxu@nao.cas.cn}

\author[]{Shuangnan Zhang}
\affiliation{\IHEP}
\email[]{zhangsn@ihep.ac.cn}

\author[]{ShiJie Zheng}
\affiliation{\IHEP}
\email[]{zhengsj@ihep.ac.cn}

\begin{abstract}

We present a comprehensive multiwavelength analysis of the long-duration gamma-ray burst (GRB) 250424A. Our dataset spans from the prompt gamma-ray emission to late-time optical monitoring, including spectra obtained with the Keck 10\,m telescope. We find that the afterglow light curves display a prominent, simultaneous shallow decay phase in both X-ray and optical bands, followed by an achromatic transition to a standard decay regime. The broadband spectral energy distributions are well-modeled by a single power-law function, indicating a common synchrotron origin for the emission across frequencies. We interpret the afterglow evolution within the framework of a relativistic forward shock refreshed by continuous energy injection. This scenario successfully reproduces the observed temporal and spectral behavior, yielding an isotropic equivalent kinetic energy of $E_{\rm K,iso} \approx 5.5 \times 10^{52}$ erg
and an injection index of $q\approx 0.34$
in a constant-density circumburst environment. The shallow decay phase is consistent with sustained energy injection lasting $\sim$ 9 ks.
Despite the relatively low redshift, late-time optical observations reveal no distinct supernova component; however, our derived upper limits do not strictly rule out the presence of a typical GRB-associated supernova.

\end{abstract}

\section{Introduction} \label{sec:intro}

Gamma-ray bursts (GRBs) represent the most luminous electromagnetic explosions in the universe, releasing isotropic equivalent energies of ($E_{\rm iso} \approx 10^{50}$--$10^{55}~\rm erg$) over timescales ranging from milliseconds to several minutes \citep{Klebesadel1973,Piran2005}. Long-duration GRBs (LGRBs; $T_{90} > 2~s$, where $T_{90}$ is defined as the interval accumulating $90\%$ of the photon fluence) are widely believed to originate from the core collapse of massive stars \citep{Woosley1993}, a process that leads to the formation of a compact central engine --- either a stellar-mass black hole or a rapidly spinning magnetar \citep{MacFadyen1999,Zhang&Meszaros2001,Woosley2006}. In this scenario, a highly collimated, ultrarelativistic jet is launched along the rotation axis of the progenitor. Energy dissipation within the jet, occurring after it propagates through and breaks out of the stellar envelope, powers the prompt $\gamma$-ray emission \citep{Rees1994,Piran1999,Kumer2015}. Subsequently, the relativistic outflow impacts the circumburst medium, driving a strong forward shock into the ambient gas. This interaction accelerates electrons and generates broadband synchrotron emission --- from X-rays to radio wavelengths --- known as the GRB afterglow \citep{Mészáros1997,Sari1998,Dai&Lu1998,Granot&Sari2002}. Detailed modeling of the temporal and spectral evolution of the afterglow serves as a powerful probe of jet dynamics, shock microphysics, central-engine properties, and the density profile of the circumburst environment \citep{Panaitescu2002}.

The advent of modern multiwavelength facilities over the past two decades \citep{swift2004,fermi,svom_wei,yuan2022} has revealed that GRB afterglow evolution is significantly more complex than the predictions of the simple relativistic blast-wave model. In particular, high-cadence monitoring has established a ``canonical'' X-ray light curve characterized by distinct evolutionary phases, including an initial steep decay, a prolonged shallow decay or ``plateau,'' and a subsequent normal decay phase \citep{Nousek2006,ZhangB2006,zhang&Liang&zhang2007,zhangb2018}. X-ray plateaus are statistically prevalent, appearing in approximately half of long GRBs \citep{Tangch2019,Dainotti2020,2026Dengc}, their achromatic counterparts (e.g., simultaneously in X-ray and optical) are relatively rare. Such achromatic behavior has been robustly identified in only a limited sample of bursts, including GRB~060614, GRB~140903A, and the exceptionally bright GRB~221009A \citep[e.g.,][]{Gehrels2006,Troja2016,Lesage2023,Williams2023,Zhengchao2024}. Crucially, the temporal evolution across these bands can be diverse, sometimes exhibiting chromatic behavior that challenges standard synchrotron interpretations \citep{Panaitescu2011,Li2012,Oates2015M}. 

A leading scenario for the shallow-decay phase is continuous energy injection into the forward shock, a process that compensates for radiative and adiabatic energy losses and decelerates the shock more gradually than in the impulsive case \citep{Dai&Lu1998,Zhang&Meszaros2001}. This sustained energy supply is typically attributed to the activity of a long-lived central engine, such as spin-down energy from a newborn millisecond magnetar \citep{Dai&Lu1998,Zhang&Meszaros2001,Rowlinson2013} or fallback accretion onto a newly formed black hole \citep{Kumar2008,Wuxf2013}.  Alternatively, a similar observational signature can arise from hydrodynamical effects without prolonged central-engine activity, specifically through ``refreshed shocks'' caused by a stratified ejecta profile where slower shells catch up with the decelerating blast wave \citep{Ree1998,Sari2000,Dereli2022NC,Geng2025}. Distinguishing between these scenarios necessitates broadband modeling to check for consistency with closure relations linking temporal decay indices ($\alpha$) and spectral indices ($\beta$).

Regarding the progenitor systems, LGRBs are firmly established to be associated with broad-lined Type Ic supernovae (SNe Ic-BL), confirming the massive-star origin of these events \citep[e.g.,][]{Galama1998Nat,Hjorth2003,Stanek2003}. The properties of the associated supernova (SN), including its peak luminosity and synthesized $^{56}$Ni mass, offer pivotal diagnostics of the explosion mechanism and the progenitor's structure \citep{Woosley2006,Cano2017}. However, the detectability of the SN component can vary significantly; while most nearby LGRBs show bright SN features, a subset of events exhibits faint or nondetected SNe, challenging the ubiquity of a simple GRB-SN scaling relation \citep{Fynbo2006Nat,Della2006}. Consequently, deep late-time optical monitoring is essential not only to confirm the association but also to probe the diversity of explosion energies and nickel production in GRB progenitors.

In this paper, we present a comprehensive multiwavelength analysis of the long GRB 250424A. Focusing on the prominent, simultaneous plateau features observed in both X-ray and optical bands, we employ a forward-shock model incorporating continuous energy injection to constrain the microphysical properties of the relativistic jet and the density profile of the circumburst medium. Furthermore, we utilize late-time optical imaging to strictly constrain the presence of any associated SN component. Throughout this work, we adopt a standard $\Lambda$CDM cosmology with H$_0 = 70\,\rm km\, s^{-1}\, Mpc^{-1}$, $\Omega_{M} = 0.3$, and $\Omega_{\Lambda} = 0.7$. All uncertainties are reported at the $1\sigma$ confidence level unless otherwise stated.

This paper is organized as follows. In Section \ref{sec: obs}, we describe the multiwavelength observation campaign and data-reduction procedures.  Section \ref{sec: result} presents the temporal and spectral analysis of the afterglow.
 We interpret in Section \ref{sec:model} the afterglow evolution within the energy-injected external-shock framework. Section \ref{sec:result_sn} discusses the physical implications of the nondetection of an associated SN. A summary and our conclusions are provided in \ref{sec: summary}.

\setcounter{footnote}{0}
\section{Observations and Data Reduction}\label{sec: obs}
\subsection{Prompt Gamma-Ray Observations}
Owing to the large effective area and the delicately designed trigger algorithm \citep{GRM_trigger}, GRB 250424A was first identified by the Gamma-Ray Monitor (GRM) onboard the {\it Space Variable Objects Monitor (SVOM)} satellite \citep{svom_wei,2026svommission} at 06:52:08 (UTC dates and times are used throughout this paper) on 2025 April 24 \citep[hereafter $T_0$;][]{grmgcn}. Approximately 20 s after the \textit{SVOM} trigger, the burst was independently detected by the Burst Alert Telescope \citep[BAT;][]{Barthelmy_2005} onboard the {\it Neil Gehrels \textit{Swift} Observatory} \citep[{\it Swift};][]{swift2004}. {\it Swift} immediately slewed to the target, facilitating rapid follow-up observations with the X-Ray Telescope \citep[XRT;][]{2005xrt} and the Ultra-Violet/Optical Telescope \citep[UVOT;][]{2005uvot}. {\it Swift} was the first to report the event in the GCN\footnote{\url{https://gcn.nasa.gov/}} \citep{swiftgcn}\footnote{The {\it Swift}/BAT trigger occurred after the main emission peak, resulting in a delay relative to the {\it SVOM}/GRM detection. We therefore adopt the {\it SVOM}/GRM trigger time as the reference epoch $T_0$ for all subsequent analysis.}. In addition to {\it SVOM} and {\it Swift}, the burst was  registered by several other space-borne missions, including {\it AstroSat} \citep{AstroSat_GCN}, {\it Konus-Wind} \citep{KW_GCN}, {\it EIRSAT-1} \citep{EIRSAT_GCN}, and {\it CALET} \citep{CALET_GCN}.

To characterize the prompt emission properties, we performed a joint temporal and spectral analysis using data from both {\it SVOM}/GRM and {\it Swift}/BAT. For the {\it SVOM} dataset, we utilized the GRM observations covering the 15 keV to 5 MeV energy range. The data were processed using the standard reduction procedure for the {\it SVOM}/GRM to extract energy-resolved light curves and spectra \citep{GRM_analysis}. For the {\it Swift}/BAT data (15--350 keV), we performed data reduction following standard procedures within the \textsc{HEASoft} package \citep{2014heasoft}. Mask-weighted light curves and spectra were generated using the task \texttt{batbinevt}, with geometric corrections applied via the corresponding response matrices. All spectral modeling was conducted using the \textsc{XSPEC} software package (v12.13; \citealp{xspec1996}).

\subsection{Afterglow Observations}\label{sec:opt-obs}

Extensive follow-up observations in the X-ray, optical, and near-infrared (NIR) bands were conducted by a broad suite of space- and ground-based facilities. Participating instruments included {\it Swift}/XRT,
the {\it SVOM} Visible Telescope (VT), the Global Rapid Advanced Network Devoted to the Multi-messenger Addicts (GRANDMA) consortium that includes Kilonova Catcher (KNC) program, the Thai Robotic Telescope (TRT), TAROT/TCH, the Euler telescope, the Rapid Eye Mount (REM) telescope, and the Skynet Robotic Telescope Network.

\subsubsection{Space-Based Observations}
{\it Swift}/XRT began observing GRB 250424A at $\rm T_0+279.5$ s, initially operating in Windowed Timing (WT) mode before switching to Photon Counting (PC) mode \citep{2005xrt}. The X-ray afterglow was promptly localized, and within its error circle, a bright optical counterpart was automatically identified by UVOT (\citealp{swiftgcn}) and was rapidly confirmed by subsequent ground-based observations in the optical. In total, XRT carried out 12 follow-up epochs extending to $T_{0}+1.1\times10^6$~s. The data were reduced using the \textsc{HEASoft} package (v6.34) and standard \texttt{XRTPIPELINE} procedures \citep{Evans2007,Evans2009}, incorporating calibration files from the latest CALDB release. 

\begin{figure*}
    \centering
    \includegraphics[width=0.99\linewidth]{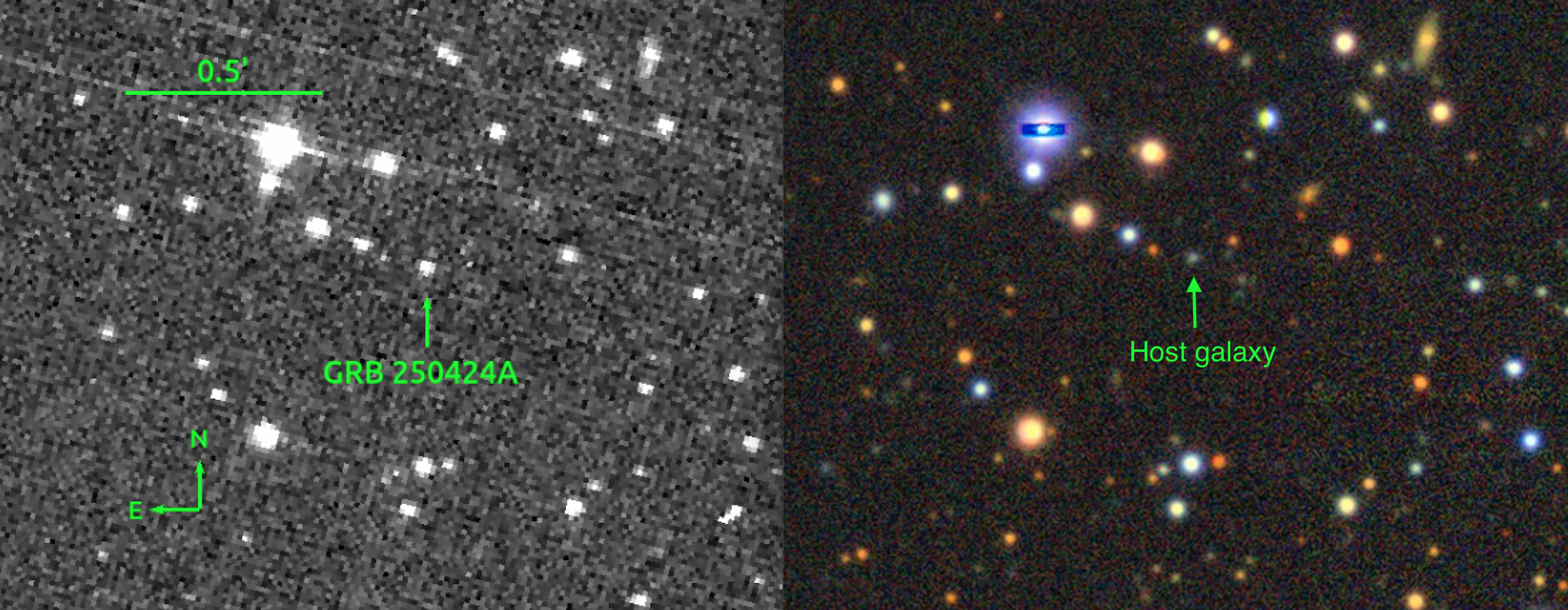}
    \caption{{\it Left:} A {\it SVOM}/VT image taken on April 24 showing the location of GRB~250424A when the afterglow is still bright. {\it Right:} The Legacy Surveys DR10 color image showing the same field with the host galaxy.}
    \label{fig:GRB250424A_fc_VT_LS}
\end{figure*}

In complement, the {\it SVOM}/VT, characterized by a 43 cm effective aperture, a $26' \times 26'$ field of view, and a pixel scale of $0.76''$, initiated Target of Opportunity (ToO) observations at 09:26:29 on 2025 April 24, approximately 154.34 min after the GRB trigger. Images were acquired simultaneously in the blue (VT$\_B$, 400--650 nm) and red (VT$\_R$, 650--1000 nm) channels. The GRB afterglow (Figure \ref{fig:GRB250424A_fc_VT_LS}, left panel) was monitored with a cadence of 1--2 days over the subsequent four weeks. All raw data were processed using the standard reduction pipeline for the \textit{SVOM}/VT instrument, which includes bias subtraction, dark-current correction, and flat-fielding. To improve the signal-to-noise ratio (S/N), individual frames from each epoch were stacked, except for the first day when the afterglow is still very bright.

From {\it SVOM}/VT images we measure the afterglow coordinates at RA (J2000) = 14$^{\rm hr}$29$^{\rm m}$59.96$^{\rm s}$ (217.4998\arcdeg), Dec. (J2000) = $-$35\arcdeg01\arcmin31.2\arcsec ($-35.0253$\arcdeg) \citep{vtgcn}. We noted (also see \citealp{hostgalaxy2025}) the presence of a cataloged galaxy (Figure \ref{fig:GRB250424A_fc_VT_LS}, right panel) from the Legacy Surveys DR10\footnote{\url{https://www.legacysurvey.org/dr10/description/}} located at $ \rm RA = 217.4999^\circ,\, Dec = -35.0252^\circ$, with a catalog magnitude of $r=21.98$ (more details in Section \ref{sec:hostgalaxy}). This source is situated $\sim 0.4''$ from the VT afterglow position, suggesting it is the host galaxy of GRB 250424A. To isolate the afterglow flux, we obtained late-time template images on 2025 June 22 and performed image subtraction on all epochs except the first night, when the afterglow was sufficiently bright to render host contamination negligible. Aperture photometry was performed on the subtracted images, and the relative photometry method was adopted with calibration against nearby field stars. Given the broad wavelength coverage of the VT filters, we calibrated the VT$\_B$ and VT$\_R$ magnitudes to the standard Bessel $V$ and $I$ systems, respectively, as the effective response wavelengths of these two filters are better matched to the VT bandpasses. We therefore denote VT$\_V$ and VT$\_I$ instead in the following Sections. The calibration catalog was adopted from The DESI Legacy Surveys Data Release DR10, whose $g,r,i,z$ magnitudes were transformed into the Landolt $B,V,R,I$-band magnitudes
using the empirical prescription presented by
Robert Lupton\footnote{http://www.sdss.org/dr7/algorithms/sdssUBVRITransform.html\#Lupton2005}. The final photometric results are presented in Table \ref{tab:optical_data}.

\subsubsection{Ground-Based Follow-up Campaigns}\label{sec:opt-telescops}
{\bf Thai Robototic Telescope, Les Makes/T60, Euler, OPD/60 and Kilonova-catcher within GRANDMA :} We monitored the evolution of the optical afterglow using the GRANDMA \citep{GRANDMA}. Participating facilities included telescopes from the Kilonova Catcher (KNC) program, the Thai Robotic Telescope (TRT) at Cerro Tololo Inter-American Observatory (CTIO), Les Makes/T60 at La Réunion, TAROT/TCH and the Euler telescope in Chile, and the OPD/60 cm in Brazil. The campaign spanned from 2025 April 24 to May 7, utilizing Bessel $B$, $V$, $R$, $I$ filters and SDSS $g$, $r$, $i$ filters to ensure broad temporal and spectral coverage. Data reduction and image subtraction were performed using the GRANDMA pipeline 
STDPipe \citep{2021ascl.soft12006K}. Photometric calibration was anchored to {\it Gaia} DR3 \citep{2023A&A...674A...1G} for Vega-based filters and SkyMapper Data Release 4 \citep[DR4;][]{2024PASA...41...61O} for the AB system \citep{Oke1983}. In total, we obtained 41 detections and 14 upper limits (see Table \ref{tab:optical_data}). Additionally, we measured the host-galaxy photometry from GRANDMA template images (details in Section \ref{sec:hostgalaxy}).

{\bf Supplement data observations from TRT outside of GRANDMA:}  A series of $r-sloan$ band images were obtained with the 0.7\,m TRT unit at CTIO between 2.3 and 2.9\,hr after the trigger. These data were reduced via the internal TRT pipeline and calibrated against the Legacy Survey DR10 catalog.

{\bf REM Telescope:} Simultaneous optical and NIR observations \citep{remgcn} were secured with the 0.6\,m Rapid Eye Mount (REM) telescope \citep{REM_Zerbi01,REM_Covino+04} at ESO La Silla, Chile. The automated sequence commenced at 06:54:31 ($ T_0 + 123$\,s) and continued for $\sim 3$\, hr in the $g$, $r$, $i$, $z$, $J$, $H$, and $K$ bands. Data were reduced using the standard REM pipeline, with nonuniformity corrections applied using flat-field frames processed via the Swift Reduction Package (SRP)\footnote{\url{http://www.me.oa-brera.inaf.it/utenti/covino/usermanual.html}}. NIR images were sky-subtracted using median-combined frames. We performed image alignment using Astroalign \citep{Astroalign} and derived astrometric solutions against {\it Gaia} DR3 stars \citep{Gaia_DR3}. The optical afterglow was clearly detected in multiple frames. We performed aperture photometry using SExtractor \citep{bertin_sextractor_1996}, calibrating the optical and NIR magnitudes against SkyMapper DR4 \citep[][]{onken_skymapper_2024} and 2MASS \citep{2Mass} sources, respectively.

{\bf Skynet Network:} 
We also utilized the 0.4\,m PROMPT-5 and PROMPT-6 telescopes, part of the Skynet Robotic Telescope Network (Skynet; \citealp{reichart_prompt_2005}) at CTIO in Chile. Scheduled automatically by the Skynet \texttt{Campaign Manager} \citep{dutton_skynets_2022}, observations began at 06:57:06 ($T_0 + \sim 5$\,min) and lasted for 3\,hr. The strategy included exposures in the $B$, $V$, $R$, and $I$ filters observations. Exposure lengths were calculated using an automated exposure-time calculator within the \texttt{Campaign Manager} software, and data reduction and astrometric calibration were performed automatically by the Skynet data pipeline. Background subtraction and aperture photometry were performed on the source using the \texttt{SExtractor} package implemented with the \texttt{SEP} Python library \citep{barbary_sep_2016}. The final photometry was calibrated against the SkyMapper DR4 catalog. Early preliminary results were reported through the GCN by \citet{dutton_grb_2025}.

\subsection{Spectroscopy}
We obtained two epochs of optical spectroscopy of the GRB 250424A host-galaxy environment using the Low Resolution Imaging Spectrometer (LRIS; \citealp{oke1995}) mounted on the 10\,m Keck I telescope on Maunakea, Hawaii. The first epoch was acquired through a ToO (PI A. V. Filippenko) on 2025 May 20 ($T_0 + 26$\,d; corresponding to $\sim 20$\,d in the rest frame) with a total integration time of 3200\,s (1500\,s + 1700\,s exposures). The second epoch was obtained on 2025 June 21 ($T_0 + 59$\,d; $\sim 45$\,d in the rest frame) with a total exposure of 3600\,s ($3 \times 1200$\,s). 

For both observations, we utilized the $1.0''$-wide slit oriented at or near the parallactic angle to minimize flux losses due to atmospheric dispersion \citep{filippenko1982}. The instrument configuration employed the 600/4000 grism on the blue side ($R \approx 1100$; $\lambda\approx 3000$--5600\,\AA) and simultaneously the 400/8500 grating on the red side ($R \approx 1100$;  $\lambda\approx 5400$--9500\,\AA), providing continuous wavelength coverage across the optical band.

Data reduction was performed using the \texttt{LPipe} pipeline \citep{Perley2019}, which handles bias subtraction, flat-fielding, cosmic-ray rejection, sky subtraction, and spectral extraction. Flux calibration was achieved using spectrophotometric standard stars observed on the same nights, at similar airmasses, and with an identical instrument configuration.

From the host emission lines, we derive a redshift of $z=0.31$ (see Section \ref{sec:result_sn}), confirming the redshift reported by \cite{gcn_redshift}, which will be used for the following analysis.

\section{Results}\label{sec: result}
\subsection{Prompt Emission}\label{sec: result_prompt}
\begin{figure}[h]
    \centering
    \includegraphics[width=0.98\linewidth]{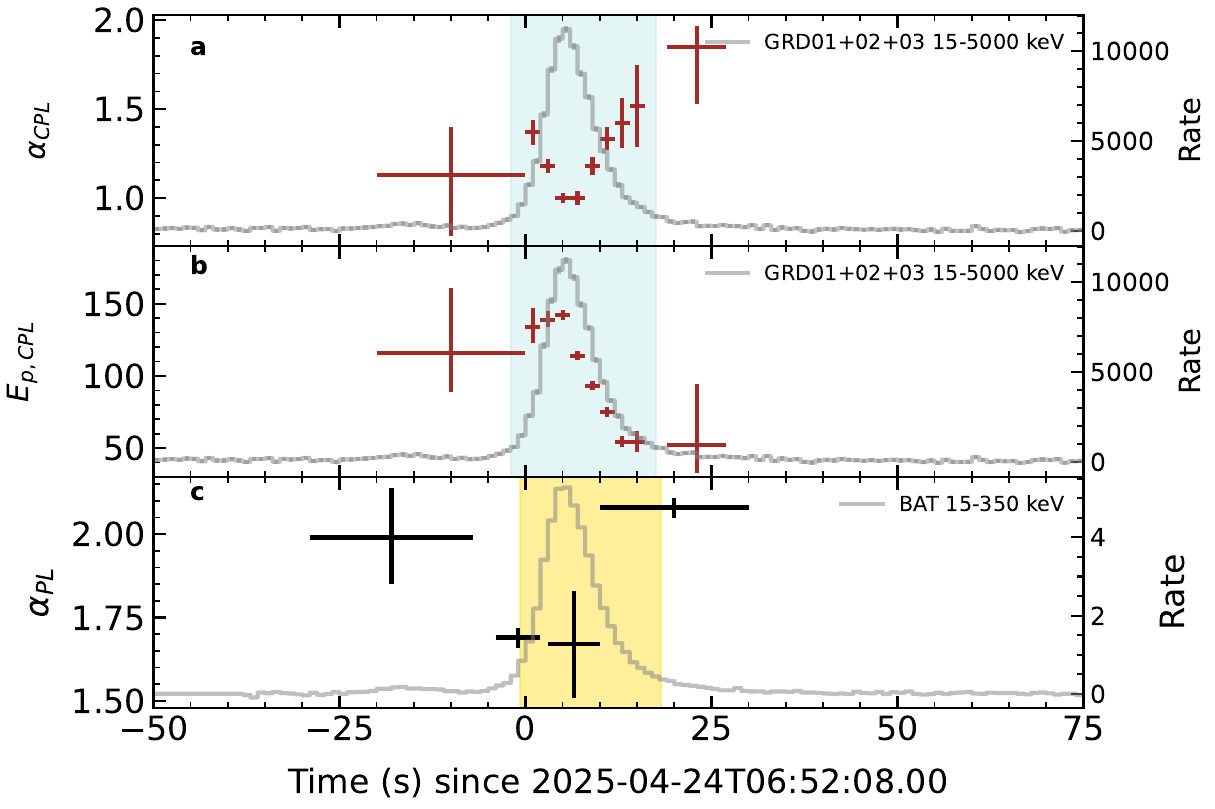}
    \caption{Temporal and spectral evolution of the prompt emission of GRB 250424A. (\textbf{a}) Background-subtracted {\it SVOM}/GRM light curve overlaid with the time-resolved photon index $\alpha$ derived from CPL fits. (\textbf{b}) Evolution of the spectral peak energy $E_{\rm p}$ obtained from the GRM CPL fits. In both panels (\textbf{a}) and (\textbf{b}), the light-blue shaded region indicates the {\it SVOM}/GRM $T_{90}$ interval ($19.5 \pm 1.0$\,s). (\textbf{c}) Background-subtracted {\it Swift}/BAT light curve shown alongside the photon-index evolution derived from simple PL fits. The bright yellow shaded region marks the BAT-measured $T_{90}$ ($19.03 \pm 1.06$\,s). The reference time $T_0$ is set to the {\it SVOM}/GRM trigger time (06:52:08.51). Error bars represent $1\sigma$ uncertainties for all spectral parameters. The background-subtracted light curves are also shown as gray step-like curves in the background of the spectral evolution panels for reference.}
    \label{fig:promp_lc}
\end{figure}
The prompt gamma-ray light curves, as displayed in Figure~\ref{fig:promp_lc}, exhibit a simple, single-pulse morphology detected simultaneously by {\it SVOM}/GRM and {\it Swift}/BAT. The pulse profiles from both instruments are well correlated, with peak emission occurring coincidentally at $T_0 + 5.0$\,s. We measured the burst duration $T_{90}$ 
using the {\it SVOM}/GRM 15--5000 keV data, yielding $T_{90} = 19.5 \pm 1.0$\,s. This value is fully consistent with the independent measurement of $T_{90} = 19.03 \pm 1.06$\,s obtained from {\it Swift}/BAT (15--350 keV), confirming the classification of GRB 250424A as a long-duration GRB.

We performed time-resolved spectroscopy to trace the spectral evolution across the GRB duration. For the wide-band {\it SVOM}/GRM data, the spectra are well modeled by a cutoff power-law (CPL) function, 
\begin{equation} 
N(E) = K E^{-\alpha_{\rm CPL}} \exp\left(-\frac{E}{E_{\rm cut}}\right), \end{equation} 
where $\alpha$ is the photon index and $E_{\rm cut}$ is the cutoff energy. The peak energy of the $\nu F_\nu$ spectrum is given by $E_{\rm p} = (2-\alpha_{\rm CPL})E_{\rm cut}$. Our analysis reveals a distinct ``hard-to-soft'' spectral evolution pattern. The peak energy $E_{\rm p}$ rises initially during the precursor phase, reaches a maximum at the onset of the main pulse, and subsequently undergoes a monotonic decay. This spectral softening is corroborated by the {\it Swift}/BAT data; although the narrower bandpass necessitates fitting with a simple power-law (PL) model ($N(E) \varpropto E^{-\alpha_{\rm PL}}$), the resulting photon index $\alpha_{\rm PL}$ exhibits a consistent steepening trend that tracks the evolutionary behavior observed in the {\it SVOM}/GRM data.

\begin{figure}
    \centering
    \includegraphics[width=0.98\linewidth]{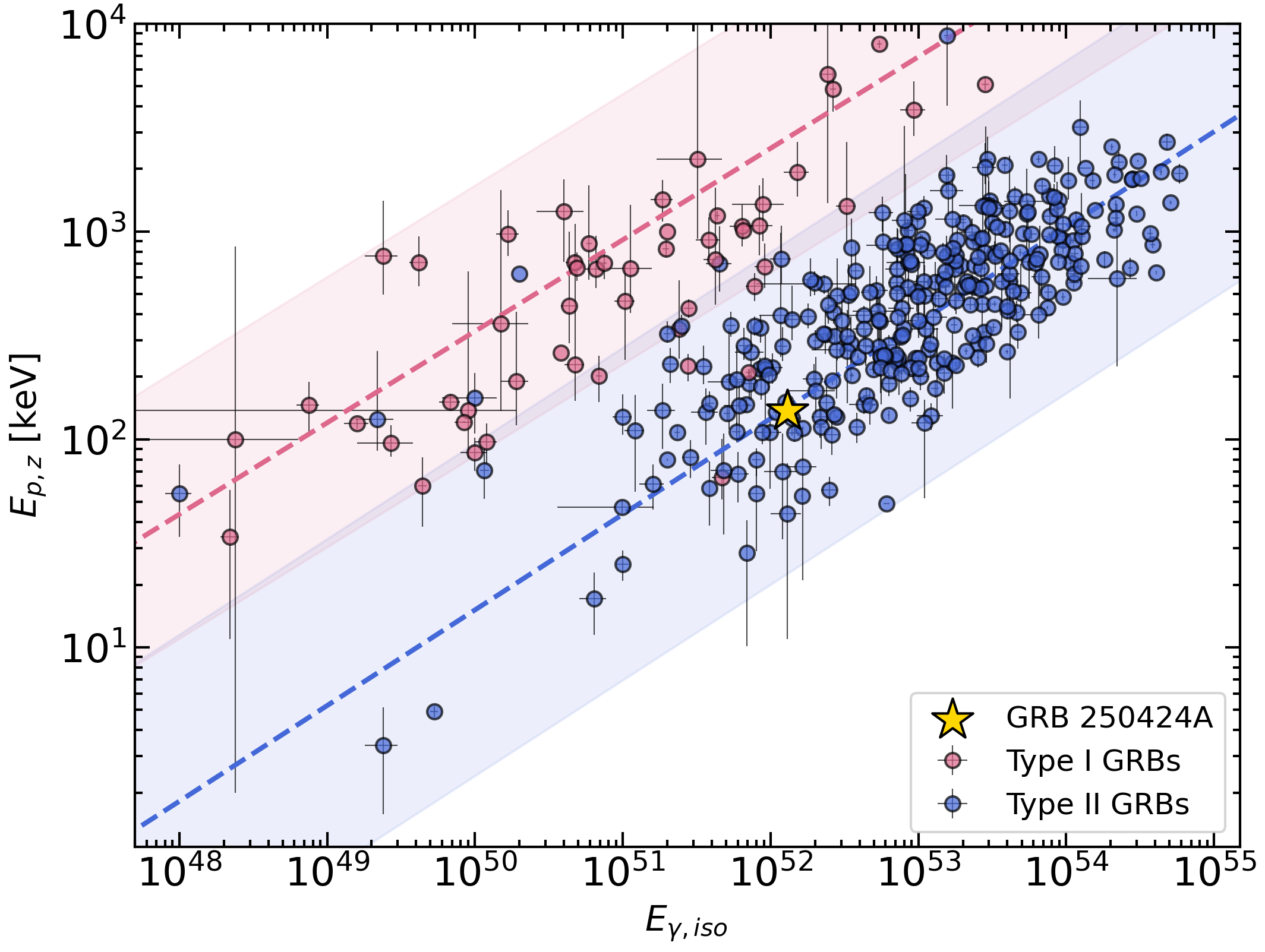}
    \caption{Location of GRB 250424A (golden star) on the $E_{{\rm p},z}$--$E_{\gamma,{\rm iso}}$ plane. The background population of typical Type II (long-duration) GRBs is shown as blue points, while Type I (short-duration) GRBs are indicated by red points (sample adapted from \citealp{2009ZhangB,Minaev2020,minaev2020b}). The shaded regions indicate the $2\sigma$ confidence intervals.}
    \label{fig:amati}
   
\end{figure}

To determine the global energetics, we performed a joint time-integrated spectral analysis using both {\it SVOM}/GRM and {\it Swift}/BAT data over the interval from $ T_0 -30\,s~{\rm to}~+50\,s$. The joint spectrum is best described by the CPL model ($\chi^{2}/{\rm dof} = 1026/636$), yielding a photon index of $\alpha_{\rm CPL} = 1.29^{+0.03}_{-0.01}$ and a cutoff energy of $E_{\rm cut} = 152^{+7.7}_{-9.0}$~keV. The corresponding intrinsic peak energy is $E_{\rm p} = 109^{+4.0}_{-3.0}$~keV. The measured bolometric fluence in the observer-frame 1--10,000\,keV band is $S = 6.8_{-0.7}^{+0.6}\times 10^{-5}~\rm erg~cm^{-2}$. The isotropic-equivalent radiated energy, $E_{\gamma,\rm iso}$, is calculated as
\begin{equation}
E_{\gamma,\rm iso} = \frac{4 \pi D_{L}^{2} K S_{\gamma}}{1+z}\, ,
\end{equation}
where $D_{L}$ is the luminosity distance
and $K$ is the $K$-correction factor \citep{2001Bloom}.
Adopting a spectroscopic redshift of $z = 0.31$ \citep{gcn_redshift}, we derive $E_{\gamma, \rm iso} = 1.7_{-0.2}^{+0.1} \times 10^{52}$~erg. The rest-frame peak energy is 
$E_{{\rm p},z}=E_{\rm p}(1+z)\approx142.8_{-3.9}^{+5.2}$~keV. As shown in Figure ~\ref{fig:amati}, the position of GRB 250424A on the $E_{{\rm p},z}$--$E_{\gamma,{\rm iso}}$ plane is fully consistent with the empirical correlation (the ``Amati relation''; \citealp{2002Amati}) for typical Type II (long) GRBs.

\subsection{Afterglow Temporal Evolution}\label{sec: result_afterg}

\begin{figure*}
    \centering
    \includegraphics[width=0.9\linewidth]{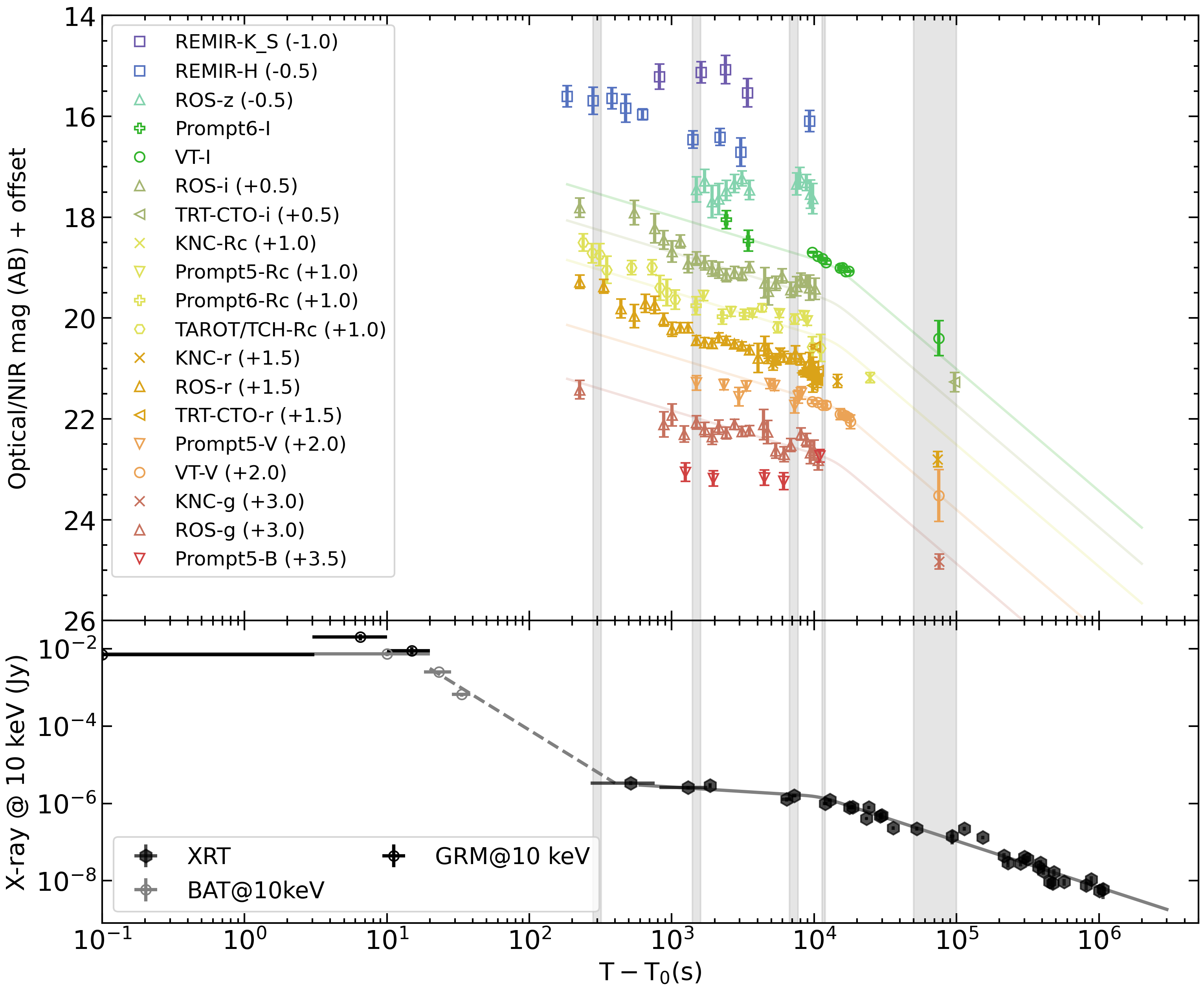}
    \caption{Multiwavelength temporal evolution of GRB~250424A. The upper panel shows optical and NIR light curves obtained from various instruments (distinguished by different symbols) and the corresponding smoothly broken PL fits (colored lines). All magnitudes are reported in the AB system and have been vertically shifted by constant factors for clarity. The lower panel presents the X-ray light curve at 10\,keV, together with a smoothly broken PL fitting shown as solid gray line. The dashed gray line marks the rapid PL decay from the ``tail'' of the prompt phase captured by {\it SVOM}/GRM and {\it Swift}/BAT (open circle). The vertical shaded bands indicate the five epochs selected for the broadband spectral energy distribution (SED) analysis.}
    \label{fig:lc}
\end{figure*}

The afterglow of GRB 250424A was monitored extensively across multiple wavelengths starting 164\,s after the burst, as illustrated in Figure \ref{fig:lc}. {\it Swift}/XRT maintained X-ray coverage until $\sim1.1\times10^6$\,s (12.60\,days). In the optical band, the afterglow was detected by {\it SVOM}/VT up to 2.5\,days post-trigger; although monitoring continued for $\sim 4$ weeks, no source was detected in subsequent epochs, providing deep upper limits at late times.

The light curves in both X-ray and optical/NIR bands exhibit a synchronized evolutionary behavior. A distinct temporal transition is evident at $t \approx 10^{4}$\,s, separating two PL decay regimes.
For the purpose of our analysis, we designate the early-time emission ($t\lesssim 10^4$\,s) as \emph{Phase~A}, characterized by a shallow decay profile. The subsequent epoch ($t\gtrsim 10^4$\,s) is designated as \emph{Phase~B}, during which the decay steepens significantly.

For both bands, we adopt the same smoothly broken power-law (SBPL) function to fit the light curves,
\begin{equation}
F(t) = F_{0}
\left[
\left(\frac{t}{t_b}\right)^{s\alpha_1}
+
\left(\frac{t}{t_b}\right)^{s\alpha_2}
\right]^{-1/s},
\end{equation}
where $F_0$ is the normalization constant, $\alpha_1$ and $\alpha_2$ are respectively the temporal decay indices before (\emph{Phase~A}) and after (\emph{Phase~B}) the break, $t_b$ denotes the break time, and $s$ controls the smoothness of the transition.

For optical/NIR bands, the best-fit result (with $s$ fixed to $0.1$) gives
$\alpha_{o,1} = 0.34 \pm 0.02$ in \emph{Phase~A}, $\alpha_{o,2} = 0.97 \pm 0.08$ in \emph{Phase~B} and $t_{\rm b} = 1.40\,(\pm 0.05)\times10^{4}\,\mathrm{s}$. While the overall trend decays as a power law, variations/fluctuations on short timescales were seen in some of the filters, especially in \emph{Phase~A}, indicating that there is 
likely central-energy activity during this period.

For the X-ray band (also with $s$ fixed to $0.1$), we obtain $\alpha_{\rm X,1} =0.23^{+0.05}_{-0.05}$, $\alpha_{\rm X,2} =1.07^{+0.03}_{-0.02}$, and $t_{b} = 1.07^{+0.11}_{-0.14} \times 10^{4}\ \mathrm{s}$, which is slightly earlier than the optical break time, but within 2$\sigma$ uncertainties.
We note that the very early X-ray emission --- including the ``tail'' of the prompt phase captured by {\it SVOM}/GRM and {\it Swift}/BAT (plotted as open circles in Figure~\ref{fig:lc}) ---displays a rapid decay that is consistent with high-latitude emission marking the cessation of the prompt activity \citep{2000Kumar}. To facilitate comparison, these high-energy data were extrapolated to the 10 keV flux density using the time-averaged prompt spectral index. Since this component is physically distinct from the external forward shock, it is excluded from the afterglow fitting procedure described above.

\subsection{Afterglow Spectral Evolution}\label{sec:spectra_evolution}

\begin{figure}[ht]
    \centering
    \includegraphics[width=0.98\linewidth]{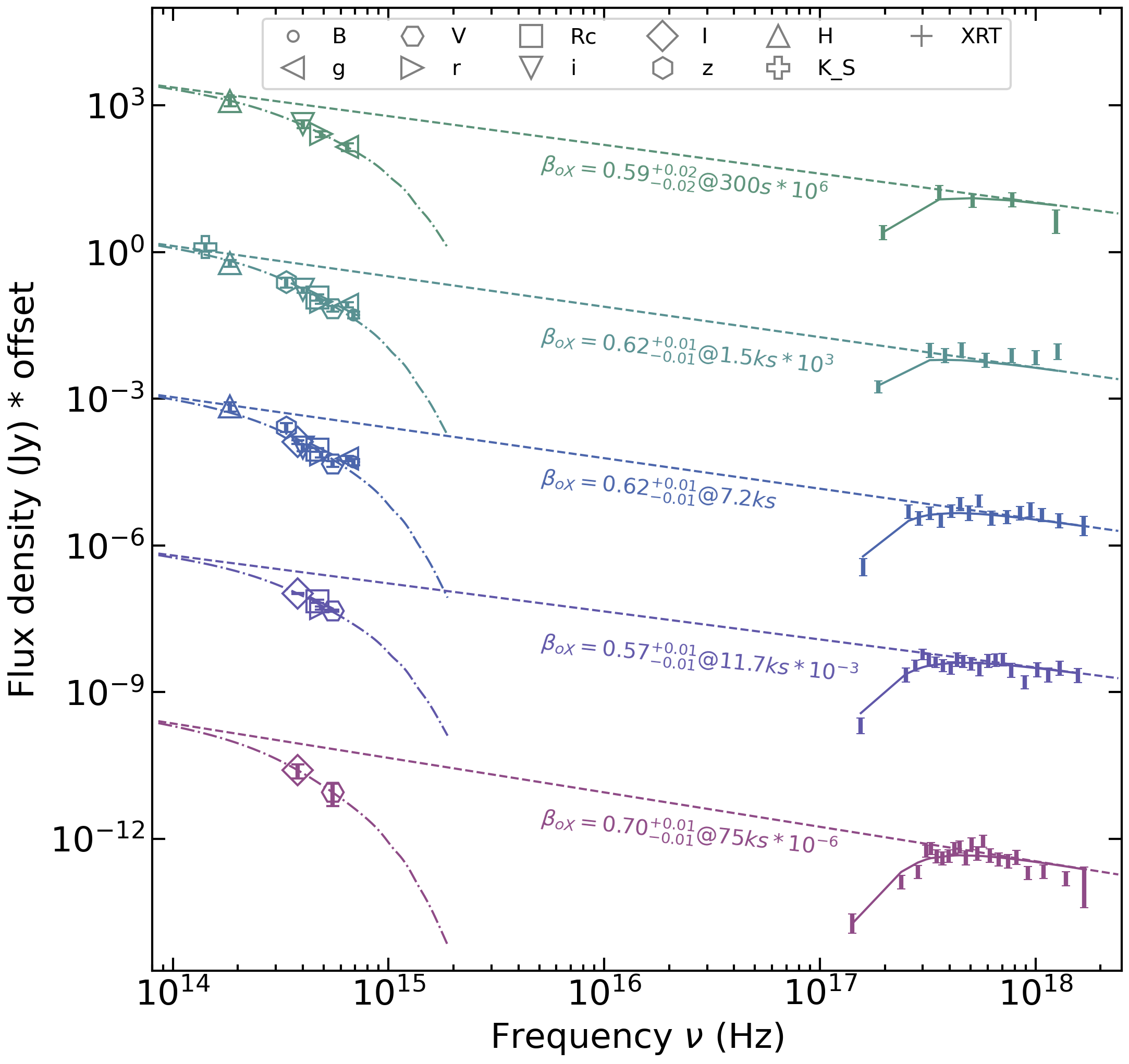}
    \caption{Evolution of the broadband SEDs of GRB 250424A. The SEDs were constructed at five representative epochs combining optical/NIR photometry and X-ray spectral data. For visual clarity, the fluxes at different epochs have been scaled by arbitrary constant factors. Optical data points in different filters are shown with different symbols. Solid and dash-dotted lines represent the best-fit single PL models modified by Galactic and host-galaxy extinction/absorption, while dashed lines indicate the corresponding intrinsic PL spectra. The derived optical-to-X-ray spectral index $\beta_{\rm OX}$ for each epoch is labeled next to the corresponding spectrum.}
    \label{fig:sed}
\end{figure}

To characterize the broadband spectral properties of the afterglow, we analyzed the SEDs at five different epochs ($300\,\mathrm{s}$, $1.5\,\mathrm{ks}$, $7.2\,\mathrm{ks}$, $11.7\,\mathrm{ks}$, and $75\,\mathrm{ks}$; see Figure~\ref{fig:sed}). These epochs were selected to sample the spectral behavior across the distinct evolutionary phases identified in the light curves: the shallow decay phase (\emph{Phase~A}; first three epochs), the transition region (epoch 4, at $t \approx 11$ ks), and the steep decay phase (\emph{Phase~B}; epoch 5).

The SEDs were generated following the methodology described by \citet{Huyd2021} and \citet{Caballero2022}, together with the analysis threads provided by the {\it Swift} team\footnote{\url{https://swift.gsfc.nasa.gov/analysis/threads/uvot_thread_spectra.html}}. For each epoch, we extracted the time-resolved 0.3--10\,keV XRT spectrum and combined it with the multiband optical/NIR photometry closest in time. 
In the joint optical-to-X-ray fitting, the intrinsic spectrum was modeled with a power law, while \texttt{tbabs} and \texttt{ztbabs} account for the X-ray absorption, and \texttt{zdust} describes the optical extinction, using the extinction module\footnote{\url{https://hetools.xyz/Extinction}} provided by \texttt{HEtools} \citep{Wangyun2023}.
In our fitting procedure, the Galactic hydrogen column density was fixed at $N_{\rm H}^{\rm Gal}=6.41\times10^{20}$~cm$^{-2}$ \citep{2016HI4PI}, the Galactic reddening was fixed at $E(B-V)_{\rm Gal}=0.0552$\,mag \citep{Schlafly&Finkbeiner2011}, and the redshift was fixed at $z=0.31$. The intrinsic absorption $N_{\rm H,\rm int}$, host-galaxy extinction $E(B-V)_{\rm host}$, and synchrotron spectral index $\beta$ were left as free parameters. 

Our analysis reveals a nonnegligible amount of extinction within the host-galaxy environment. Across the five epochs, the derived $\rm E(B-V)_{ host}$ ranges from 0.35 to 0.51\,mag, indicating a moderately high extinction from the host galaxy. 
Since the dust content along the line of sight is not expected to evolve over the timescale of the afterglow, we adopt the value derived from the 7.2\,ks epoch --- which benefits from the most comprehensive multiband coverage --- as the representative extinction for the host galaxy with $\rm E(B-V)_{ host}=0.47$\,mag. This value is applied to correct for extinction and absorption in all subsequent analyses. Similarly, we find (from the same epoch) an intrinsic X-ray absorption column density of $N_{\rm H,\rm int} \approx 7.6\times 10^{21}~\rm cm^{-2}$, which is considered to be moderately high compared to a GRB sample (see Figure 3 of \citealp{Zheng09}), consistent with the moderately high extinction found in the host galaxy.

After correcting for absorption and extinction, the broadband SEDs at all epochs are well described by a single PL function extending from optical to X-ray frequencies. We find no evidence of spectral curvature or a cooling break between these bands. The derived spectral indices show mild evolution over time. During \emph{Phase~A} (before and around the break), the typical optical-to-X-ray spectral index is $\beta_{\rm OX} \approx 0.6$ (epochs 1--4). In the post-break \emph{Phase~B} (epoch 5), the spectrum softens slightly to $\beta_{\rm OX} = 0.7^{+0.01}_{-0.01}$. The result of $\beta_{\rm OX} = 0.6$--0.7 is commonly seen in GRB afterglows \cite{Zheng09}, though it tends to the lower value end (optically gray).
The consistency of the single PL fit suggests that both the optical and X-ray emission originate from the same segment of the synchrotron spectrum

\subsection{Closure Relations and Physical Interpretation}\label{sec:closure}
We interpret the multiwavelength evolution of GRB 250424A within the standard synchrotron external-shock framework \citep{Sari1998,Granot&Sari2002}. In this scenario, a relativistic blast wave decelerates into the circumburst medium (CSM), accelerating electrons into a PL energy distribution $N(\gamma_e) \propto \gamma_e^{-p}$. The resulting temporal decay index $\alpha$ and spectral index $\beta$ are dictated by the hydrodynamics of the shock (e.g., constant energy vs. energy injection), the density profile of the CSM (constant-density ISM vs. $r^{-2}$ wind), and the spectral regime relative to the characteristic frequencies ($\nu_m$ and $\nu_c$).

We first examine the late-time evolution (\emph{Phase~B}; $t \gtrsim 10^4$\,s), where the light curves exhibit a ``standard'' decay behavior. During this phase, the X-ray and optical bands decay with $\alpha_B \approx 1.0$, while the optical-to-X-ray spectral index measured from the late-time (\emph{Phase B}) SED is $\beta_{\rm OX} \approx 0.7$. We test the standard closure relations for an adiabatic forward shock with constant energy ($E_{\rm K} = \rm{const}$). For a constant-density ISM environment in the slow-cooling regime ($\nu_m < \nu_{\rm O} < \nu_{\rm X} < \nu_c$), the predicted closure relations are $\alpha = 3\beta/2$ or theoretically $\alpha = 3(p-1)/4$ and $\beta = (p-1)/2$ \citep{ZhangB2006}. Using the observed $\beta_{\rm OX} \approx 0.7$, we infer an electron spectral index of $p = 2\beta + 1 \approx 2.4$, which is consistent with expected values between 2 and 3 for standard GRB afterglows \citep{Piran2004,Zhang&Meszaros2004}. The corresponding predicted temporal decay is $\alpha = 3(2.4-1)/4 = 1.05$. This value is in excellent agreement with the observed decay rate ($\alpha_{\rm O}=0.97^{+0.08}_{-0.08}$ and $\alpha_{\rm X}=1.07^{+0.03}_{-0.02}$). In contrast, a wind-medium environment would predict a much steeper decay ($\alpha = (3p-1)/4 \approx 1.55$) for the same spectral index, which is inconsistent with our data. Thus, we conclude that the late-time afterglow is produced by a standard forward shock propagating into a constant-density ISM.

Turning to the early-time evolution (\emph{Phase~A}; $t \lesssim 10^4$\,s), the shallow decay ($\alpha_{\rm A} \approx 0.3$) deviates significantly from the standard adiabatic prediction. However, the spectral index remains stable at $\beta_{\rm OX} \approx 0.6$, corresponding to $p = 2\beta + 1 \approx 2.2$. 
This shallow decay phase is naturally explained by continuous energy injection into the blast wave. 
Assuming a luminosity injection profile of $L_{\rm inj}(t) \propto A_{\rm inj}t^{-q}$, the observed shallow decay can be interpreted as the result of continuous energy supply to the forward shock.
In this energy-injection framework (ISM case, $\nu_m < \nu < \nu_c$), the closure relation is modified to $\alpha = (q-1) + (2+q)\beta/2$  \citep{ZhangB2006}. Substituting the observed values $\alpha \approx 0.3$ and $\beta \approx 0.6$, we derive an injection index of $q\approx 0.5$. This indicates a sustained energy supply ($L \propto t^{-0.5}$) that significantly refreshes the forward shock and slows down the temporal decay. This injection phase lasts until the break time $t_b \approx 10^4$\,s, after which the energy supply likely ceases or becomes negligible, and the system transitions to the standard decay phase (\emph{Phase~B}).

During transforming from \emph{Phase~A} to \emph{Phase~B}, the afterglow only displays slight change in spectral indices (from $\beta_{A} \approx 0.6$ to from $\beta_{B} \approx 0.7$), and the similarity of the inferred electron indices ($p=2.4$ to $p=2.2$) suggests that the emission in both phases originates from the same synchrotron component generated by a forward shock in the same spectral regime ($\nu_m < \nu < \nu_c$), except that the energy injection ceased after \emph{Phase~B}.

\section{NUMERICAL MODELING AND PHYSICAL PARAMETERS}
\label{sec:model}

\subsection{Modeling Methodology and Global Constraints}
To rigorously characterize the afterglow evolution and constrain the energy injection process, we modeled the broadband data using the \texttt{ASGARD} code \citep{asgard}. This code numerically solves the relativistic hydrodynamics of the blast wave and the time-dependent continuity equation for the electron energy distribution. It self-consistently calculates the synchrotron and synchrotron self-Compton (SSC) emission, accounting for nonlinear effects such as synchrotron self-absorption (SSA) and electron-positron pair production ($\gamma\gamma$ annihilation).

\begin{figure}
    \centering
    \includegraphics[width=0.99\linewidth]{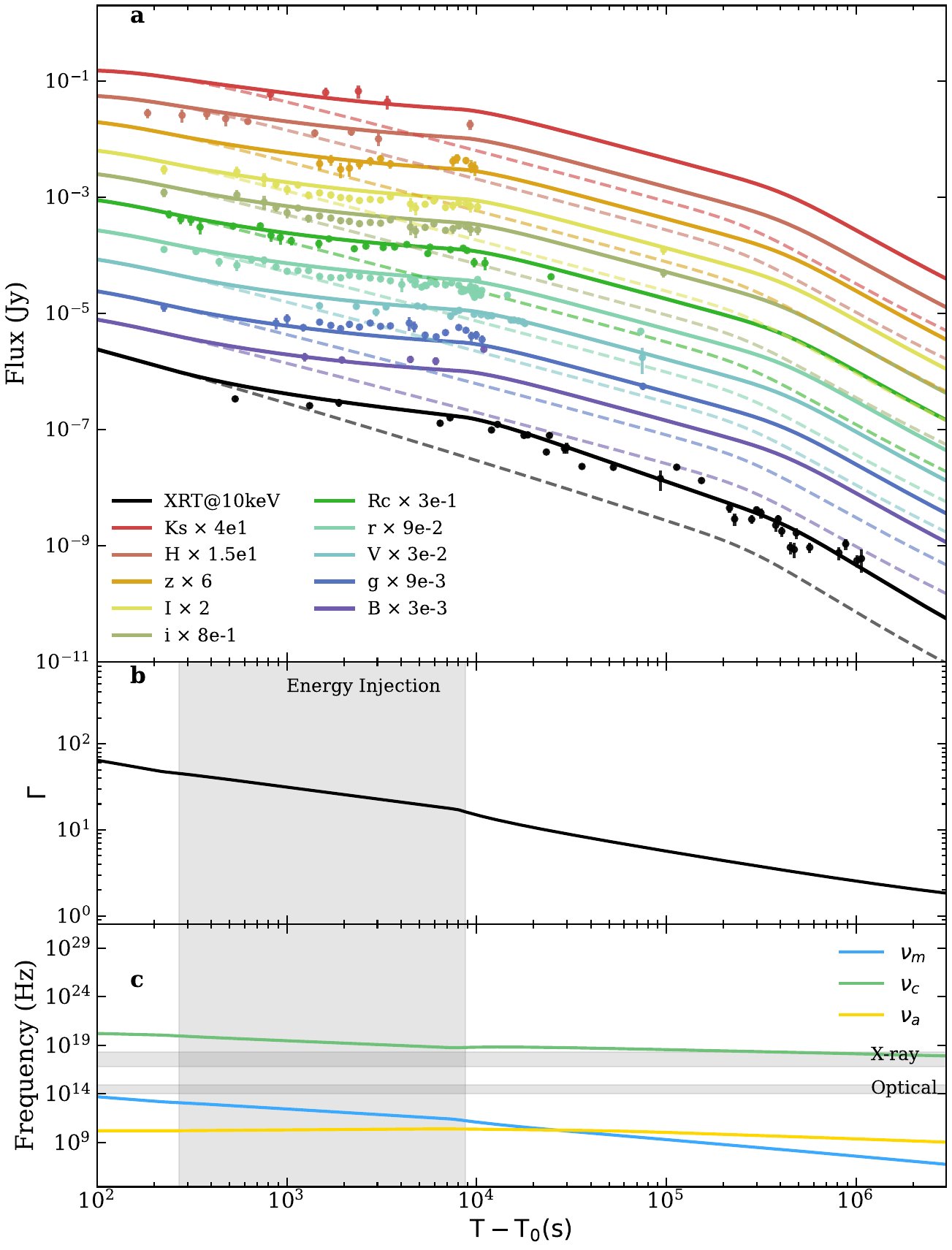}
    \caption{Afterglow model fitting and dynamical evolution with energy injection.
    \textbf{(a)} Multiband afterglow light curves. Solid curves show the best-fit total model including energy injection, with different colors corresponding to different observing bands. Data points denote the observed fluxes. Dashed curves represent the model without energy injection for comparison.
    \textbf{(b)} Temporal evolution of the bulk Lorentz factor $\Gamma$ of the jet.
    \textbf{(c)} Temporal evolution of the characteristic synchrotron break frequencies, including the self-absorption frequency $\nu_a$, the injection frequency $\nu_m$, and the cooling frequency $\nu_c$.
    The horizontal shaded bands mark the observing frequency ranges of the optical bands ($B$--$Ks$) and the X-ray band (0.3--10\,keV). 
    The vertical shaded regions in panels \textbf{b} and \textbf{c}  indicate the time intervals during which energy injection is active.}
    \label{fig:model}
\end{figure}

\begin{figure*}[ht]
    \centering
    \includegraphics[width=0.98\linewidth]{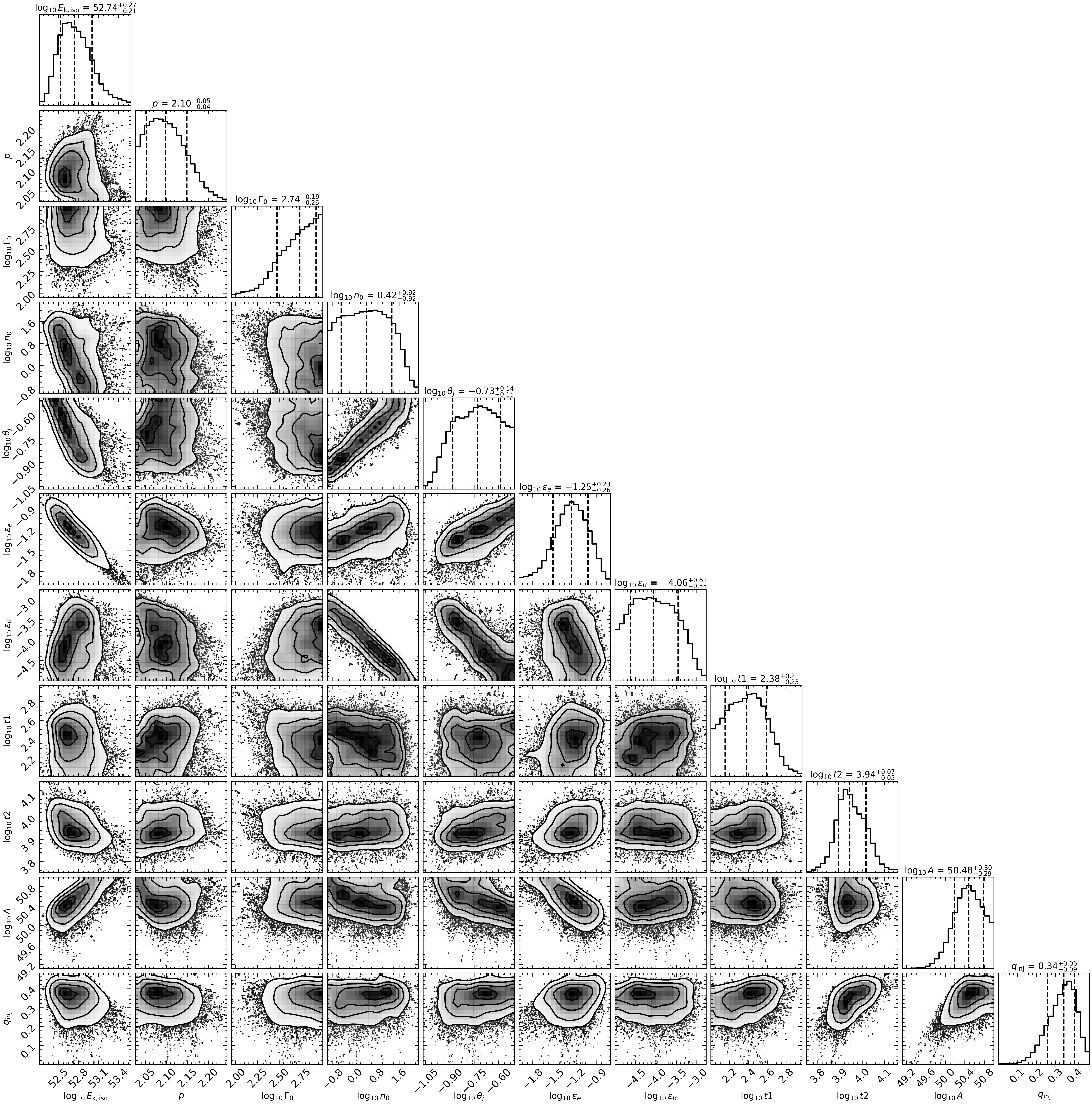}
    \caption{Corner plot showing the posterior distributions of the model parameters inferred from the MCMC fitting to the multiband afterglow data using the forward-shock plus energy-injection model. }
    \label{fig:corner}
\end{figure*}

We adopt a standard external forward-shock model, assuming a uniform (top-hat) jet viewed on-axis in a constant-density medium. The prior parameter ranges used in the Markov Chain Monte Carlo (MCMC) fitting are summarized in Table~\ref{tab:placeholder}.
The parameter space was explored with 30,000 steps to ensure convergence. The resulting best-fit light curves are presented in Figure~\ref{fig:model}, and the posterior distributions are displayed in the corner plot (Figure~\ref{fig:corner}). The median values and $1\sigma$ credible intervals of the inferred parameters are summarized in Table~\ref{tab:placeholder}.

The modeling yields an initial kinetic energy of the blast wave of $E_{\rm K,iso} \approx 5.5 \times 10^{52}$ erg. This value is comparable to the prompt gamma-ray energy release ($E_{\gamma, \rm iso} \approx 2 \times 10^{52}$ erg), implying a radiative efficiency of $\eta = E_\gamma / (E_\gamma + E_{\rm K,iso}) \approx 27\%$, lies within the typical range inferred for long GRBs from broadband afterglow modeling (e.g., \citealp[]{Zhang2007,Beniamini2015,Beniamini2016}). The value indicates a reasonably efficient conversion of the outflow energy into prompt gamma-ray emission. The initial bulk Lorentz factor is constrained to $\rm log_{10}\Gamma_0=2.74_{-0.26}^{+0.19}$, corresponding to $\Gamma_0\approx550$ with a 68\% credible interval of [302,851]. This value falls within the range commonly inferred for long GRBs (e.g., \citealp[]{Molinari2007,Liangew2010,Ghirlanda2018}), although it lies toward the upper end of the observed distribution. It rules out extremely high Lorentz factors ($\gtrsim 1000$) at a 68\% confidence level.

Regarding the jet geometry, no achromatic jet break is detected within the observational window.
The late-time light curve exhibits a possible steepening around $t_{\rm break} \approx $$6 \times 10^5$\,s, which could be indicative of a jet break. However, the current data are too sparse to robustly constrain the break time, and no firm conclusion can be drawn.
Consequently, the jet half-opening angle is only weakly constrained; the value $\theta_j \approx 0.19$ rad ($\approx 11^\circ$) should be regarded as a representative fit rather than a well-determined parameter. This corresponds to a collimation-corrected kinetic energy of $E_{\rm K} \sim 10^{51}$ erg, which lies well within the typical range for long GRBs and does not indicate an exceptionally energetic event.

The CSM is best described by a constant-density profile with $n_0 \approx 2.6$ cm$^{-3}$. This value is consistent with the expected origin of a massive-star progenitor. The shock microphysics is well constrained, with an electron energy fraction $\epsilon_e \approx 0.06$ and a magnetic energy fraction $\epsilon_B \approx 10^{-4}$. These values fall within the typical range inferred for GRB afterglows \citep{Panaitescu2002}. Although the small value of $\epsilon_B$ formally implies a large Compton parameter, the synchrotron self-Compton component does not significantly affect the observed bands. This is likely because the SSC emission peaks at higher energies and may also be partially suppressed by Klein–Nishina effects.

\subsection{Energy Injection Properties}\label{sec:discu_model}

The parameters governing the energy injection phase are tightly constrained by the duration and slope of the shallow-decay phase. Our modeling indicates that the injection commences at $t_1 \approx 2.0\times10^{2}$\,s and terminates at $t_2 \approx 8.1\times10^{3}$\,s. This interval closely matches the temporal extent of ``\emph{Phase~A}'' identified in the observed optical and X-ray light curves. 
The inferred injection energy is $E_{\rm inj}\approx 10^{53}$ erg. After applying a beaming correction, the injected energy remains relatively large, though its absolute value is subject to model uncertainties.
The temporal evolution of the injected power follows a relatively shallow decay, with a power-law index of $q_{\rm inj} \approx 0.34 $, implying a sustained but gradually declining energy supply to the forward shock spanning nearly two decades in time.

Such a prolonged energy injection profile is inconsistent with an impulsive energy release ($E = \rm{const}$). Instead, it strongly favors a scenario in which the blast wave is continuously ``refreshed.'' Physically, this behavior may arise from long-lived central-engine activity (e.g., a magnetar or fallback accretion) or from a stratified ejecta structure where trailing, slower shells catch up with the decelerating forward shock \citep{Dai&Lu1998,Ree1998,Zhang&Meszaros2001}. We note that the injection index derived from the numerical modeling ($q_{\rm inj} \approx 0.34$) is broadly consistent with that inferred from the closure relations in Section 3.4 ($q_{\rm inj} \approx 0.5$). This consistency indicates that the observed shallow-decay phase can be self-consistently explained within the external shock framework with energy injection.

In addition to the global smooth evolution, small-amplitude fluctuations are observed during \emph{Phase~A}, particularly in the optical bands. These short-timescale features are not reproduced by the smooth external-shock model and likely reflect residual central-engine intermittency or density inhomogeneities in the CSM. However, since these fluctuations do not significantly affect the overall energetics or the bulk temporal evolution of the afterglow, we do not model them explicitly in this work.

\begin{table*}
    \centering
    \caption{Best-fitting parameters obtained by the MCMC method for the energy injection model.}
    \begin{tabular}{ccccc}
    \hline\hline
     Parameter &  Parameter Bound & Posterior Value \\ \hline
     $\rm log_{10}(E_{K,iso}/\rm erg)$&  [50, 54] &  $52.74^{+0.27}_{-0.21}$ \\
     $p$  &  [1.9, 2.4] & $2.10^{+0.05}_{-0.04}$ \\ 
     $\rm log_{10}(\Gamma_0)$ & [1.0, 3.0]&  $2.74^{+0.19}_{-0.26}$\\ 
     $\rm log_{10}(n_0/\rm cm^{-3})$ & [-3.0, 3.0] & $0.42^{+0.92}_{-0.92}$ \\ 
     $\rm log_{10}(\epsilon_e)$  & [-3.0, -0.1]& $-1.25^{+0.23}_{-0.26}$ \\ 
     $\rm log_{10}(\epsilon_B)$  & [-4.5, -0.1]& $-4.06^{+0.61}_{-0.55}$\\  
     $\rm log_{10}(\theta_j/\rm rad)$  & [-2.5, -0.1]& $-0.73^{+0.14}_{-0.15}$\\ 
     $\rm log_{10}(t_1/\rm s)$  & [2.0, 3.5] & $2.38^{+0.21}_{-0.23}$\\ 
     $\rm log_{10}(t_2/\rm s)$  & [3.5, 5.0] & $3.94^{+0.07}_{-0.05}$\\ 
     $\rm log_{10}(A_{inj}/\rm erg\,s^{-1})$ & [48.0, 51.0] & $50.48^{+0.30}_{-0.29}$\\ 
     $\rm log_{10}(q_{inj})$  & [0.0, 2.0] & $0.34^{+0.06}_{-0.09}$\\ \hline 
    \end{tabular}
    \\
    \vspace{2mm}
    {\footnotesize
    \textit{Note.} $ A_{\rm inj}$ is the normalization constant in the energy injection model $ L(t)=A_{\rm inj} t^{q_{\rm inj}}$.
    }
    \label{tab:placeholder}
 
\end{table*}

\subsection{Implications of Broadband Spectral and Dynamical Evolution}\label{sec:discu_sed}

The broadband SED analysis presented in Section~\ref{sec:spectra_evolution} serves as a crucial validation of our physical interpretation.  The fact that the optical-to-X-ray spectra at all five representative epochs are well described by a single PL function ($\beta_{\rm OX} \approx 0.6$--0.7) imposes a strict constraint on the spectral regime: the cooling frequency $\nu_c$ must lie above the X-ray band throughout the energy injection time window. This confirms that the optical and X-ray emission originate from the same synchrotron spectral segment ($\nu_m < \nu < \nu_c$), simplifying the modeling by ruling out complex spectral break crossings between the two bands.

A key result of our analysis is the consistency between the analytical and numerical constraints on the energy injection process. The closure-relation analysis (Section \ref{sec:closure}), based on the observed asymptotic decay slopes, suggests an injection index of $q_{\rm inj} \approx 0.5$. In comparison, our detailed numerical modeling (Section \ref{sec:discu_model}) yields a slightly shallower index of $q_{\rm inj} \approx 0.39$. This minor discrepancy is expected: analytical closure relations rely on simplified PL approximations and instantaneous transitions, whereas the \texttt{ASGARD} numerical code self-consistently accounts for the continuous evolution of the blast-wave dynamics, radiative cooling, and non-PL effects near spectral breaks. The broad agreement between these two independent methods reinforces the conclusion that sustained energy injection is the dominant driver of the \emph{Phase~A} plateau.

The dynamical evolution of the blast wave, depicted in the lower panels of
Figure~\ref{fig:model}, provides further insight into this process. The bulk Lorentz factor $\Gamma(t)$ exhibits a modified decay profile: during the injection phase, $\Gamma(t)$ decays more gradually than the standard $\Gamma \propto t^{-3/8}$ solution expected for an adiabatic ISM expansion. This ``flattening'' in the deceleration reflects the continuous replenishment of kinetic energy to the shock. Simultaneously, the characteristic synchrotron frequencies evolve smoothly, maintaining the ordering $\nu_m < \nu_{\rm opt} < \nu_{\rm X} < \nu_c$ throughout \emph{Phase~A}, which naturally explains the stability of the spectral index $\beta_{\rm OX}$.

In summary, combining the microphysical parameters derived from the MCMC fit (Table \ref{tab:placeholder}) with the multiwavelength temporal behavior, we conclude that the early afterglow of GRB~250424A is regulated by a prolonged phase of energy injection ($L \propto t^{-0.4}$), likely arising from long-lived central-engine activity or refreshed shocks. Once this injection ceases, the system transitions seamlessly into a standard energy-conserving forward shock.

\section{Constraints on an Associated Supernova and Host Galaxy}\label{sec:result_sn}
\subsection{Supernova}
The association between long-duration GRBs and broad-lined Type Ic supernovae (SNe Ic-BL) is well established at low redshifts \citep[e.g.,][]{Galama1998Nat,Hjorth2003,Stanek2003}. At $z=0.31$, GRB 250424A is a prime candidate for detecting such an association. An SN comparable in luminosity to the prototype SN 1998bw would be expected to reach a peak apparent magnitude of $V/R/I \approx 22$--23 (assuming negligible extinction). Motivated by this expectation, we conducted a dedicated search for SN signatures in both our photometric and spectroscopic follow-up campaigns.

\begin{figure}[ht]
    \centering
    \includegraphics[width=0.98\linewidth]{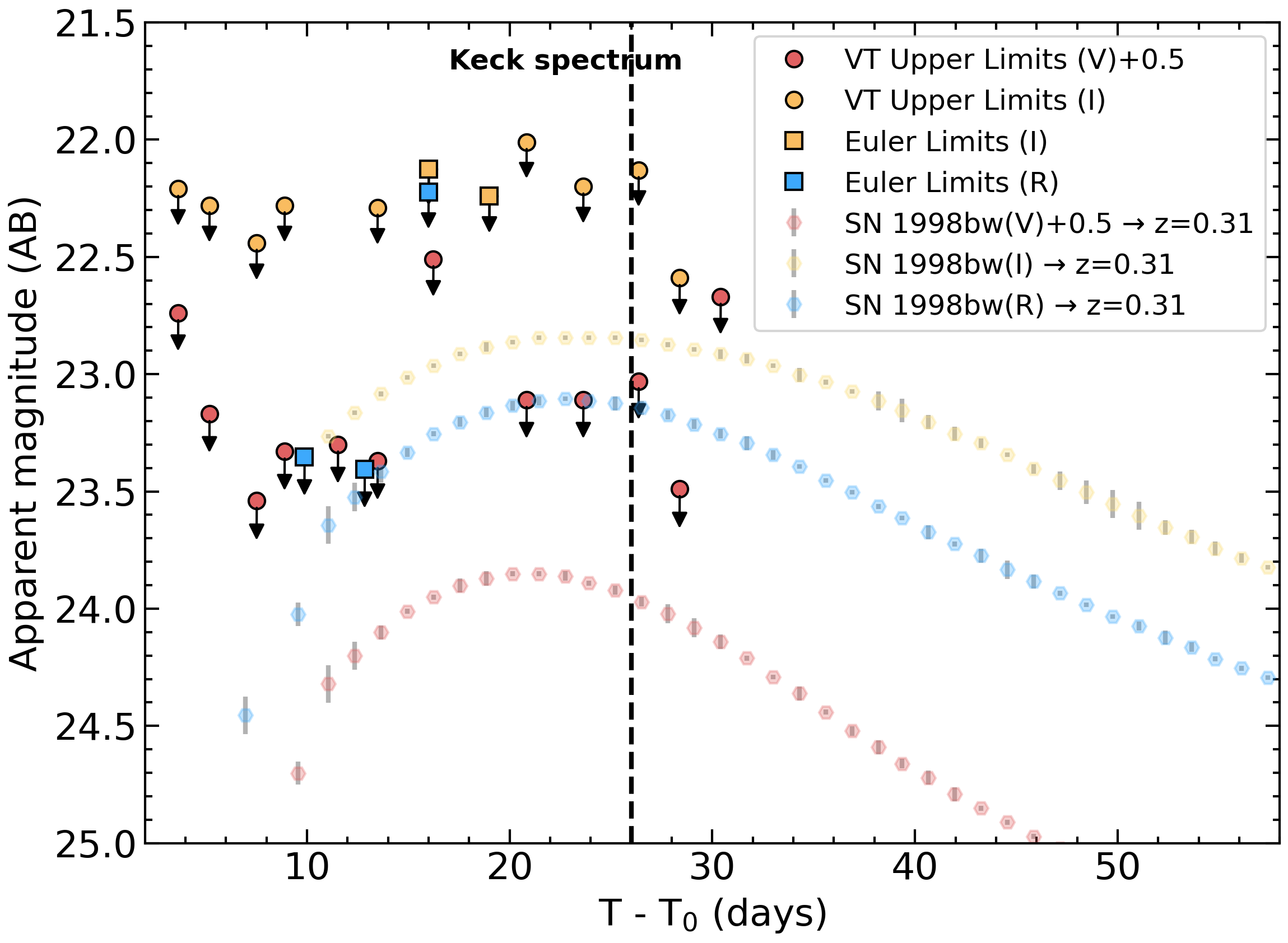}
    \caption{Optical upper limits compared to the expected SN light curves. The plot displays the template light curves of the prototypical Type Ic-BL SN 1998bw, redshifted to $z=0.31$ and attenuated by the host-galaxy extinction derived in Section \ref{sec:discu_sed}. The SN light curves for the \textit{V}, \textit{R}, and \textit{I} bands are represented by red, blue, and yellow hexagons, respectively. Observational upper limits are shown in matching colors, with circles denoting {\it SVOM}/VT data and squares denoting Euler telescope data. The vertical dashed line marks the epoch of the first Keck-I spectroscopic observation ($+26$\,d observer frame).}

    \label{fig:SN_ul}
\end{figure}

The {\it SVOM}/VT monitored the field with a 1--2 day cadence for $\sim 4$ weeks post-burst, covering the expected observer-frame peak time of a potential SN ($\sim 20$ days at $z=0.31$), similar to the peak evolution observed in SN 1998bw-like GRB-associated supernovae \citep{Cano2013,Cano2017}. We performed image subtraction using late-time template images taken 45 days after the burst to remove the host-galaxy contribution. No residual transient source was detected in either the VT$\_V$ or VT$\_I$ bands. The resulting upper limits are presented in Figure \ref{fig:SN_ul}. Complementary observations from the Euler telescope in the \textit{I} and \textit{R} bands also yielded nondetections. As shown in Figure \ref{fig:SN_ul}, our upper limits constrain the optical flux to \textit{V} $\gtrsim$ 22.5--23.0 mag, \textit{I} $\gtrsim$ 22.0--22.5 mag, and \textit{R} $\gtrsim$ 22.0--23.5 mag. These suggest that the SN, if it exists, would be fainter than 23.0 (\textit{I}) mag. 

These nondetections, while seemingly deep, must be interpreted in the context of the substantial host-galaxy extinction derived in Section \ref{sec:discu_sed}. Our SED analysis indicates a host reddening of $E(B-V)_{\rm host} \approx 0.47$\,mag. Assuming a standard Milky Way extinction law ($R_V = 3.1$), this corresponds to a significant visual extinction of $A_V \approx 1.5$\,mag and $A_I \approx 0.9$\,mag. To assess the detectability of a standard SN, we transformed the light curves of SN 1998bw to be at $z=0.31$.  We applied the necessary $K$-corrections, time dilation, and luminosity distance dimming, and crucially, attenuated the model flux by the derived host extinction. The resulting curves are overplotted in Figure \ref{fig:SN_ul}. It is evident that, once corrected for the heavy line-of-sight extinction, the predicted brightness of an SN 1998bw-like event falls below the sensitivity limits of our monitoring campaign in all bands. Therefore, the absence of a photometric SN detection does not imply an intrinsically ``dark'' or SN-less event. Rather, it is consistent with a standard GRB-SN explosion that was obscured by dust within the host environment.

\begin{figure*}
    \centering
    \includegraphics[width=0.9\linewidth]{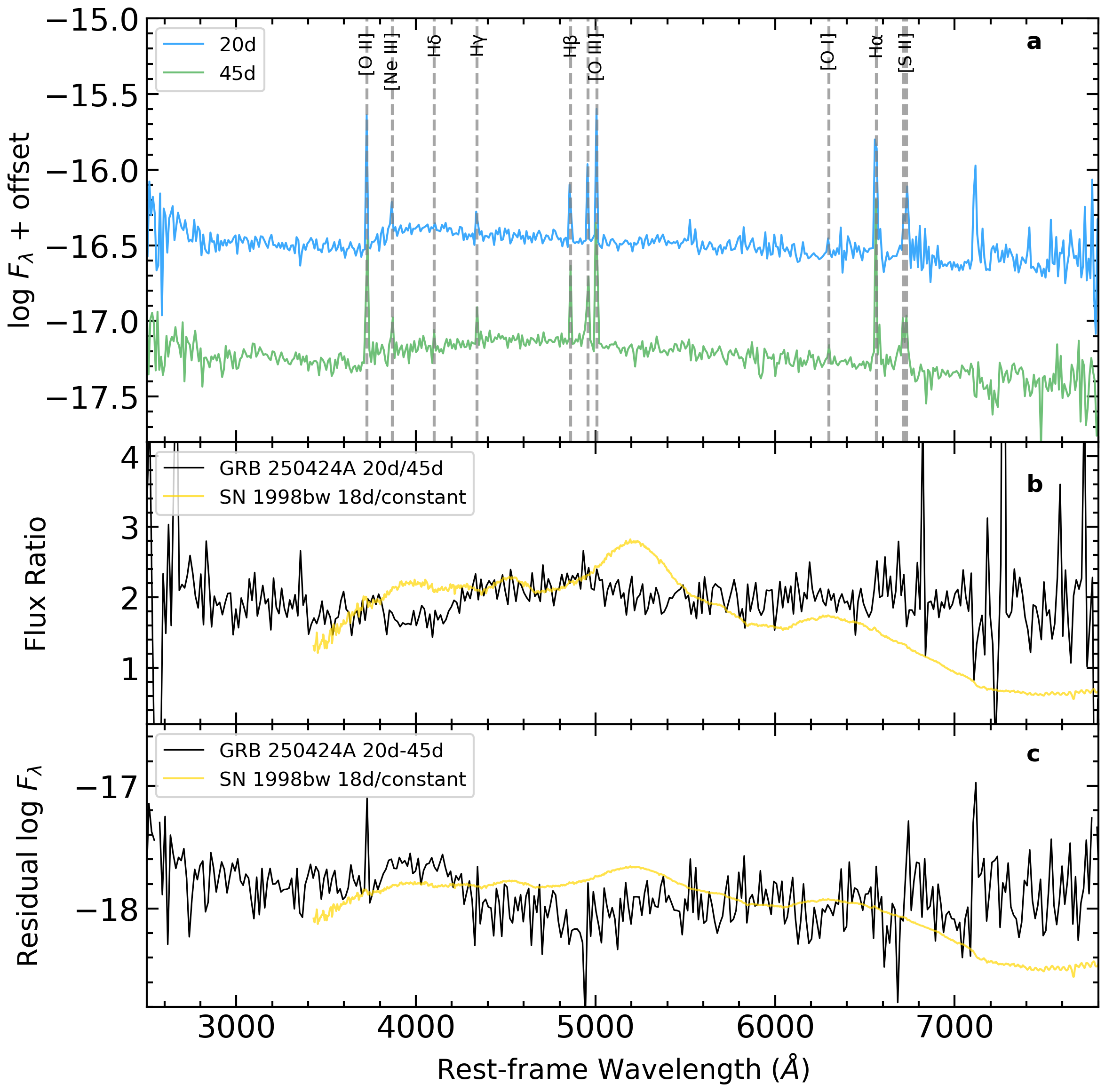}
    \caption{Rest-frame optical spectra of GRB~250424A and a comparison with SN~1998bw. 
    (\textbf{a}) The rebinned (10~\AA) logarithmic flux spectra at +20~d (blue) and +45~d (green) in the rest frame, with the +20~d spectrum vertically offset for clarity. Prominent host-galaxy emission lines are marked.  (\textbf{b}) The ratio spectrum between the +45~d and +20~d observations, together with the scaled spectral shape of SN~1998bw at +18~d in the rest frame. (\textbf{c}) The residual spectrum obtained by subtracting the +45~d spectrum from the +20~d spectrum, compared with the scaled SN~1998bw spectrum. No obvious SN feature are seen in both panels (\textbf{b}) and (\textbf{c}).}
    \label{fig:spectrum_keck&98bw}
\end{figure*}

To probe for SN features more directly, we obtained two epochs of Keck-I/LRIS spectroscopy. The first epoch was acquired 26 days after the burst (20 days in the rest frame), coinciding with the expected peak of a SN 1998bw-like event. As shown in the top panel of Figure~\ref{fig:spectrum_keck&98bw}, the spectrum is dominated by a flat continuum superimposed with strong nebular emission lines from the host galaxy, including [O~II]~$\lambda3727$, [Ne~III]~$\lambda3868$, Balmer lines from H$\alpha$ to H$\delta$, as well as [O~III], [O~I], and [S~II] emission. These lines securely establish the redshift at $z=0.31$, confirming the value reported by \cite{gcn_redshift}.
No clear SN feature can be identified and the spectrum appears to be  dominated by the host-galaxy contribution.

To rigorously search for underlying SN features masked by the host galaxy, we utilized the second-epoch spectrum taken at 59 days ($45$ days in the rest frame). We employed two different methods to explore the possible SN component. First, we divided the first epoch spectrum at 20\,d (rest frame) by the second epoch spectrum at 45\,d. The residual ratio provides a sensitive diagnostic for broad spectral features associated with Type Ic-BL SN by suppressing the smooth afterglow continuum and narrow host-galaxy emission lines.
However, the ratio spectrum remains largely featureless across the observed wavelength range and does not exhibit the broad undulations or absorption structures commonly observed in GRB-associated SN (see  Figure~\ref{fig:spectrum_keck&98bw}b). For comparison, we overlay the spectral shape of SN~1998bw at $\sim 18$ days (rest frame) after explosion, obtained from the WISeREP~\footnote{https://www.wiserep.org/object/933} database, which exhibits prominent broad features characteristic of a luminous Type Ic-BL SN.
No analogous structures are present in the spectrum of GRB~250424A, indicating the absence of detectable spectroscopic signatures of an accompanying SN.

Second, we subtract the 45\,d spectrum from the 20\,d spectrum, assuming the 20\,d spectrum has SN contribution while 45\,d spectrum is purely host-galaxy spectrum. However, the subtracted residual spectrum remains largely featureless (see  Figure~\ref{fig:spectrum_keck&98bw}c), similar to the ratio spectrum, and does not exhibit the broad undulations  compared to  SN~1998bw.
Since both methods show no features from an SN~Ic-BL, we conclude that we did not detect the SN component even in our first-epoch spectrum at 20\,d.

Deeper and more extended late-time observations would be required to either detect an SN component or place meaningful upper limits on its peak luminosity. Within the sensitivity of our data, the absence of a detected SN is still consistent with a typical GRB–SN association, and does not indicate any deviation from the canonical collapsar scenario.

\subsection{Host Galaxy}\label{sec:hostgalaxy}

\begin{figure*}[ht]
    \centering
    \includegraphics[width=0.9\linewidth]{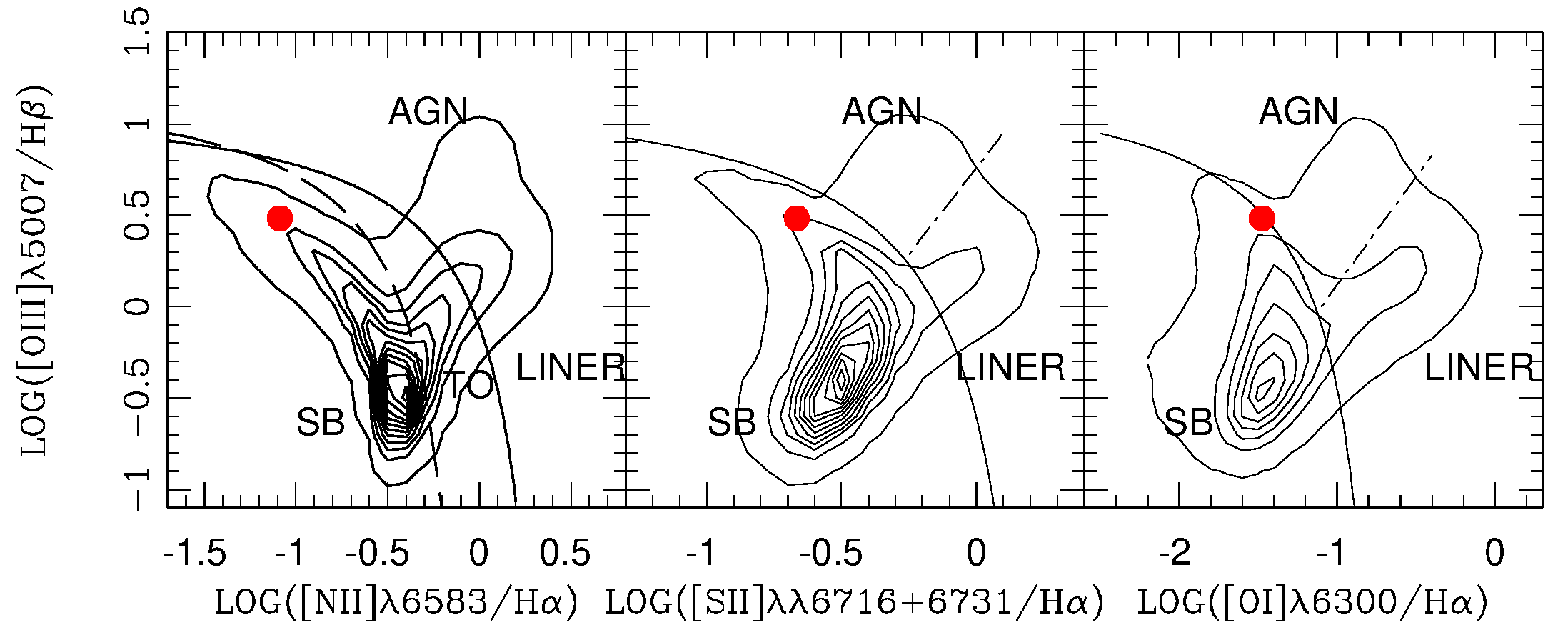}
    \caption{BPT diagnostic diagrams for the host galaxy of GRB~250424A (the red point). The underlying density contours denote a typical distribution of the narrow-line galaxies studied by \citet{Heckman2004} and \citet{Kauffmann2003}). Only the galaxies with signal-to-noise ratio $>$20 and emission lines detected with at least 3$\sigma$ significance are plotted. The theoretical demarcation lines separating active galactic nuclei (AGNs) from star-forming galaxies proposed by \citet{Kewley2001} are presented by the solid lines. The long-dashed line in the left panel marks the empirical demarcation line separating ``pure'' star-forming galaxies \citep{Kauffmann2003}. The dashed-dotted lines in the middle and right panels are the empirical demarcation lines separating Seyfert galaxies and Low-Ionization Nuclear Emission-line Regions(LINERs) \citep{Kewley2006}. The acronyms SB and TO represent Star burst, Transition Objects.} 
    \label{fig:host_BPT}
\end{figure*}

The host galaxy has catalog magnitudes of $g=22.60$, $r=21.98$, $i=22.05$, and $z=21.70$ from the Legacy Surveys DR10. We also measured the host-galaxy photometry from GRANDMA template images, yielding magnitudes of SDSS-$g=22.61\pm0.17$ (AB), SDSS-$r=22.13\pm0.10$ (AB), SDSS-$i=22.41\pm0.15$ (AB),
Bessel-$R=21.78\pm0.04$ (Vega), Bessel-$B=23.29\pm0.12$ (Vega), and
Bessel-$I=21.76\pm0.08$ (Vega).

In order to study the properties of the host galaxy, the Keck-I spectra 
were analyzed by following our previous standard routines. At the beginning,
the continuum was modeled by a starlight component that consists of a
linear combination of the first seven eigenspectra that are built through the principal-component analysis method \citep{Francis1992,Haol2005,Wang&wei2008} from the standard single stellar
population spectral library developed by \citep{Bruzual2003}, along with an intrinsic extinction described by the Milky Way extinction 
curve with $R_V=3.1$  \citep{Cardelli1989}. After removing the modeled continuum, 
we modeled each line profile with a 
Gaussain function by the IRAF/SPECFIT task\footnote{IRAF is distributed by NOAO, which is operated by AURA, Inc., under
cooperative agreement with the U.S. National Science Foundation (NSF)} \citep{Kriss1994}. The measured line fluxes
are given in Table \ref{tab:line}, where the reported uncertainties
correspond to the 1$\sigma$ significance level caused
by the fitting, rather than by the removal of the stellar continuum.

Figure \ref{fig:host_BPT} shows the location of the host galaxy of GRB~250424A
in the traditional Baldwin–Phillips–Terlevich (BPT) diagnostic diagram (e.g.,
\citep{Veilleux&Osterbrock1987}). One can see that the host galaxy of GRB~250424A islocated in the star-forming galaxy region in all three panels. 
The electron density of the gas in the host is calculated to be  $\mathrm{110\ cm^{-3}}$
according to the [\ion{S}{2}] doublet ratio. An intermediate strong intrinsic extinction 
of $A_V = 0.27$\,mag\footnote{This extinction value is significantly lower than the value we derived from the afterglow SED fitting in Section \ref{sec:spectra_evolution}. A possible explanation is that the afterglow SED fitting method is performed directly to the afterglow itself specifically at the GRB location, which is slightly away from the host-galaxy center, while the $\mathrm{H\alpha/H\beta}$ method is derived from the host spectrum averaged from the whole galaxy.} is obtained by the flux ratio
$\mathrm{H\alpha/H\beta}$ by assuming the Balmer decrement of standard Case B
recombination and a Galactic extinction curve with $R_V = 3.1$. After a 
correction of this intrinsic extinction, the gas metallicity is estimated to be 
$\mathrm{12+\log(O/H) = 7.58}$ by the well-documented $R_{23}$ calibration \citep{Maiolino2008}. Adopting the solar value of $\mathrm{12+\log(O/H) = 8.69}$
yields a value of $Z\approx0.07\,Z_\odot$ when the solar abundance ratio is assumed. This places the host at the low-metallicity end of the observed long-GRB host population, consistent with the tendency of long GRBs to occur preferentially in metal-poor star-forming environments (e.g., \citealp[]{Levesque2010,Graham2023}).

With the estimated metallicity, the calibration given by \citep{Kewley2004}, 
\begin{equation}
  \mathrm{SFR} = 7.9\bigg(\frac{L_{\mathrm{[O~II], 42}}}{16.73-1.75[12+\log\mathrm{O/H}]}\bigg)\,\mathrm{M_{\odot}yr^{-1}}\, ,  
\end{equation}
returns a current star-formation rate (SFR) of $1.6\ \mathrm{M_\odot\ yr^{-1}}$, where $L_{\mathrm{[O~II]},42}$ is the luminosity of [\ion{O}{2}] $\lambda$3727 line in units of $10^{42}\ \mathrm{erg\ s^{-1}}$.

\begin{table}
        \centering
        \caption{Results of Spectral Analysis}
        \label{tab:line}
        \begin{tabular}{cc} 
        \hline
        \hline
        Line & Flux \\
             & ($10^{-16}\ \mathrm{erg\ s^{-1}\ cm^{-2}}$)  \\
        (1) & (2) \\
        \hline
        $\mathrm{[O~II]~\lambda3727}$   &  $6.15\pm0.08$ \\
        $\mathrm{[Ne~III]~\lambda3869}$ &  $0.57\pm0.07$ \\
        $\mathrm{H\epsilon}$ &  $0.34\pm0.06$ \\
        $\mathrm{[Ne~III]~\lambda3869}$ &  $0.42\pm0.08$ \\
        $\mathrm{H\delta}$ &  $0.56\pm0.07$ \\
        $\mathrm{H\gamma}$ &  $0.96\pm0.05$ \\
        $\mathrm{H\beta}$ &  $2.78\pm0.06$ \\
        $\mathrm{[O~III]~\lambda5007}$ &  $8.45\pm0.06$ \\
        $\mathrm{[O~I]~\lambda6300}$ &  $0.29\pm0.06$ \\
        $\mathrm{H\alpha}$ &  $8.74\pm0.06$ \\
        $\mathrm{[N~II]~\lambda6583}$ &  $0.72\pm0.06$ \\
        $\mathrm{[S~II]~\lambda6716}$ &  $1.07\pm0.06$ \\
        $\mathrm{[S~II]~\lambda6731}$ &  $0.82\pm0.05$ \\
        \hline
        \end{tabular}

        \label{tab:properties}
\end{table}

\section{Summary and Conclusions}\label{sec: summary}

We have presented a comprehensive multiwavelength analysis of the long-duration GRB 250424A, a strong burst detected by {\it Swift} and {\it SVOM} and rapidly followed up by a broad suite of ground-based facilities. Our dataset spans from the prompt gamma-ray emission to late-time optical spectroscopy. By combining high-cadence light curves with broadband SEDs and performing detailed numerical modeling, we have constrained the intrinsic properties of the relativistic jet and probed the nature of its central engine. Our main conclusions are summarized as follows.

{\bf i. Anatomy of the Plateau:} The afterglow light curves exhibit a pronounced, achromatic shallow-decay phase (\emph{Phase~A}; $t \lesssim 10^4$\,s) simultaneously in the X-ray and optical bands. This is followed by a distinct transition to a standard decay regime (\emph{Phase~B}; $t \gtrsim 10^4$\,s). Broadband SED analysis confirms that the optical and X-ray emission originate from the same synchrotron spectral segment ($\nu_m < \nu_{\rm opt} < \nu_{\rm X} < \nu_c$) throughout the evolution, ruling out spectral break crossings as the cause of the temporal break.

{\bf ii. Evidence for Sustained Energy Injection:} We modeled the afterglow dynamics using the \texttt{ASGARD} numerical code, which self-consistently evolves the electron energy distribution under radiative cooling. The observed shallow decay is naturally reproduced by a prolonged phase of energy injection into the forward shock, lasting from $\sim 10^{2.31}$\,s to $\sim 10^{3.91}$\,s. The inferred injection luminosity follows a PL decay with an index of $q_{\rm inj} \approx 0.4$, implying a central engine (or stratified ejecta) that provides a sustained energy supply to the blast wave. This numerical result is in excellent agreement with the independent constraints derived from analytical closure relations ($q_{\rm inj} \approx 0.5$).

{\bf iii. Jet Properties and Environment}: The late-time evolution is consistent with a standard forward shock propagating into a constant-density interstellar medium (ISM) with $n_0 \approx 2.6$ cm$^{-3}$. The inferred isotropic-equivalent kinetic energy is $E_{\rm K,iso} \approx 5.5 \times 10^{52}$ erg, comparable to the radiated gamma-ray energy, indicating a high radiative efficiency. No jet break is detected within the observational window, yielding a lower limit on the jet opening angle of $\theta_j \gtrsim 0.19$ rad ($\sim 11^\circ$). This implies a large collimation-corrected energy reservoir ($E_{\rm K} \approx 10^{51}$\,erg), classifying GRB 250424A as a highly energetic event.

{\bf iv. The Hidden Supernova:} Despite the low redshift ($z=0.31$) and deep late-time monitoring, no SN component was detected in either photometric imaging or Keck spectroscopy. However, our SED analysis reveals significant dust extinction within the host galaxy ($A_V \approx 1.5$\,mag). We demonstrate that a prototypical SN 1998bw-like event, if subjected to this level of extinction, would fall below our detection limits. Thus, the nondetection is consistent with the standard collapsar model, highlighting the critical role of local environment obscuration in GRB-SN studies.

In summary, GRB 250424A serves as a ``textbook'' example of an energy-injection-dominated afterglow. The high-quality data from {\it SVOM}, {\it Swift}, and complementary ground-based telescopes have allowed us to isolate the effects of central-engine activity from geometric and environmental factors. As {\it SVOM} continues its operations, the detection and detailed characterization of such events will be pivotal in mapping the diversity of central-engine lifetimes and the connection between GRBs and SNe in the local universe.

\begin{acknowledgments}

We thank the anonymous referee for helpful suggestions and comments.

We thank JJG, QLW and BW for helpful discussions.

The {\it Space-based multiband Variable Objects Monitor (SVOM)} is a joint Chinese-French mission led by the Chinese National Space Administration (CNSA), the French Space Agency (CNES), and the Chinese Academy of Sciences (CAS). We gratefully acknowledge the unwavering support of NSSC, IAMCAS, XIOPM, NAOC, IHEP, CNES, CEA, and CNRS.

This work is supported by the following funding programs:
the National Natural Science Foundation of China (grants 12393813, 12373052, 12494572, 12321003),
the National Key R\&D Program of China (2024YFA1611703), 
the Strategic Priority Research Program of the Chinese Academy of Sciences (grant 
XDB0550300), 
and the CAS Project for Young Scientists in
Basic Research (grant YSBR-063).
A.V.F.'s group at U.C.  Berkeley is grateful for financial assistance from Gary and Cynthia Bengier, Clark and Sharon Winslow, Alan Eustace and Kathy Kwan (W.Z. is a Bengier-Winslow-Eustace 
Specialist in Astronomy), and many other donors.
R.B. acknowledges funding from the Italian Space Agency, contract ASI/INAF no. I/004/11/6. Y.W. is supported by the Jiangsu Funding Program for Excellent Postdoctoral Talent (grant No. 2024ZB110), the Postdoctoral Fellowship Program (grant No. GZC20241916) and the General Fund (grant No. 2024M763531) of the China Postdoctoral Science Foundation. H.Z. is supported by Basic Research Program of Jiangsu (No. BK20251707), and the Postdoctoral Innovation Talents Support Program (No. BX20250159). Y.H. issupported by the National Key R\&D Program of China (2021Y FA0718500), and the Xinjiang Tianchi Program.

This work is also supported by the National Key R\&D Program of China (grant No. 2024YFA1611600, 2024YFA161170* and 2024YFA1611700), the Strategic Priority Research Program of the Chinese Academy of Sciences (NFSC, grant No. XDB0550401), and the National Natural Science Foundation of China (NSFC, grant No. 12494571 and 12494570, 12494573, 12133003). This research was partly supported by Natural Science Foundation of Xinjiang Uygur Autonomous Region (grant No. 2024D01D32), Tianshan Talent Training Program (grant No. 2023TSYCLJ0053), and Tianshan Innovation Team Program (grant No. 2024D14015). The authors are thankful for support from the Strategic Priority Research Program of the Chinese Academy of Sciences(Grant No.XDB0550401, XDB0550100).

We acknowledge the following observational support for this work: TRT data were based on observations made with the Thai Robotic Telescopes (under program IDs TRTC12B\_007 and TRTC12A\_003), which are operated by the National Astronomical Research Institute of Thailand (Public Organization).
Some of the data presented herein were obtained at the W. M. Keck
Observatory, which is operated as a scientific partnership among the
California Institute of Technology, the University of California, and
NASA; the observatory was made possible by the generous financial
support of the W. M. Keck Foundation.

\end{acknowledgments}

\vspace{5mm}

\software{Astropy~\citep{astropy:2013, astropy:2018, astropy:2022}; IRAF~\citep{Tody1986,Tody1993}}

\bibliography{0424.bib}{}
\bibliographystyle{aasjournal}

\appendix

\input{tboptical_data}

\end{sloppypar}
\end{document}

%% file: tboptical_data.tex
\begin{longtable}{ccccc}
\caption{Optical photometries.}\label{tab:optical_data}\\ 

\hline\hline
T$_m$-T$_0$ & Exposure & Instrument & Band & Mag \\
(sec) & (s) & & & (AB) \\
\hline\hline
\endfirsthead

\hline\hline
T$_m$-T$_0$ & Exposure & Instrument & Band & Mag \\
(sec) & (s) & & & (AB) \\
\hline\hline
\endhead

\hline
\endfoot

\hline
\endlastfoot

185 & 10 & REMIR & $H$ & 16.10 $\pm$ 0.21 \\ 
225 & 10 & ROS & $r$ & 17.78 $\pm$ 0.13 \\ 
225 & 10 & ROS & $i$ & 17.31 $\pm$ 0.19 \\ 
225 & 10 & ROS & $g$ & 18.42 $\pm$ 0.18 \\ 
239 & 30 & TAROT/TCH & $Rc$ & 17.50 $\pm$ 0.17 \\ 
276 & 30 & TAROT/TCH & $Rc$ & 17.71 $\pm$ 0.19 \\ 
281 & 10 & REMIR & $H$ & 16.19 $\pm$ 0.27 \\ 
312 & 30 & TAROT/TCH & $Rc$ & 17.74 $\pm$ 0.22 \\ 
332 & 10 & ROS & $r$ & 17.87 $\pm$ 0.14 \\ 
349 & 30 & TAROT/TCH & $Rc$ & 18.04 $\pm$ 0.27 \\ 
379 & 10 & REMIR & $H$ & 16.14 $\pm$ 0.21 \\ 
439 & 10 & ROS & $r$ & 18.31 $\pm$ 0.19 \\ 
476 & 10 & REMIR & $H$ & 16.34 $\pm$ 0.28 \\ 
492 & 10 & ROS & $z$ & $>17.2$ \\ 
521 & 90 & TAROT/TCH & $Rc$ & 18.00 $\pm$ 0.14 \\ 
546 & 10 & ROS & $r$ & 18.46 $\pm$ 0.23 \\ 
546 & 10 & ROS & $i$ & 17.41 $\pm$ 0.24 \\ 
623 & 30 & REMIR & $H$ & 16.46 $\pm$ 0.11 \\ 
653 & 10 & ROS & $r$ & 18.21 $\pm$ 0.18 \\ 
725 & 90 & TAROT/TCH & $Rc$ & 17.99 $\pm$ 0.15 \\ 
759 & 10 & ROS & $r$ & 18.24 $\pm$ 0.17 \\ 
759 & 10 & ROS & $i$ & 17.72 $\pm$ 0.29 \\ 
820 & 30 & REMIR & $K_S$ & 16.21 $\pm$ 0.25 \\ 
824 & 90 & TAROT/TCH & $Rc$ & 18.40 $\pm$ 0.24 \\ 
877 & 30 & ROS & $r$ & 18.53 $\pm$ 0.13 \\ 
877 & 30 & ROS & $i$ & 17.95 $\pm$ 0.18 \\ 
877 & 30 & ROS & $g$ & 19.11 $\pm$ 0.25 \\ 
924 & 90 & TAROT/TCH & $Rc$ & 18.49 $\pm$ 0.26 \\ 
1003 & 30 & ROS & $g$ & 18.92 $\pm$ 0.22 \\ 
1003 & 30 & ROS & $r$ & 18.72 $\pm$ 0.14 \\ 
1003 & 30 & ROS & $i$ & 18.18 $\pm$ 0.21 \\ 
1014 & 30 & REMIR & $J$ & $>17.0$ \\ 
1055 & 180 & TAROT/TCH & $Rc$ & 18.63 $\pm$ 0.19 \\ 
1144 & 60 & ROS & $r$ & 18.69 $\pm$ 0.10 \\ 
1144 & 60 & ROS & $i$ & 17.98 $\pm$ 0.13 \\ 
1222 & 60 & ROS & $g$ & 19.30 $\pm$ 0.16 \\ 
1244 & \nodata & Prompt5 & $B$ & 19.56 $\pm$ 0.18 \\ 
1302 & 60 & ROS & $r$ & 18.69 $\pm$ 0.10 \\ 
1302 & 60 & ROS & $i$ & 18.42 $\pm$ 0.18 \\ 
1404 & 30 & REMIR & $H$ & 16.96 $\pm$ 0.17 \\ 
1475 & \nodata & Prompt6 & $Rc$ & 18.76 $\pm$ 0.18 \\ 
1487 & \nodata & Prompt5 & $V$ & 19.29 $\pm$ 0.14 \\ 
1488 & 120 & ROS & $r$ & 18.94 $\pm$ 0.09 \\ 
1488 & 120 & ROS & $g$ & 19.07 $\pm$ 0.13 \\ 
1488 & 120 & ROS & $z$ & 17.95 $\pm$ 0.25 \\ 
1488 & 120 & ROS & $i$ & 18.32 $\pm$ 0.13 \\ 
1602 & 30 & REMIR & $K_S$ & 16.13 $\pm$ 0.21 \\ 
1669 & \nodata & Prompt5 & $Rc$ & 18.55 $\pm$ 0.10 \\ 
1705 & 120 & ROS & $z$ & 17.78 $\pm$ 0.23 \\ 
1705 & 120 & ROS & $r$ & 18.99 $\pm$ 0.10 \\ 
1705 & 120 & ROS & $i$ & 18.41 $\pm$ 0.14 \\ 
1705 & 120 & ROS & $g$ & 19.21 $\pm$ 0.14 \\ 
1921 & 120 & ROS & $i$ & 18.52 $\pm$ 0.15 \\ 
1921 & 120 & ROS & $z$ & 18.19 $\pm$ 0.32 \\ 
1921 & 120 & ROS & $r$ & 19.00 $\pm$ 0.10 \\ 
1921 & 120 & ROS & $g$ & 19.35 $\pm$ 0.16 \\ 
1950 & \nodata & Prompt5 & $B$ & 19.68 $\pm$ 0.15 \\ 
2138 & 120 & ROS & $i$ & 18.55 $\pm$ 0.16 \\ 
2138 & 120 & ROS & $g$ & 19.16 $\pm$ 0.14 \\ 
2138 & 120 & ROS & $z$ & 18.14 $\pm$ 0.31 \\ 
2138 & 120 & ROS & $r$ & 18.89 $\pm$ 0.10 \\ 
2186 & 30 & REMIR & $H$ & 16.91 $\pm$ 0.17 \\ 
2268 & \nodata & Prompt6 & $Rc$ & 18.98 $\pm$ 0.15 \\ 
2324 & \nodata & Prompt5 & $V$ & 19.32 $\pm$ 0.11 \\ 
2383 & 30 & REMIR & $K_S$ & 16.07 $\pm$ 0.28 \\ 
2415 & 240 & ROS & $z$ & 17.97 $\pm$ 0.20 \\ 
2415 & 240 & ROS & $r$ & 18.94 $\pm$ 0.08 \\ 
2415 & 240 & ROS & $i$ & 18.65 $\pm$ 0.13 \\ 
2415 & 240 & ROS & $g$ & 19.28 $\pm$ 0.12 \\ 
2419 & \nodata & Prompt6 & $I$ & 18.05 $\pm$ 0.18 \\ 
2610 & \nodata & Prompt5 & $Rc$ & 18.87 $\pm$ 0.10 \\ 
2752 & 240 & ROS & $g$ & 19.11 $\pm$ 0.10 \\ 
2752 & 240 & ROS & $i$ & 18.60 $\pm$ 0.12 \\ 
2752 & 240 & ROS & $z$ & 17.83 $\pm$ 0.18 \\ 
2752 & 240 & ROS & $r$ & 19.01 $\pm$ 0.08 \\ 
2951 & \nodata & Prompt5 & $V$ & 19.56 $\pm$ 0.18 \\ 
3043 & 60 & REMIR & $H$ & 17.21 $\pm$ 0.28 \\ 
3119 & 300 & ROS & $z$ & 17.73 $\pm$ 0.15 \\ 
3119 & 300 & ROS & $r$ & 19.05 $\pm$ 0.08 \\ 
3119 & 300 & ROS & $i$ & 18.63 $\pm$ 0.13 \\ 
3119 & 300 & ROS & $g$ & 19.25 $\pm$ 0.10 \\ 
3222 & \nodata & Prompt6 & $Rc$ & 18.93 $\pm$ 0.10 \\ 
3335 & \nodata & Prompt5 & $V$ & 19.35 $\pm$ 0.10 \\ 
3390 & 60 & REMIR & $K_S$ & 16.53 $\pm$ 0.28 \\ 
3436 & \nodata & Prompt6 & $I$ & 18.47 $\pm$ 0.21 \\ 
3515 & 300 & ROS & $i$ & 18.49 $\pm$ 0.11 \\ 
3515 & 300 & ROS & $z$ & 17.96 $\pm$ 0.19 \\ 
3515 & 300 & ROS & $r$ & 19.13 $\pm$ 0.09 \\ 
3515 & 300 & ROS & $g$ & 19.23 $\pm$ 0.10 \\ 
3702 & \nodata & Prompt5 & $Rc$ & 18.91 $\pm$ 0.08 \\ 
3734 & 60 & REMIR & $J$ & $>17.1$ \\ 
4035 & 10 & ROS & $r$ & 19.29 $\pm$ 0.29 \\ 
4288 & 180 & TAROT/TCH & $Rc$ & 18.79 $\pm$ 0.08 \\ 
4420 & 30 & ROS & $g$ & 19.11 $\pm$ 0.30 \\ 
4478 & \nodata & Prompt5 & $B$ & 19.67 $\pm$ 0.15 \\ 
4482 & 30 & ROS & $r$ & 19.05 $\pm$ 0.19 \\ 
4482 & 30 & ROS & $i$ & 18.80 $\pm$ 0.30 \\ 
4688 & 60 & ROS & $g$ & 19.26 $\pm$ 0.23 \\ 
4742 & 300 & KNC & $r$ & 19.30 $\pm$ 0.10 \\ 
4766 & 60 & ROS & $r$ & 19.15 $\pm$ 0.15 \\ 
4766 & 60 & ROS & $i$ & 18.97 $\pm$ 0.27 \\ 
4881 & \nodata & Prompt5 & $V$ & 19.30 $\pm$ 0.10 \\ 
5138 & 300 & KNC & $r$ & 19.40 $\pm$ 0.13 \\ 
5283 & \nodata & Prompt5 & $V$ & 19.34 $\pm$ 0.10 \\ 
5356 & 120 & ROS & $g$ & 19.63 $\pm$ 0.15 \\ 
5356 & 120 & ROS & $i$ & 18.81 $\pm$ 0.14 \\ 
5356 & 120 & ROS & $r$ & 19.32 $\pm$ 0.11 \\ 
5441 & 300 & KNC & $r$ & 19.34 $\pm$ 0.07 \\ 
5533 & 180 & TAROT/TCH & $Rc$ & 19.19 $\pm$ 0.11 \\ 
5685 & \nodata & Prompt5 & $Rc$ & 18.91 $\pm$ 0.07 \\ 
5744 & 300 & KNC & $r$ & 19.18 $\pm$ 0.08 \\ 
5958 & 240 & ROS & $i$ & 18.66 $\pm$ 0.14 \\ 
6095 & \nodata & Prompt5 & $B$ & 19.74 $\pm$ 0.17 \\ 
6125 & 240 & ROS & $r$ & 19.27 $\pm$ 0.11 \\ 
6125 & 240 & ROS & $g$ & 19.70 $\pm$ 0.15 \\ 
6858 & 300 & ROS & $r$ & 19.31 $\pm$ 0.10 \\ 
6858 & 300 & ROS & $i$ & 18.94 $\pm$ 0.15 \\ 
6858 & 300 & ROS & $g$ & 19.52 $\pm$ 0.13 \\ 
7285 & 180 & TAROT/TCH & $Rc$ & 19.02 $\pm$ 0.10 \\ 
7301 & \nodata & Prompt5 & $V$ & 19.73 $\pm$ 0.15 \\ 
7397 & 120 & ROS & $r$ & 19.21 $\pm$ 0.16 \\ 
7505 & 120 & ROS & $i$ & 18.87 $\pm$ 0.19 \\ 
7505 & 120 & ROS & $z$ & 17.84 $\pm$ 0.22 \\ 
7514 & 60 & REMIR & $H$ & $>18.0$ \\ 
7703 & \nodata & Prompt5 & $V$ & 19.56 $\pm$ 0.13 \\ 
7848 & 60 & REMIR & $Ks$ & $>16.8$ \\ 
7891 & 240 & ROS & $z$ & 17.71 $\pm$ 0.20 \\ 
8058 & 240 & ROS & $r$ & 19.32 $\pm$ 0.11 \\ 
8058 & 240 & ROS & $i$ & 18.75 $\pm$ 0.14 \\ 
8058 & 240 & ROS & $g$ & 19.30 $\pm$ 0.12 \\ 
8104 & \nodata & Prompt5 & $V$ & 19.49 $\pm$ 0.12 \\ 
8218 & 60 & REMIR & $J$ & $>17.5$ \\ 
8315 & 200 & TRT-CTO & $r$ & 19.59 $\pm$ 0.04 \\ 
8507 & \nodata & Prompt5 & $Rc$ & 18.95 $\pm$ 0.08 \\ 
8530 & 200 & TRT-CTO & $r$ & 19.58 $\pm$ 0.05 \\ 
8594 & 300 & ROS & $r$ & 19.52 $\pm$ 0.15 \\ 
8742 & 200 & TRT-CTO & $r$ & 19.54 $\pm$ 0.05 \\ 
8792 & 300 & ROS & $g$ & 19.42 $\pm$ 0.13 \\ 
8792 & 300 & ROS & $z$ & 17.81 $\pm$ 0.16 \\ 
8792 & 300 & ROS & $i$ & 18.77 $\pm$ 0.12 \\ 
8910 & \nodata & Prompt5 & $Rc$ & 19.05 $\pm$ 0.09 \\ 
9249 & 30 & REMIR & $H$ & 16.59 $\pm$ 0.21 \\ 
9297 & 120 & ROS & $r$ & 19.37 $\pm$ 0.19 \\ 
9297 & 120 & ROS & $i$ & 18.90 $\pm$ 0.25 \\ 
9340 & 200 & TRT-CTO & $r$ & 19.55 $\pm$ 0.05 \\ 
9405 & 120 & ROS & $z$ & 18.04 $\pm$ 0.28 \\ 
9405 & 120 & ROS & $g$ & 19.67 $\pm$ 0.22 \\ 
9551 & 200 & TRT-CTO & $r$ & 19.68 $\pm$ 0.05 \\ 
9711 & 430 & VT & $I$ & 18.69 $\pm$ 0.01 \\ 
9711 & 430 & VT & $V$ & 19.66 $\pm$ 0.04 \\ 
9714 & 180 & TAROT/TCH & $Rc$ & 19.57 $\pm$ 0.20 \\ 
9756 & 60 & TRT-CTO & $r$ & 19.84 $\pm$ 0.13 \\ 
9791 & 240 & ROS & $r$ & 19.44 $\pm$ 0.16 \\ 
9791 & 240 & ROS & $z$ & 18.13 $\pm$ 0.30 \\ 
9826 & 60 & TRT-CTO & $r$ & 19.46 $\pm$ 0.09 \\ 
9898 & 60 & TRT-CTO & $r$ & 19.45 $\pm$ 0.09 \\ 
9958 & 240 & ROS & $g$ & 19.62 $\pm$ 0.20 \\ 
9968 & 60 & TRT-CTO & $r$ & 19.71 $\pm$ 0.11 \\ 
10039 & 60 & TRT-CTO & $r$ & 19.69 $\pm$ 0.12 \\ 
10127 & 240 & ROS & $i$ & 18.92 $\pm$ 0.21 \\ 
10139 & 120 & TRT-CTO & $r$ & 19.07 $\pm$ 0.04 \\ 
10270 & 120 & TRT-CTO & $r$ & 19.73 $\pm$ 0.08 \\ 
10401 & 120 & TRT-CTO & $r$ & 19.70 $\pm$ 0.08 \\ 
10495 & 300 & ROS & $r$ & 19.70 $\pm$ 0.18 \\ 
10530 & 120 & TRT-CTO & $r$ & 19.73 $\pm$ 0.08 \\ 
10581 & 430 & VT & $I$ & 18.77 $\pm$ 0.01 \\ 
10581 & 430 & VT & $V$ & 19.68 $\pm$ 0.03 \\ 
10662 & 120 & TRT-CTO & $r$ & 19.54 $\pm$ 0.07 \\ 
10693 & 300 & ROS & $g$ & 19.82 $\pm$ 0.19 \\ 
10890 & 300 & ROS & $z$ & $>18.3$ \\ 
10927 & \nodata & Prompt5 & $B$ & 19.23 $\pm$ 0.12 \\ 
11076 & 180 & TAROT/TCH & $Rc$ & 19.59 $\pm$ 0.27 \\ 
11451 & 430 & VT & $I$ & 18.82 $\pm$ 0.01 \\ 
11464 & 430 & VT & $V$ & 19.73 $\pm$ 0.03 \\ 
12082 & 430 & VT & $V$ & 19.72 $\pm$ 0.07 \\ 
12099 & 430 & VT & $I$ & 18.90 $\pm$ 0.03 \\ 
14537 & 120 & KNC & $r$ & 19.75 $\pm$ 0.13 \\ 
15204 & 430 & VT & $V$ & 19.91 $\pm$ 0.11 \\ 
15265 & 430 & VT & $I$ & 19.01 $\pm$ 0.05 \\ 
15775 & 430 & VT & $I$ & 19.00 $\pm$ 0.02 \\ 
15775 & 430 & VT & $V$ & 19.91 $\pm$ 0.04 \\ 
16645 & 430 & VT & $I$ & 19.07 $\pm$ 0.02 \\ 
16645 & 430 & VT & $V$ & 19.94 $\pm$ 0.04 \\ 
17484 & 430 & VT & $V$ & 19.98 $\pm$ 0.04 \\ 
17484 & 430 & VT & $I$ & 19.07 $\pm$ 0.02 \\ 
17994 & 430 & VT & $V$ & 20.05 $\pm$ 0.14 \\ 
24705 & 180 & KNC & $Rc$ & 20.18 $\pm$ 0.10 \\ 
73433 & 300 & KNC & $r$ & 21.30 $\pm$ 0.15 \\ 
74820 & \nodata & VT & $V$ & 21.52 $\pm$ 0.52 \\ 
74820 & \nodata & VT & $I$ & 20.40 $\pm$ 0.35 \\ 
75309 & 300 & KNC & $g$ & 21.83 $\pm$ 0.15 \\ 
96929 & 300 & TRT-CTO & $i$ & 20.77 $\pm$ 0.19 \\ 
177468 & 180 & KNC & $r$ & $>21.5$ \\ 
182772 & 300 & TRT-CTO & $V$ & $>21.8$ \\ 
214955 & \nodata & VT & $V$ & $>23.3$ \\ 
214955 & \nodata & VT & $I$ & $>22.2$ \\ 
237184 & 300 & TRT-CTO & $r$ & $>22.6$ \\ 
239930 & 300 & TRT-CTO & $i$ & $>21.8$ \\ 
314231 & \nodata & VT & $I$ & $>22.2$ \\ 
314231 & \nodata & VT & $V$ & $>22.2$ \\ 
405601 & 600 & TRT-CTO & $r$ & $>22.6$ \\ 
410468 & 300 & TRT-CTO & $V$ & $>22.3$ \\ 
447613 & \nodata & VT & $V$ & $>22.7$ \\ 
447613 & \nodata & VT & $I$ & $>22.3$ \\ 
648930 & \nodata & VT & $V$ & $>23.0$ \\ 
648930 & \nodata & VT & $I$ & $>22.4$ \\ 
768036 & \nodata & VT & $V$ & $>22.8$ \\ 
768036 & \nodata & VT & $I$ & $>22.3$ \\ 
851907 & 180 & Euler & $B$ & $>23.4$ \\ 
853083 & 180 & Euler & $Rc$ & $>23.4$ \\ 
853971 & 180 & Euler & $I$ & $>22.4$ \\ 
995554 & \nodata & VT & $I$ & $>20.9$ \\ 
995554 & \nodata & VT & $V$ & $>22.8$ \\ 
1032266 & 180 & KNC & $g$ & $>22.2$ \\ 
1108431 & 180 & Euler & $B$ & $>22.7$ \\ 
1109599 & 180 & Euler & $Rc$ & $>23.4$ \\ 
1110768 & 180 & Euler & $I$ & $>23.7$ \\ 
1113767 & 300 & KNC & $r$ & $>22.0$ \\ 
1164405 & \nodata & VT & $V$ & $>22.9$ \\ 
1164405 & \nodata & VT & $I$ & $>22.3$ \\ 
1286552 & 180 & KNC & $r$ & $>21.2$ \\ 
1292701 & 11700 & KNC & $r$ & $>21.1$ \\ 
1379980 & 180 & Euler & $B$ & $>21.8$ \\ 
1381150 & 180 & Euler & $Rc$ & $>22.2$ \\ 
1382612 & 540 & Euler & $I$ & $>22.1$ \\ 
1394029 & 24300 & KNC & $r$ & $>21.0$ \\ 
1402122 & \nodata & VT & $V$ & $>22.0$ \\ 
1402122 & \nodata & VT & $I$ & $>21.6$ \\ 
1468402 & 180 & KNC & $r$ & $>20.8$ \\ 
1490737 & \nodata & VT & $I$ & $>21.8$ \\ 
1490737 & \nodata & VT & $V$ & $>22.6$ \\ 
1627213 & 180 & Euler & $B$ & $>21.6$ \\ 
1628755 & 180 & Euler & $Rc$ & $>22.2$ \\ 
1641155 & 180 & Euler & $I$ & $>22.2$ \\ 
1799668 & \nodata & VT & $V$ & $>22.6$ \\ 
1799668 & \nodata & VT & $I$ & $>22.0$ \\ 
1970623 & 180 & KNC & $r$ & $>21.4$ \\ 
2041571 & \nodata & VT & $V$ & $>22.6$ \\ 
2041571 & \nodata & VT & $I$ & $>22.3$ \\ 
2238299 & 180 & KNC & $r$ & $>21.7$ \\ 
2277478 & \nodata & VT & $V$ & $>22.5$ \\ 
2277478 & \nodata & VT & $I$ & $>22.1$ \\ 
2452195 & \nodata & VT & $V$ & $>23.0$ \\ 
2452195 & \nodata & VT & $I$ & $>22.6$ \\ 
2626972 & \nodata & VT & $V$ & $>22.2$ \\ 
2626972 & \nodata & VT & $I$ & $>21.0$ \\

\end{longtable}

\tablecomments{Telescope abbreviations used in this table are defined in Section \ref{sec:opt-obs}.}

%% file: 0424.bib
@ARTICLE{grmgcn,
       author = {{Svom/Grm Team} and {Zhang}, Jin-Peng and {Wang}, Chen-Wei and {Huang}, Yue and {Zheng}, Shi-Jie and {Xiong}, Shao-Lin and {Zhang}, Shuang-Nan and {Svom/Eclairs Team} and {Atteia}, Jean-Luc and {Godet}, Olivier and {Schanne}, St{\'e}phane and {Daigne}, Fr{\'e}d{\'e}ric and {Svom Team:}},
        title = "{GRB 250424A: SVOM/GRM observation of a long burst}",
      journal = {GRB Coordinates Network},
         year = 2025,
        month = apr,
       volume = {40252},
        pages = {1},
       adsurl = {https://ui.adsabs.harvard.edu/abs/2025GCN.40252....1S}
}

@ARTICLE{Oke1983,
       author = {{Oke}, J.~B. and {Gunn}, J.~E.},
        title = "{Secondary standard stars for absolute spectrophotometry.}",
      journal = {\apj},
         year = 1983,
        month = mar,
       volume = {266},
        pages = {713-717},
          doi = {10.1086/160817},
       adsurl = {https://ui.adsabs.harvard.edu/abs/1983ApJ...266..713O}
}

@ARTICLE{swiftgcn,
       author = {{Cenko}, S.~B. and {D'Elia}, V. and {DeLaunay}, J.~J. and {Gupta}, R. and {Melandri}, A. and {Page}, K.~L. and {Siegel}, M.~H. and {Neil Gehrels Swift Observatory Team}},
        title = "{GRB 250424A: Swift detection of a burst with an optical counterpart}",
      journal = {GRB Coordinates Network},
         year = 2025,
        month = apr,
       volume = {40224},
        pages = {1},
       adsurl = {https://ui.adsabs.harvard.edu/abs/2025GCN.40224....1C}
}

@ARTICLE{2005xrt,
       author = {{Burrows}, David N. and {Hill}, J.~E. and {Nousek}, J.~A. and {Kennea}, J.~A. and {Wells}, A. and {Osborne}, J.~P. and {Abbey}, A.~F. and {Beardmore}, A. and {Mukerjee}, K. and {Short}, A.~D.~T. and {Chincarini}, G. and {Campana}, S. and {Citterio}, O. and {Moretti}, A. and {Pagani}, C. and {Tagliaferri}, G. and {Giommi}, P. and {Capalbi}, M. and {Tamburelli}, F. and {Angelini}, L. and {Cusumano}, G. and {Br{\"a}uninger}, H.~W. and {Burkert}, W. and {Hartner}, G.~D.},
        title = "{The Swift X-Ray Telescope}",
      journal = {\ssr},
         year = 2005,
        month = oct,
       volume = {120},
       number = {3-4},
        pages = {165-195},
          doi = {10.1007/s11214-005-5097-2},
archivePrefix = {arXiv},
       eprint = {astro-ph/0508071},
 primaryClass = {astro-ph},
       adsurl = {https://ui.adsabs.harvard.edu/abs/2005SSRv..120..165B}
}

@ARTICLE{2005uvot,
       author = {{Roming}, Peter W.~A. and {Kennedy}, Thomas E. and {Mason}, Keith O. and {Nousek}, John A. and {Ahr}, Lindy and {Bingham}, Richard E. and {Broos}, Patrick S. and {Carter}, Mary J. and {Hancock}, Barry K. and {Huckle}, Howard E. and {Hunsberger}, S.~D. and {Kawakami}, Hajime and {Killough}, Ronnie and {Koch}, T. Scott and {McLelland}, Michael K. and {Smith}, Kelly and {Smith}, Philip J. and {Soto}, Juan Carlos and {Boyd}, Patricia T. and {Breeveld}, Alice A. and {Holland}, Stephen T. and {Ivanushkina}, Mariya and {Pryzby}, Michael S. and {Still}, Martin D. and {Stock}, Joseph},
        title = "{The Swift Ultra-Violet/Optical Telescope}",
      journal = {\ssr},
         year = 2005,
        month = oct,
       volume = {120},
       number = {3-4},
        pages = {95-142},
          doi = {10.1007/s11214-005-5095-4},
archivePrefix = {arXiv},
       eprint = {astro-ph/0507413},
 primaryClass = {astro-ph},
       adsurl = {https://ui.adsabs.harvard.edu/abs/2005SSRv..120...95R}
}

@ARTICLE{svom_wei,
       author = {{Wei}, J. and {Cordier}, B. and {Antier}, S. and {Antilogus}, P. and {Atteia}, J. -L. and {Bajat}, A. and {Basa}, S. and {Beckmann}, V. and {Bernardini}, M.~G. and {Boissier}, S. and {Bouchet}, L. and {Burwitz}, V. and {Claret}, A. and {Dai}, Z. -G. and {Daigne}, F. and {Deng}, J. and {Dornic}, D. and {Feng}, H. and {Foglizzo}, T. and {Gao}, H. and {Gehrels}, N. and {Godet}, O. and {Goldwurm}, A. and {Gonzalez}, F. and {Gosset}, L. and {G{\"o}tz}, D. and {Gouiffes}, C. and {Grise}, F. and {Gros}, A. and {Guilet}, J. and {Han}, X. and {Huang}, M. and {Huang}, Y. -F. and {Jouret}, M. and {Klotz}, A. and {La Marle}, O. and {Lachaud}, C. and {Le Floch}, E. and {Lee}, W. and {Leroy}, N. and {Li}, L. -X. and {Li}, S.~C. and {Li}, Z. and {Liang}, E. -W. and {Lyu}, H. and {Mercier}, K. and {Migliori}, G. and {Mochkovitch}, R. and {O'Brien}, P. and {Osborne}, J. and {Paul}, J. and {Perinati}, E. and {Petitjean}, P. and {Piron}, F. and {Qiu}, Y. and {Rau}, A. and {Rodriguez}, J. and {Schanne}, S. and {Tanvir}, N. and {Vangioni}, E. and {Vergani}, S. and {Wang}, F. -Y. and {Wang}, J. and {Wang}, X. -G. and {Wang}, X. -Y. and {Watson}, A. and {Webb}, N. and {Wei}, J.~J. and {Willingale}, R. and {Wu}, C. and {Wu}, X. -F. and {Xin}, L. -P. and {Xu}, D. and {Yu}, S. and {Yu}, W. -F. and {Yu}, Y. -W. and {Zhang}, B. and {Zhang}, S. -N. and {Zhang}, Y. and {Zhou}, X.~L.},
        title = "{The Deep and Transient Universe in the SVOM Era: New Challenges and Opportunities - Scientific prospects of the SVOM mission}",
      journal = {arXiv e-prints},
         year = 2016,
        month = oct,
          eid = {arXiv:1610.06892},
        pages = {arXiv:1610.06892},
          doi = {10.48550/arXiv.1610.06892},
archivePrefix = {arXiv},
       eprint = {1610.06892},
 primaryClass = {astro-ph.IM},
       adsurl = {https://ui.adsabs.harvard.edu/abs/2016arXiv161006892W}
}

@article{Barthelmy_2005,
   title={The Burst Alert Telescope (BAT) on the SWIFT Midex Mission},
   volume={120},
   ISSN={1572-9672},
   url={http://dx.doi.org/10.1007/s11214-005-5096-3},
   DOI={10.1007/s11214-005-5096-3},
   number={3–4},
   journal={Space Science Reviews},
   publisher={Springer Science and Business Media LLC},
   author={Barthelmy, Scott D. and Barbier, Louis M. and Cummings, Jay R. and Fenimore, Ed E. and Gehrels, Neil and Hullinger, Derek and Krimm, Hans A. and Markwardt, Craig B. and Palmer, David M. and Parsons, Ann and Sato, Goro and Suzuki, Masaya and Takahashi, Tadayuki and Tashiro, Makota and Tueller, Jack},
   year={2005},
   month=oct, pages={143–164} }

@ARTICLE{2001Bloom,
       author = {{Bloom}, Joshua S. and {Frail}, Dale A. and {Sari}, Re'em},
        title = "{The Prompt Energy Release of Gamma-Ray Bursts using a Cosmological k-Correction}",
      journal = {\aj},
         year = 2001,
        month = jun,
       volume = {121},
       number = {6},
        pages = {2879-2888},
          doi = {10.1086/321093},
archivePrefix = {arXiv},
       eprint = {astro-ph/0102371},
 primaryClass = {astro-ph},
       adsurl = {https://ui.adsabs.harvard.edu/abs/2001AJ....121.2879B}
}

@ARTICLE{2002Amati,
       author = {{Amati}, L. and {Frontera}, F. and {Tavani}, M. and {in't Zand}, J.~J.~M. and {Antonelli}, A. and {Costa}, E. and {Feroci}, M. and {Guidorzi}, C. and {Heise}, J. and {Masetti}, N. and {Montanari}, E. and {Nicastro}, L. and {Palazzi}, E. and {Pian}, E. and {Piro}, L. and {Soffitta}, P.},
        title = "{Intrinsic spectra and energetics of BeppoSAX Gamma-Ray Bursts with known redshifts}",
      journal = {\aap},
         year = 2002,
        month = jul,
       volume = {390},
        pages = {81-89},
          doi = {10.1051/0004-6361:20020722},
archivePrefix = {arXiv},
       eprint = {astro-ph/0205230},
 primaryClass = {astro-ph},
       adsurl = {https://ui.adsabs.harvard.edu/abs/2002A&A...390...81A}
}

@article{dutton_skynets_2022,
	title = {Skynet's {New} {Observing} {Mode}: {The} {Campaign} {Manager}},
	volume = {134},
	issn = {0004-6280, 1538-3873},
	shorttitle = {Skynet's {New} {Observing} {Mode}},
	url = {http://arxiv.org/abs/2210.08613},
	doi = {10.1088/1538-3873/ac3f7c},
	number = {1031},
	urldate = {2025-01-08},
	journal = {PASP},
	author = {Dutton, Dylan A. and Reichart, Daniel E. and Haislip, Joshua B. and Kouprianov, Vladimir V. and Shaban, Omar H. and Soto, Alan Vasquez},
	month = jan,
	year = {2022},
	note = {arXiv:2210.08613 [astro-ph]},
	pages = {015001},
}

@article{reichart_prompt_2005,
	title = {{PROMPT}: {Panchromatic} {Robotic} {Optical} {Monitoring} and {Polarimetry} {Telescopes}},
	volume = {28},
	issn = {0390-5551},
	shorttitle = {{PROMPT}},
	url = {https://ui.adsabs.harvard.edu/abs/2005NCimC..28..767R},
	doi = {10.1393/ncc/i2005-10149-6},
	urldate = {2025-02-16},
	journal = {Nuovo Cimento C Geophysics Space Physics C},
	author = {Reichart, D. and Nysewander, M. and Moran, J. and Bartelme, J. and Bayliss, M. and Foster, A. and Clemens, J. C. and Price, P. and Evans, C. and Salmonson, J. and Trammell, S. and Carney, B. and Keohane, J. and Gotwals, R.},
	month = jul,
	year = {2005},
	note = {ADS Bibcode: 2005NCimC..28..767R},
	pages = {767},
}

@article{dutton_grb_2025,
	title = {{GRB} {250424A}: {Skynet} {Optical} {Observations}},
	volume = {40241},
	shorttitle = {{GRB} {250424A}},
	url = {https://ui.adsabs.harvard.edu/abs/2025GCN.40241....1D},
	urldate = {2025-09-29},
	journal = {GRB Coordinates Network},
	author = {Dutton, Dylan and Reichart, Daniel and Haislip, Joshua and Kouprianov, Vladimir and Schlekat, Donovan and {Skynet Robotic Telescope Network}},
	month = apr,
	year = {2025},
	note = {ADS Bibcode: 2025GCN.40241....1D},
	pages = {1},
}

@article{barbary_sep_2016,
	title = {{SEP}: {Source} {Extractor} as a library},
	volume = {1},
	issn = {2475-9066},
	shorttitle = {{SEP}},
	url = {https://joss.theoj.org/papers/10.21105/joss.00058},
	doi = {10.21105/joss.00058},
	language = {en},
	number = {6},
	urldate = {2025-09-29},
	journal = {Journal of Open Source Software},
	author = {Barbary, Kyle},
	month = oct,
	year = {2016},
	pages = {58},
}

@article{bertin_sextractor_1996,
	title = {{SExtractor}: {Software} for source extraction.},
	volume = {117},
	issn = {0365-01380004-6361},
	shorttitle = {{SExtractor}},
	url = {https://ui.adsabs.harvard.edu/abs/1996A&AS..117..393B},
	doi = {10.1051/aas:1996164},
	urldate = {2025-09-29},
	journal = {Astronomy and Astrophysics Supplement Series},
	author = {Bertin, E. and Arnouts, S.},
	month = jun,
	year = {1996},
	note = {Publisher: EDP
ADS Bibcode: 1996A\&AS..117..393B},
	pages = {393--404},
}

@article{onken_skymapper_2024,
	title = {{SkyMapper} {Southern} {Survey}: {Data} release 4},
	volume = {41},
	issn = {1323-3580},
	shorttitle = {{SkyMapper} {Southern} {Survey}},
	url = {https://ui.adsabs.harvard.edu/abs/2024PASA...41...61O},
	doi = {10.1017/pasa.2024.53},
	urldate = {2025-09-29},
	journal = {Publications of the Astronomical Society of Australia},
	author = {Onken, Christopher A. and Wolf, Christian and Bessell, Michael S. and Chang, Seo-Won and Luvaul, Lance C. and Tonry, John L. and White, Marc C. and Da Costa, Gary S.},
	month = oct,
	year = {2024},
	note = {ADS Bibcode: 2024PASA...41...61O},
	pages = {e061},
}

@ARTICLE{remgcn,
       author = {{Brivio}, R. and {Melandri}, A. and {Ferro}, M. and {D'Avanzo}, P. and {Covino}, S. and {Tagliaferri}, G. and {Fugazza}, D. and {REM Team}},
        title = "{GRB 250424A: REM optical/NIR afterglow detection}",
      journal = {GRB Coordinates Network},
         year = 2025,
        month = apr,
       volume = {40225},
        pages = {1},
       adsurl = {https://ui.adsabs.harvard.edu/abs/2025GCN.40225....1B}
}

@ARTICLE{vtgcn,
       author = {{Hu}, Y.~D. and {Zhang}, L. and {Chen}, X.~L. and {Xin}, L.~P. and {Qiu}, Y.~L. and {Li}, H.~L. and {Wu}, C. and {Yao}, Z.~H. and {Han}, X.~H. and {Xu}, Y. and {Wang}, J. and {Zhang}, P.~P. and {Xie}, W.~J. and {Xiao}, Y.~J. and {Cai}, H.~B. and {Lan}, L. and {Deng}, J.~S. and {Wei}, J.~Y. and {SVOM/VT Team}},
        title = "{GRB 250424A: SVOM/VT optical observation}",
      journal = {GRB Coordinates Network},
         year = 2025,
        month = apr,
       volume = {40246},
        pages = {1},
       adsurl = {https://ui.adsabs.harvard.edu/abs/2025GCN.40246....1H}
}

@ARTICLE{2Mass,
       author = {{Skrutskie}, M.~F. and {Cutri}, R.~M. and {Stiening}, R. and {Weinberg}, M.~D. and {Schneider}, S. and {Carpenter}, J.~M. and {Beichman}, C. and {Capps}, R. and {Chester}, T. and {Elias}, J. and {Huchra}, J. and {Liebert}, J. and {Lonsdale}, C. and {Monet}, D.~G. and {Price}, S. and {Seitzer}, P. and {Jarrett}, T. and {Kirkpatrick}, J.~D. and {Gizis}, J.~E. and {Howard}, E. and {Evans}, T. and {Fowler}, J. and {Fullmer}, L. and {Hurt}, R. and {Light}, R. and {Kopan}, E.~L. and {Marsh}, K.~A. and {McCallon}, H.~L. and {Tam}, R. and {Van Dyk}, S. and {Wheelock}, S.},
        title = "{The Two Micron All Sky Survey (2MASS)}",
      journal = {\aj},
         year = 2006,
        month = feb,
       volume = {131},
       number = {2},
        pages = {1163-1183},
          doi = {10.1086/498708},
       adsurl = {https://ui.adsabs.harvard.edu/abs/2006AJ....131.1163S}
}

@ARTICLE{REM_Zerbi01,
       author = {{Zerbi}, R.~M. and {Chincarini}, G. and {Ghisellini}, G. and {Rondon{\'o}}, M. and {Tosti}, G. and {Antonelli}, L.~A. and {Conconi}, P. and {Covino}, S. and {Cutispoto}, G. and {Molinari}, E. and {Nicastro}, L. and {Palazzi}, E. and {Akerlof}, C. and {Burderi}, L. and {Campana}, S. and {Crimi}, G. and {Danzinger}, J. and {di Paola}, A. and {Fernandez-Soto}, A. and {Fiore}, F. and {Frontera}, F. and {Fugazza}, D. and {Gentile}, G. and {Goldoni}, P. and {Israel}, G. and {Jordan}, B. and {Lorenzetti}, D. and {McBreen}, B. and {Martinetti}, E. and {Mazzoleni}, R. and {Masetti}, N. and {Messina}, S. and {Meurs}, E. and {Monfardini}, A. and {Nucciarelli}, G. and {Orlandini}, M. and {Paul}, J. and {Pian}, E. and {Saracco}, P. and {Sardone}, S. and {Stella}, L. and {Tagliaferri}, L. and {Tavani}, M. and {Testa}, V. and {Vitali}, F.},
        title = "{The REM telescope: detecting the near infra-red counterparts of Gamma-Ray Bursts and the prompt behavior of their optical continuum}",
      journal = {Astronomische Nachrichten},
         year = 2001,
        month = dec,
       volume = {322},
        pages = {275-285},
          doi = {10.1002/1521-3994(200112)322:5/6<275::AID-ASNA275>3.0.CO;2-N},
       adsurl = {https://ui.adsabs.harvard.edu/abs/2001AN....322..275Z}
}

@INPROCEEDINGS{REM_Covino+04,
       author = {{Covino}, Stefano and {Stefanon}, Mauro and {Sciuto}, Giorgio and {Fernandez-Soto}, Alberto and {Tosti}, Gino and {Zerbi}, Filippo M. and {Chincarini}, Guido and {Antonelli}, Lucio A. and {Conconi}, Paolo and {Cutispoto}, Giuseppe and {Molinari}, Emilio and {Nicastro}, Luciano and {Rodono}, Marcello},
        title = "{REM: a fully robotic telescope for GRB observations}",
    booktitle = {Ground-based Instrumentation for Astronomy},
         year = 2004,
       editor = {{Moorwood}, Alan F.~M. and {Iye}, Masanori},
       series = {Society of Photo-Optical Instrumentation Engineers (SPIE) Conference Series},
       volume = {5492},
        month = sep,
        pages = {1613-1622},
          doi = {10.1117/12.551532},
       adsurl = {https://ui.adsabs.harvard.edu/abs/2004SPIE.5492.1613C}
}

@ARTICLE{Astroalign,
       author = {{Beroiz}, M. and {Cabral}, J.~B. and {Sanchez}, B.},
        title = "{Astroalign: A Python module for astronomical image registration}",
      journal = {Astronomy and Computing},
         year = 2020,
        month = jul,
       volume = {32},
          eid = {100384},
        pages = {100384},
          doi = {10.1016/j.ascom.2020.100384},
archivePrefix = {arXiv},
       eprint = {1909.02946},
 primaryClass = {astro-ph.IM},
       adsurl = {https://ui.adsabs.harvard.edu/abs/2020A&C....3200384B}
}

@ARTICLE{Gaia_DR3,
       author = {{Gaia Collaboration} and {Vallenari}, A. and {Brown}, A.~G.~A. and {Prusti}, T. and {de Bruijne}, J.~H.~J. and {Arenou}, F. and {Babusiaux}, C. and {Biermann}, M. and {Creevey}, O.~L. and {Ducourant}, C. and {Evans}, D.~W. and {Eyer}, L. and {Guerra}, R. and {Hutton}, A. and {Jordi}, C. and {Klioner}, S.~A. and {Lammers}, U.~L. and {Lindegren}, L. and {Luri}, X. and {Mignard}, F. and {Panem}, C. and {Pourbaix}, D. and {Randich}, S. and {Sartoretti}, P. and {Soubiran}, C. and {Tanga}, P. and {Walton}, N.~A. and {Bailer-Jones}, C.~A.~L. and {Bastian}, U. and {Drimmel}, R. and {Jansen}, F. and {Katz}, D. and {Lattanzi}, M.~G. and {van Leeuwen}, F. and {Bakker}, J. and {Cacciari}, C. and {Casta{\~n}eda}, J. and {De Angeli}, F. and {Fabricius}, C. and {Fouesneau}, M. and {Fr{\'e}mat}, Y. and {Galluccio}, L. and {Guerrier}, A. and {Heiter}, U. and {Masana}, E. and {Messineo}, R. and {Mowlavi}, N. and {Nicolas}, C. and {Nienartowicz}, K. and {Pailler}, F. and {Panuzzo}, P. and {Riclet}, F. and {Roux}, W. and {Seabroke}, G.~M. and {Sordo}, R. and {Th{\'e}venin}, F. and {Gracia-Abril}, G. and {Portell}, J. and {Teyssier}, D. and {Altmann}, M. and {Andrae}, R. and {Audard}, M. and {Bellas-Velidis}, I. and {Benson}, K. and {Berthier}, J. and {Blomme}, R. and {Burgess}, P.~W. and {Busonero}, D. and {Busso}, G. and {C{\'a}novas}, H. and {Carry}, B. and {Cellino}, A. and {Cheek}, N. and {Clementini}, G. and {Damerdji}, Y. and {Davidson}, M. and {de Teodoro}, P. and {Nu{\~n}ez Campos}, M. and {Delchambre}, L. and {Dell'Oro}, A. and {Esquej}, P. and {Fern{\'a}ndez-Hern{\'a}ndez}, J. and {Fraile}, E. and {Garabato}, D. and {Garc{\'\i}a-Lario}, P. and {Gosset}, E. and {Haigron}, R. and {Halbwachs}, J. -L. and {Hambly}, N.~C. and {Harrison}, D.~L. and {Hern{\'a}ndez}, J. and {Hestroffer}, D. and {Hodgkin}, S.~T. and {Holl}, B. and {Jan{\ss}en}, K. and {Jevardat de Fombelle}, G. and {Jordan}, S. and {Krone-Martins}, A. and {Lanzafame}, A.~C. and {L{\"o}ffler}, W. and {Marchal}, O. and {Marrese}, P.~M. and {Moitinho}, A. and {Muinonen}, K. and {Osborne}, P. and {Pancino}, E. and {Pauwels}, T. and {Recio-Blanco}, A. and {Reyl{\'e}}, C. and {Riello}, M. and {Rimoldini}, L. and {Roegiers}, T. and {Rybizki}, J. and {Sarro}, L.~M. and {Siopis}, C. and {Smith}, M. and {Sozzetti}, A. and {Utrilla}, E. and {van Leeuwen}, M. and {Abbas}, U. and {{\'A}brah{\'a}m}, P. and {Abreu Aramburu}, A. and {Aerts}, C. and {Aguado}, J.~J. and {Ajaj}, M. and {Aldea-Montero}, F. and {Altavilla}, G. and {{\'A}lvarez}, M.~A. and {Alves}, J. and {Anders}, F. and {Anderson}, R.~I. and {Anglada Varela}, E. and {Antoja}, T. and {Baines}, D. and {Baker}, S.~G. and {Balaguer-N{\'u}{\~n}ez}, L. and {Balbinot}, E. and {Balog}, Z. and {Barache}, C. and {Barbato}, D. and {Barros}, M. and {Barstow}, M.~A. and {Bartolom{\'e}}, S. and {Bassilana}, J. -L. and {Bauchet}, N. and {Becciani}, U. and {Bellazzini}, M. and {Berihuete}, A. and {Bernet}, M. and {Bertone}, S. and {Bianchi}, L. and {Binnenfeld}, A. and {Blanco-Cuaresma}, S. and {Blazere}, A. and {Boch}, T. and {Bombrun}, A. and {Bossini}, D. and {Bouquillon}, S. and {Bragaglia}, A. and {Bramante}, L. and {Breedt}, E. and {Bressan}, A. and {Brouillet}, N. and {Brugaletta}, E. and {Bucciarelli}, B. and {Burlacu}, A. and {Butkevich}, A.~G. and {Buzzi}, R. and {Caffau}, E. and {Cancelliere}, R. and {Cantat-Gaudin}, T. and {Carballo}, R. and {Carlucci}, T. and {Carnerero}, M.~I. and {Carrasco}, J.~M. and {Casamiquela}, L. and {Castellani}, M. and {Castro-Ginard}, A. and {Chaoul}, L. and {Charlot}, P. and {Chemin}, L. and {Chiaramida}, V. and {Chiavassa}, A. and {Chornay}, N. and {Comoretto}, G. and {Contursi}, G. and {Cooper}, W.~J. and {Cornez}, T. and {Cowell}, S. and {Crifo}, F. and {Cropper}, M. and {Crosta}, M. and {Crowley}, C. and {Dafonte}, C. and {Dapergolas}, A. and {David}, M. and {David}, P. and {de Laverny}, P. and {De Luise}, F. and {De March}, R.},
        title = "{Gaia Data Release 3. Summary of the content and survey properties}",
      journal = {\aap},
         year = 2023,
        month = jun,
       volume = {674},
          eid = {A1},
        pages = {A1},
          doi = {10.1051/0004-6361/202243940},
archivePrefix = {arXiv},
       eprint = {2208.00211},
 primaryClass = {astro-ph.GA},
       adsurl = {https://ui.adsabs.harvard.edu/abs/2023A&A...674A...1G}
}

@ARTICLE{Huyd2021,
       author = {{Hu}, Y.-D. and {Castro-Tirado}, A.~J. and {Kumar}, A. and {Gupta}, R. and {Valeev}, A.~F. and {Pandey}, S.~B. and {Kann}, D.~A. and {Castell{\'o}n}, A. and {Agudo}, I. and {Aryan}, A. and {Caballero-Garc{\'\i}a}, M.~D. and {Guziy}, S. and {Martin-Carrillo}, A. and {Oates}, S.~R. and {Pian}, E. and {S{\'a}nchez-Ram{\'\i}rez}, R. and {Sokolov}, V.~V. and {Zhang}, B.-B.},
        title = "{10.4 m GTC observations of the nearby VHE-detected GRB 190829A/SN 2019oyw}",
      journal = {\aap},
         year = 2021,
        month = feb,
       volume = {646},
          eid = {A50},
        pages = {A50},
          doi = {10.1051/0004-6361/202039349},
archivePrefix = {arXiv},
       eprint = {2009.04021},
 primaryClass = {astro-ph.HE},
       adsurl = {https://ui.adsabs.harvard.edu/abs/2021A&A...646A..50H}
}

@ARTICLE{Caballero2022,
       author = {{Caballero-Garc{\'\i}a}, M.~D. and {Gupta}, Rahul and {Pandey}, S.~B. and {Oates}, S.~R. and {Marisaldi}, M. and {Ramsli}, A. and {Hu}, Y.-D. and {Castro-Tirado}, A.~J. and {S{\'a}nchez-Ram{\'\i}rez}, R. and {Connell}, P.~H. and {Christiansen}, F. and {Ror}, A. Kumar and {Aryan}, A. and {Bai}, J.-M. and {Castro-Tirado}, M.~A. and {Fan}, Y.-F. and {Fern{\'a}ndez-Garc{\'\i}a}, E. and {Kumar}, A. and {Lindanger}, A. and {Mezentsev}, A. and {Navarro-Gonz{\'a}lez}, J. and {Neubert}, T. and {{\O}stgaard}, N. and {P{\'e}rez-Garc{\'\i}a}, I. and {Reglero}, V. and {Sarria}, D. and {Sun}, T.~R. and {Xiong}, D.-R. and {Yang}, J. and {Yang}, Y.-H. and {Zhang}, B.-B.},
        title = "{Multiwavelength study of the luminous GRB 210619B observed with Fermi and ASIM}",
      journal = {\mnras},
         year = 2023,
        month = mar,
       volume = {519},
       number = {3},
        pages = {3201-3226},
          doi = {10.1093/mnras/stac3629},
archivePrefix = {arXiv},
       eprint = {2205.07790},
 primaryClass = {astro-ph.HE},
       adsurl = {https://ui.adsabs.harvard.edu/abs/2023MNRAS.519.3201C}
}

@ARTICLE{Sari1998,
       author = {{Sari}, Re'em and {Piran}, Tsvi and {Narayan}, Ramesh},
        title = "{Spectra and Light Curves of Gamma-Ray Burst Afterglows}",
      journal = {\apjl},
         year = 1998,
        month = apr,
       volume = {497},
       number = {1},
        pages = {L17-L20},
          doi = {10.1086/311269},
archivePrefix = {arXiv},
       eprint = {astro-ph/9712005},
 primaryClass = {astro-ph},
       adsurl = {https://ui.adsabs.harvard.edu/abs/1998ApJ...497L..17S}
}

@ARTICLE{Schlafly&Finkbeiner2011,
       author = {{Schlafly}, Edward F. and {Finkbeiner}, Douglas P.},
        title = "{Measuring Reddening with Sloan Digital Sky Survey Stellar Spectra and Recalibrating SFD}",
      journal = {\apj},
         year = 2011,
        month = aug,
       volume = {737},
       number = {2},
          eid = {103},
        pages = {103},
          doi = {10.1088/0004-637X/737/2/103},
archivePrefix = {arXiv},
       eprint = {1012.4804},
 primaryClass = {astro-ph.GA},
       adsurl = {https://ui.adsabs.harvard.edu/abs/2011ApJ...737..103S}
}

@ARTICLE{Wangyun2023,
       author = {{Wang}, Yun and {Xia}, Zi-Qing and {Zheng}, Tian-Ci and {Ren}, Jia and {Fan}, Yi-Zhong},
        title = "{A Broken ``{\ensuremath{\alpha}}-intensity'' Relation Caused by the Evolving Photosphere Emission and the Nature of the Extraordinarily Bright GRB 230307A}",
      journal = {\apjl},
         year = 2023,
        month = aug,
       volume = {953},
       number = {1},
          eid = {L8},
        pages = {L8},
          doi = {10.3847/2041-8213/ace7d4},
archivePrefix = {arXiv},
       eprint = {2303.11083},
 primaryClass = {astro-ph.HE},
       adsurl = {https://ui.adsabs.harvard.edu/abs/2023ApJ...953L...8W}
}

@ARTICLE{Woosley1993,
       author = {{Woosley}, S.~E.},
        title = "{Gamma-Ray Bursts from Stellar Mass Accretion Disks around Black Holes}",
      journal = {\apj},
         year = 1993,
        month = mar,
       volume = {405},
        pages = {273},
          doi = {10.1086/172359},
       adsurl = {https://ui.adsabs.harvard.edu/abs/1993ApJ...405..273W}
}

@ARTICLE{Klebesadel1973,
       author = {{Klebesadel}, Ray W. and {Strong}, Ian B. and {Olson}, Roy A.},
        title = "{Observations of Gamma-Ray Bursts of Cosmic Origin}",
      journal = {\apjl},
         year = 1973,
        month = jun,
       volume = {182},
        pages = {L85},
          doi = {10.1086/181225},
       adsurl = {https://ui.adsabs.harvard.edu/abs/1973ApJ...182L..85K}
}

@ARTICLE{MacFadyen1999,
       author = {{MacFadyen}, A.~I. and {Woosley}, S.~E.},
        title = "{Collapsars: Gamma-Ray Bursts and Explosions in ``Failed Supernovae''}",
      journal = {\apj},
         year = 1999,
        month = oct,
       volume = {524},
       number = {1},
        pages = {262-289},
          doi = {10.1086/307790},
archivePrefix = {arXiv},
       eprint = {astro-ph/9810274},
 primaryClass = {astro-ph},
       adsurl = {https://ui.adsabs.harvard.edu/abs/1999ApJ...524..262M}
}

@ARTICLE{Zhang&Meszaros2001,
       author = {{Zhang}, Bing and {M{\'e}sz{\'a}ros}, Peter},
        title = "{Gamma-Ray Burst Afterglow with Continuous Energy Injection: Signature of a Highly Magnetized Millisecond Pulsar}",
      journal = {\apjl},
         year = 2001,
        month = may,
       volume = {552},
       number = {1},
        pages = {L35-L38},
          doi = {10.1086/320255},
archivePrefix = {arXiv},
       eprint = {astro-ph/0011133},
 primaryClass = {astro-ph},
       adsurl = {https://ui.adsabs.harvard.edu/abs/2001ApJ...552L..35Z}
}

@ARTICLE{Rees1994,
       author = {{Rees}, M.~J. and {Meszaros}, P.},
        title = "{Unsteady Outflow Models for Cosmological Gamma-Ray Bursts}",
      journal = {\apjl},
         year = 1994,
        month = aug,
       volume = {430},
        pages = {L93},
          doi = {10.1086/187446},
archivePrefix = {arXiv},
       eprint = {astro-ph/9404038},
 primaryClass = {astro-ph},
       adsurl = {https://ui.adsabs.harvard.edu/abs/1994ApJ...430L..93R}
}

@ARTICLE{Kumer2015,
       author = {{Kumar}, Pawan and {Zhang}, Bing},
        title = "{The physics of gamma-ray bursts \& relativistic jets}",
      journal = {\physrep},
         year = 2015,
        month = feb,
       volume = {561},
        pages = {1-109},
          doi = {10.1016/j.physrep.2014.09.008},
archivePrefix = {arXiv},
       eprint = {1410.0679},
 primaryClass = {astro-ph.HE},
       adsurl = {https://ui.adsabs.harvard.edu/abs/2015PhR...561....1K}
}

@ARTICLE{Dai&Lu1998,
       author = {{Dai}, Z.~G. and {Lu}, T.},
        title = "{Gamma-ray burst afterglows and evolution of postburst fireballs with energy injection from strongly magnetic millisecond pulsars}",
      journal = {\aap},
         year = 1998,
        month = may,
       volume = {333},
        pages = {L87-L90},
          doi = {10.48550/arXiv.astro-ph/9810402},
archivePrefix = {arXiv},
       eprint = {astro-ph/9810402},
 primaryClass = {astro-ph},
       adsurl = {https://ui.adsabs.harvard.edu/abs/1998A&A...333L..87D}
}

@ARTICLE{oke1995,
       author = {{Oke}, J.~B. and {Cohen}, J.~G. and {Carr}, M. and {Cromer}, J. and {Dingizian}, A. and {Harris}, F.~H. and {Labrecque}, S. and {Lucinio}, R. and {Schaal}, W. and {Epps}, H. and {Miller}, J.},
        title = "{The Keck Low-Resolution Imaging Spectrometer}",
      journal = {\pasp},
         year = 1995,
        month = apr,
       volume = {107},
        pages = {375},
          doi = {10.1086/133562},
       adsurl = {https://ui.adsabs.harvard.edu/abs/1995PASP..107..375O}
}

@ARTICLE{filippenko1982,
       author = {{Filippenko}, A.~V.},
        title = "{The importance of atmospheric differential refraction in spectrophotometry.}",
      journal = {\pasp},
         year = 1982,
        month = aug,
       volume = {94},
        pages = {715-721},
          doi = {10.1086/131052},
       adsurl = {https://ui.adsabs.harvard.edu/abs/1982PASP...94..715F}
}

@ARTICLE{Perley2019,
       author = {{Perley}, Daniel A.},
        title = "{Fully Automated Reduction of Longslit Spectroscopy with the Low Resolution Imaging Spectrometer at the Keck Observatory}",
      journal = {\pasp},
         year = 2019,
        month = aug,
       volume = {131},
       number = {1002},
        pages = {084503},
          doi = {10.1088/1538-3873/ab215d},
archivePrefix = {arXiv},
       eprint = {1903.07629},
 primaryClass = {astro-ph.IM},
       adsurl = {https://ui.adsabs.harvard.edu/abs/2019PASP..131h4503P}
}

@ARTICLE{ZhangB2006,
       author = {{Zhang}, Bing and {Fan}, Y.~Z. and {Dyks}, Jaroslaw and {Kobayashi}, Shiho and {M{\'e}sz{\'a}ros}, Peter and {Burrows}, David N. and {Nousek}, John A. and {Gehrels}, Neil},
        title = "{Physical Processes Shaping Gamma-Ray Burst X-Ray Afterglow Light Curves: Theoretical Implications from the Swift X-Ray Telescope Observations}",
      journal = {\apj},
         year = 2006,
        month = may,
       volume = {642},
       number = {1},
        pages = {354-370},
          doi = {10.1086/500723},
archivePrefix = {arXiv},
       eprint = {astro-ph/0508321},
 primaryClass = {astro-ph},
       adsurl = {https://ui.adsabs.harvard.edu/abs/2006ApJ...642..354Z}
}

@ARTICLE{swift2004,
       author = {{Gehrels}, N. and {Chincarini}, G. and {Giommi}, P. and {Mason}, K.~O. and {Nousek}, J.~A. and {Wells}, A.~A. and {White}, N.~E. and {Barthelmy}, S.~D. and {Burrows}, D.~N. and {Cominsky}, L.~R. and {Hurley}, K.~C. and {Marshall}, F.~E. and {M{\'e}sz{\'a}ros}, P. and {Roming}, P.~W.~A. and {Angelini}, L. and {Barbier}, L.~M. and {Belloni}, T. and {Campana}, S. and {Caraveo}, P.~A. and {Chester}, M.~M. and {Citterio}, O. and {Cline}, T.~L. and {Cropper}, M.~S. and {Cummings}, J.~R. and {Dean}, A.~J. and {Feigelson}, E.~D. and {Fenimore}, E.~E. and {Frail}, D.~A. and {Fruchter}, A.~S. and {Garmire}, G.~P. and {Gendreau}, K. and {Ghisellini}, G. and {Greiner}, J. and {Hill}, J.~E. and {Hunsberger}, S.~D. and {Krimm}, H.~A. and {Kulkarni}, S.~R. and {Kumar}, P. and {Lebrun}, F. and {Lloyd-Ronning}, N.~M. and {Markwardt}, C.~B. and {Mattson}, B.~J. and {Mushotzky}, R.~F. and {Norris}, J.~P. and {Osborne}, J. and {Paczynski}, B. and {Palmer}, D.~M. and {Park}, H.-S. and {Parsons}, A.~M. and {Paul}, J. and {Rees}, M.~J. and {Reynolds}, C.~S. and {Rhoads}, J.~E. and {Sasseen}, T.~P. and {Schaefer}, B.~E. and {Short}, A.~T. and {Smale}, A.~P. and {Smith}, I.~A. and {Stella}, L. and {Tagliaferri}, G. and {Takahashi}, T. and {Tashiro}, M. and {Townsley}, L.~K. and {Tueller}, J. and {Turner}, M.~J.~L. and {Vietri}, M. and {Voges}, W. and {Ward}, M.~J. and {Willingale}, R. and {Zerbi}, F.~M. and {Zhang}, W.~W.},
        title = "{The Swift Gamma-Ray Burst Mission}",
      journal = {\apj},
         year = 2004,
        month = aug,
       volume = {611},
       number = {2},
        pages = {1005-1020},
          doi = {10.1086/422091},
archivePrefix = {arXiv},
       eprint = {astro-ph/0405233},
 primaryClass = {astro-ph},
       adsurl = {https://ui.adsabs.harvard.edu/abs/2004ApJ...611.1005G}
}

@INCOLLECTION{yuan2022,
       author = {{Yuan}, Weimin and {Zhang}, Chen and {Chen}, Yong and {Ling}, Zhixing},
        title = "{The Einstein Probe Mission}",
    booktitle = {Handbook of X-ray and Gamma-ray Astrophysics},
         year = 2022,
       editor = {{Bambi}, Cosimo and {Sangangelo}, Andrea},
          eid = {86},
        pages = {86},
          doi = {10.1007/978-981-16-4544-0_151-1},
       adsurl = {https://ui.adsabs.harvard.edu/abs/2022hxga.book...86Y}
}

@ARTICLE{Li2012,
       author = {{Li}, Liang and {Liang}, En-Wei and {Tang}, Qing-Wen and {Chen}, Jie-Min and {Xi}, Shao-Qiang and {L{\"u}}, Hou-Jun and {Gao}, He and {Zhang}, Bing and {Zhang}, Jin and {Yi}, Shuang-Xi and {Lu}, Rui-Jing and {L{\"u}}, Lian-Zhong and {Wei}, Jian-Yan},
        title = "{A Comprehensive Study of Gamma-Ray Burst Optical Emission. I. Flares and Early Shallow-decay Component}",
      journal = {\apj},
         year = 2012,
        month = oct,
       volume = {758},
       number = {1},
          eid = {27},
        pages = {27},
          doi = {10.1088/0004-637X/758/1/27},
archivePrefix = {arXiv},
       eprint = {1203.2332},
 primaryClass = {astro-ph.HE},
       adsurl = {https://ui.adsabs.harvard.edu/abs/2012ApJ...758...27L}
}

@BOOK{zhangb2018,
       author = {{Zhang}, Bing},
        title = "{The Physics of Gamma-Ray Bursts}",
         year = 2018,
          doi = {10.1017/9781139226530},
       adsurl = {https://ui.adsabs.harvard.edu/abs/2018pgrb.book.....Z}
}

@ARTICLE{Rowlinson2013,
       author = {{Rowlinson}, A. and {O'Brien}, P.~T. and {Metzger}, B.~D. and {Tanvir}, N.~R. and {Levan}, A.~J.},
        title = "{Signatures of magnetar central engines in short GRB light curves}",
      journal = {\mnras},
         year = 2013,
        month = apr,
       volume = {430},
       number = {2},
        pages = {1061-1087},
          doi = {10.1093/mnras/sts683},
archivePrefix = {arXiv},
       eprint = {1301.0629},
 primaryClass = {astro-ph.HE},
       adsurl = {https://ui.adsabs.harvard.edu/abs/2013MNRAS.430.1061R}
}

@ARTICLE{Geng2025,
       author = {{Geng}, Jin-Jun and {Hu}, Ding-Fang and {Gao}, Hao-Xuan and {Liang}, Yi-Fang and {Hua}, Yan-Long and {Zhang}, Guo-Rui and {Sun}, Tian-Rui and {Li}, Bing and {Liu}, Yuan-Qi and {Xu}, Fan and {Deng}, Chen and {Hu}, Chen-Ran and {Xu}, Ming and {Huang}, Yong-Feng and {Zhang}, Miao-Miao and {Fang}, Min and {Yan}, Jing-Zhi and {An}, Tao and {Wu}, Xue-Feng},
        title = "{Gamma-Ray Burst Timing: Decoding the Hidden Slow Jets in GRB 060729}",
      journal = {\apjl},
         year = 2025,
        month = may,
       volume = {984},
       number = {2},
          eid = {L65},
        pages = {L65},
          doi = {10.3847/2041-8213/add00e},
archivePrefix = {arXiv},
       eprint = {2503.17766},
 primaryClass = {astro-ph.HE},
       adsurl = {https://ui.adsabs.harvard.edu/abs/2025ApJ...984L..65G}
}

@ARTICLE{Dereli2022NC,
       author = {{Dereli-B{\'e}gu{\'e}}, H{\"u}sne and {Pe'er}, Asaf and {Ryde}, Felix and {Oates}, Samantha R. and {Zhang}, Bing and {Dainotti}, Maria G.},
        title = "{A wind environment and Lorentz factors of tens explain gamma-ray bursts X-ray plateau}",
      journal = {Nature Communications},
         year = 2022,
        month = sep,
       volume = {13},
          eid = {5611},
        pages = {5611},
          doi = {10.1038/s41467-022-32881-1},
archivePrefix = {arXiv},
       eprint = {2207.11066},
 primaryClass = {astro-ph.HE},
       adsurl = {https://ui.adsabs.harvard.edu/abs/2022NatCo..13.5611D}
}

@ARTICLE{Woosley2006,
       author = {{Woosley}, S.~E. and {Bloom}, J.~S.},
        title = "{The Supernova Gamma-Ray Burst Connection}",
      journal = {\araa},
         year = 2006,
        month = sep,
       volume = {44},
       number = {1},
        pages = {507-556},
          doi = {10.1146/annurev.astro.43.072103.150558},
archivePrefix = {arXiv},
       eprint = {astro-ph/0609142},
 primaryClass = {astro-ph},
       adsurl = {https://ui.adsabs.harvard.edu/abs/2006ARA&A..44..507W}
}

@ARTICLE{Cano2017,
       author = {{Cano}, Zach and {Wang}, Shan-Qin and {Dai}, Zi-Gao and {Wu}, Xue-Feng},
        title = "{The Observer's Guide to the Gamma-Ray Burst Supernova Connection}",
      journal = {Advances in Astronomy},
         year = 2017,
        month = jan,
       volume = {2017},
          eid = {8929054},
        pages = {8929054},
          doi = {10.1155/2017/8929054},
archivePrefix = {arXiv},
       eprint = {1604.03549},
 primaryClass = {astro-ph.HE},
       adsurl = {https://ui.adsabs.harvard.edu/abs/2017AdAst2017E...5C}
}

@ARTICLE{Granot&Sari2002,
       author = {{Granot}, Jonathan and {Sari}, Re'em},
        title = "{The Shape of Spectral Breaks in Gamma-Ray Burst Afterglows}",
      journal = {\apj},
         year = 2002,
        month = apr,
       volume = {568},
       number = {2},
        pages = {820-829},
          doi = {10.1086/338966},
archivePrefix = {arXiv},
       eprint = {astro-ph/0108027},
 primaryClass = {astro-ph},
       adsurl = {https://ui.adsabs.harvard.edu/abs/2002ApJ...568..820G}
}

@ARTICLE{asgard,
       author = {{Ren}, Jia and {Wang}, Yun and {Dai}, Zi-Gao},
       title = "{Jet Structure and Burst Environment of GRB 221009A}",
       journal = {\apj},
         year = 2024,
        month = feb,
       volume = {962},
       number = {2},
          eid = {115},
        pages = {115},
          doi = {10.3847/1538-4357/ad1bcd},
       archivePrefix = {arXiv},
       eprint = {2310.15886},
       primaryClass = {astro-ph.HE},
       adsurl = {https://ui.adsabs.harvard.edu/abs/2024ApJ...962..115R}
}

@ARTICLE{Nousek2006,
       author = {{Nousek}, J.~A. and {Kouveliotou}, C. and {Grupe}, D. and {Page}, K.~L. and {Granot}, J. and {Ramirez-Ruiz}, E. and {Patel}, S.~K. and {Burrows}, D.~N. and {Mangano}, V. and {Barthelmy}, S. and {Beardmore}, A.~P. and {Campana}, S. and {Capalbi}, M. and {Chincarini}, G. and {Cusumano}, G. and {Falcone}, A.~D. and {Gehrels}, N. and {Giommi}, P. and {Goad}, M.~R. and {Godet}, O. and {Hurkett}, C.~P. and {Kennea}, J.~A. and {Moretti}, A. and {O'Brien}, P.~T. and {Osborne}, J.~P. and {Romano}, P. and {Tagliaferri}, G. and {Wells}, A.~A.},
        title = "{Evidence for a Canonical Gamma-Ray Burst Afterglow Light Curve in the Swift XRT Data}",
      journal = {\apj},
         year = 2006,
        month = may,
       volume = {642},
       number = {1},
        pages = {389-400},
          doi = {10.1086/500724},
archivePrefix = {arXiv},
       eprint = {astro-ph/0508332},
 primaryClass = {astro-ph},
       adsurl = {https://ui.adsabs.harvard.edu/abs/2006ApJ...642..389N}
}

@ARTICLE{astropy:2018,
   author = {{Astropy Collaboration} and {Price-Whelan}, A.~M. and {Sip{\H o}cz}, B.~M. and
	{G{\"u}nther}, H.~M. and {Lim}, P.~L. and {Crawford}, S.~M. and
	{Conseil}, S. and {Shupe}, D.~L. and {Craig}, M.~W. and {Dencheva}, N. and
	{Ginsburg}, A. and {VanderPlas}, J.~T. and {Bradley}, L.~D. and
	{P{\'e}rez-Su{\'a}rez}, D. and {de Val-Borro}, M. and {Paper Contributors}, (. and
	{Aldcroft}, T.~L. and {Cruz}, K.~L. and {Robitaille}, T.~P. and
	{Tollerud}, E.~J. and {Coordination Committee}, (. and {Ardelean}, C. and
	{Babej}, T. and {Bach}, Y.~P. and {Bachetti}, M. and {Bakanov}, A.~V. and
	{Bamford}, S.~P. and {Barentsen}, G. and {Barmby}, P. and {Baumbach}, A. and
	{Berry}, K.~L. and {Biscani}, F. and {Boquien}, M. and {Bostroem}, K.~A. and
	{Bouma}, L.~G. and {Brammer}, G.~B. and {Bray}, E.~M. and {Breytenbach}, H. and
	{Buddelmeijer}, H. and {Burke}, D.~J. and {Calderone}, G. and
	{Cano Rodr{\'{\i}}guez}, J.~L. and {Cara}, M. and {Cardoso}, J.~V.~M. and
	{Cheedella}, S. and {Copin}, Y. and {Corrales}, L. and {Crichton}, D. and
	{D{\'A}vella}, D. and {Deil}, C. and {Depagne}, {\'E}. and
	{Dietrich}, J.~P. and {Donath}, A. and {Droettboom}, M. and
	{Earl}, N. and {Erben}, T. and {Fabbro}, S. and {Ferreira}, L.~A. and
	{Finethy}, T. and {Fox}, R.~T. and {Garrison}, L.~H. and {Gibbons}, S.~L.~J. and
	{Goldstein}, D.~A. and {Gommers}, R. and {Greco}, J.~P. and
	{Greenfield}, P. and {Groener}, A.~M. and {Grollier}, F. and
	{Hagen}, A. and {Hirst}, P. and {Homeier}, D. and {Horton}, A.~J. and
	{Hosseinzadeh}, G. and {Hu}, L. and {Hunkeler}, J.~S. and {Ivezi{\'c}}, {\v Z}. and
	{Jain}, A. and {Jenness}, T. and {Kanarek}, G. and {Kendrew}, S. and
	{Kern}, N.~S. and {Kerzendorf}, W.~E. and {Khvalko}, A. and
	{King}, J. and {Kirkby}, D. and {Kulkarni}, A.~M. and {Kumar}, A. and
	{Lee}, A. and {Lenz}, D. and {Littlefair}, S.~P. and {Ma}, Z. and
	{Macleod}, D.~M. and {Mastropietro}, M. and {McCully}, C. and
	{Montagnac}, S. and {Morris}, B.~M. and {Mueller}, M. and {Mumford}, S.~J. and
	{Muna}, D. and {Murphy}, N.~A. and {Nelson}, S. and {Nguyen}, G.~H. and
	{Ninan}, J.~P. and {N{\"o}the}, M. and {Ogaz}, S. and {Oh}, S. and
	{Parejko}, J.~K. and {Parley}, N. and {Pascual}, S. and {Patil}, R. and
	{Patil}, A.~A. and {Plunkett}, A.~L. and {Prochaska}, J.~X. and
	{Rastogi}, T. and {Reddy Janga}, V. and {Sabater}, J. and {Sakurikar}, P. and
	{Seifert}, M. and {Sherbert}, L.~E. and {Sherwood-Taylor}, H. and
	{Shih}, A.~Y. and {Sick}, J. and {Silbiger}, M.~T. and {Singanamalla}, S. and
	{Singer}, L.~P. and {Sladen}, P.~H. and {Sooley}, K.~A. and
	{Sornarajah}, S. and {Streicher}, O. and {Teuben}, P. and {Thomas}, S.~W. and
	{Tremblay}, G.~R. and {Turner}, J.~E.~H. and {Terr{\'o}n}, V. and
	{van Kerkwijk}, M.~H. and {de la Vega}, A. and {Watkins}, L.~L. and
	{Weaver}, B.~A. and {Whitmore}, J.~B. and {Woillez}, J. and
	{Zabalza}, V. and {Contributors}, (.},
    title = "{The Astropy Project: Building an Open-science Project and Status of the v2.0 Core Package}",
  journal = {\aj},
archivePrefix = "arXiv",
   eprint = {1801.02634},
 primaryClass = "astro-ph.IM",
     year = 2018,
    month = sep,
   volume = 156,
      eid = {123},
    pages = {123},
      doi = {10.3847/1538-3881/aabc4f},
   adsurl = {https://ui.adsabs.harvard.edu/abs/2018AJ....156..123T}
}

@ARTICLE{astropy:2013,
   author = {{Astropy Collaboration} and {Robitaille}, T.~P. and {Tollerud}, E.~J. and
    {Greenfield}, P. and {Droettboom}, M. and {Bray}, E. and {Aldcroft}, T. and
    {Davis}, M. and {Ginsburg}, A. and {Price-Whelan}, A.~M. and
    {Kerzendorf}, W.~E. and {Conley}, A. and {Crighton}, N. and
    {Barbary}, K. and {Muna}, D. and {Ferguson}, H. and {Grollier}, F. and
    {Parikh}, M.~M. and {Nair}, P.~H. and {Unther}, H.~M. and {Deil}, C. and
    {Woillez}, J. and {Conseil}, S. and {Kramer}, R. and {Turner}, J.~E.~H. and
    {Singer}, L. and {Fox}, R. and {Weaver}, B.~A. and {Zabalza}, V. and
    {Edwards}, Z.~I. and {Azalee Bostroem}, K. and {Burke}, D.~J. and
    {Casey}, A.~R. and {Crawford}, S.~M. and {Dencheva}, N. and
    {Ely}, J. and {Jenness}, T. and {Labrie}, K. and {Lian Lim}, P. and
    {Pierfederici}, F. and {Pontzen}, A. and {Ptak}, A. and {Refsdal}, B. and
    {Servillat}, M. and {Streicher}, O.},
    title = "{Astropy: A community Python package for astronomy}",
  journal = {\aap},
     year = 2013,
    month = oct,
   volume = 558,
      eid = {A33},
    pages = {A33},
      doi = {10.1051/0004-6361/201322068},
   adsurl = {https://ui.adsabs.harvard.edu/abs/2013A%26A...558A..33A}
}

@ARTICLE{astropy:2022,
       author = {{Astropy Collaboration} and {Price-Whelan}, Adrian M. and {Lim}, Pey Lian and {Earl}, Nicholas and {Starkman}, Nathaniel and {Bradley}, Larry and {Shupe}, David L. and {Patil}, Aarya A. and {Corrales}, Lia and {Brasseur}, C.~E. and {N{"o}the}, Maximilian and {Donath}, Axel and {Tollerud}, Erik and {Morris}, Brett M. and {Ginsburg}, Adam and {Vaher}, Eero and {Weaver}, Benjamin A. and {Tocknell}, James and {Jamieson}, William and {van Kerkwijk}, Marten H. and {Robitaille}, Thomas P. and {Merry}, Bruce and {Bachetti}, Matteo and {G{"u}nther}, H. Moritz and {Aldcroft}, Thomas L. and {Alvarado-Montes}, Jaime A. and {Archibald}, Anne M. and {B{'o}di}, Attila and {Bapat}, Shreyas and {Barentsen}, Geert and {Baz{'a}n}, Juanjo and {Biswas}, Manish and {Boquien}, M{'e}d{'e}ric and {Burke}, D.~J. and {Cara}, Daria and {Cara}, Mihai and {Conroy}, Kyle E. and {Conseil}, Simon and {Craig}, Matthew W. and {Cross}, Robert M. and {Cruz}, Kelle L. and {D'Eugenio}, Francesco and {Dencheva}, Nadia and {Devillepoix}, Hadrien A.~R. and {Dietrich}, J{"o}rg P. and {Eigenbrot}, Arthur Davis and {Erben}, Thomas and {Ferreira}, Leonardo and {Foreman-Mackey}, Daniel and {Fox}, Ryan and {Freij}, Nabil and {Garg}, Suyog and {Geda}, Robel and {Glattly}, Lauren and {Gondhalekar}, Yash and {Gordon}, Karl D. and {Grant}, David and {Greenfield}, Perry and {Groener}, Austen M. and {Guest}, Steve and {Gurovich}, Sebastian and {Handberg}, Rasmus and {Hart}, Akeem and {Hatfield-Dodds}, Zac and {Homeier}, Derek and {Hosseinzadeh}, Griffin and {Jenness}, Tim and {Jones}, Craig K. and {Joseph}, Prajwel and {Kalmbach}, J. Bryce and {Karamehmetoglu}, Emir and {Ka{l}uszy{'n}ski}, Miko{l}aj and {Kelley}, Michael S.~P. and {Kern}, Nicholas and {Kerzendorf}, Wolfgang E. and {Koch}, Eric W. and {Kulumani}, Shankar and {Lee}, Antony and {Ly}, Chun and {Ma}, Zhiyuan and {MacBride}, Conor and {Maljaars}, Jakob M. and {Muna}, Demitri and {Murphy}, N.~A. and {Norman}, Henrik and {O'Steen}, Richard and {Oman}, Kyle A. and {Pacifici}, Camilla and {Pascual}, Sergio and {Pascual-Granado}, J. and {Patil}, Rohit R. and {Perren}, Gabriel I. and {Pickering}, Timothy E. and {Rastogi}, Tanuj and {Roulston}, Benjamin R. and {Ryan}, Daniel F. and {Rykoff}, Eli S. and {Sabater}, Jose and {Sakurikar}, Parikshit and {Salgado}, Jes{'u}s and {Sanghi}, Aniket and {Saunders}, Nicholas and {Savchenko}, Volodymyr and {Schwardt}, Ludwig and {Seifert-Eckert}, Michael and {Shih}, Albert Y. and {Jain}, Anany Shrey and {Shukla}, Gyanendra and {Sick}, Jonathan and {Simpson}, Chris and {Singanamalla}, Sudheesh and {Singer}, Leo P. and {Singhal}, Jaladh and {Sinha}, Manodeep and {Sip{H{o}}cz}, Brigitta M. and {Spitler}, Lee R. and {Stansby}, David and {Streicher}, Ole and {{{S}}umak}, Jani and {Swinbank}, John D. and {Taranu}, Dan S. and {Tewary}, Nikita and {Tremblay}, Grant R. and {Val-Borro}, Miguel de and {Van Kooten}, Samuel J. and {Vasovi{'c}}, Zlatan and {Verma}, Shresth and {de Miranda Cardoso}, Jos{'e} Vin{'i}cius and {Williams}, Peter K.~G. and {Wilson}, Tom J. and {Winkel}, Benjamin and {Wood-Vasey}, W.~M. and {Xue}, Rui and {Yoachim}, Peter and {Zhang}, Chen and {Zonca}, Andrea and {Astropy Project Contributors}},
        title = "{The Astropy Project: Sustaining and Growing a Community-oriented Open-source Project and the Latest Major Release (v5.0) of the Core Package}",
      journal = {\apj},
         year = 2022,
        month = aug,
       volume = {935},
       number = {2},
          eid = {167},
        pages = {167},
          doi = {10.3847/1538-4357/ac7c74},
archivePrefix = {arXiv},
       eprint = {2206.14220},
 primaryClass = {astro-ph.IM},
       adsurl = {https://ui.adsabs.harvard.edu/abs/2022ApJ...935..167A}
}

@ARTICLE{Kumar2008,
       author = {{Kumar}, Pawan and {Narayan}, Ramesh and {Johnson}, Jarrett L.},
        title = "{Mass fall-back and accretion in the central engine of gamma-ray bursts}",
      journal = {\mnras},
         year = 2008,
        month = aug,
       volume = {388},
       number = {4},
        pages = {1729-1742},
          doi = {10.1111/j.1365-2966.2008.13493.x},
archivePrefix = {arXiv},
       eprint = {0807.0441},
 primaryClass = {astro-ph},
       adsurl = {https://ui.adsabs.harvard.edu/abs/2008MNRAS.388.1729K}
}

@ARTICLE{Hjorth2003,
       author = {{Hjorth}, Jens and {Sollerman}, Jesper and {M{\o}ller}, Palle and {Fynbo}, Johan P.~U. and {Woosley}, Stan E. and {Kouveliotou}, Chryssa and {Tanvir}, Nial R. and {Greiner}, Jochen and {Andersen}, Michael I. and {Castro-Tirado}, Alberto J. and {Castro Cer{\'o}n}, Jos{\'e} Mar{\'\i}a and {Fruchter}, Andrew S. and {Gorosabel}, Javier and {Jakobsson}, P{\'a}ll and {Kaper}, Lex and {Klose}, Sylvio and {Masetti}, Nicola and {Pedersen}, Holger and {Pedersen}, Kristian and {Pian}, Elena and {Palazzi}, Eliana and {Rhoads}, James E. and {Rol}, Evert and {van den Heuvel}, Edward P.~J. and {Vreeswijk}, Paul M. and {Watson}, Darach and {Wijers}, Ralph A.~M.~J.},
        title = "{A very energetic supernova associated with the {\ensuremath{\gamma}}-ray burst of 29 March 2003}",
      journal = {\nat},
         year = 2003,
        month = jun,
       volume = {423},
       number = {6942},
        pages = {847-850},
          doi = {10.1038/nature01750},
archivePrefix = {arXiv},
       eprint = {astro-ph/0306347},
 primaryClass = {astro-ph},
       adsurl = {https://ui.adsabs.harvard.edu/abs/2003Natur.423..847H}
}

@ARTICLE{Oates2015M,
       author = {{Oates}, S.~R. and {Racusin}, J.~L. and {De Pasquale}, M. and {Page}, M.~J. and {Castro-Tirado}, A.~J. and {Gorosabel}, J. and {Smith}, P.~J. and {Breeveld}, A.~A. and {Kuin}, N.~P.~M.},
        title = "{Exploring the canonical behaviour of long gamma-ray bursts using an intrinsic multiwavelength afterglow correlation}",
      journal = {\mnras},
         year = 2015,
        month = nov,
       volume = {453},
       number = {4},
        pages = {4121-4135},
          doi = {10.1093/mnras/stv1956},
archivePrefix = {arXiv},
       eprint = {1508.06567},
 primaryClass = {astro-ph.HE},
       adsurl = {https://ui.adsabs.harvard.edu/abs/2015MNRAS.453.4121O}
}

@ARTICLE{2023A&A...674A...1G,
       author = {{Gaia Collaboration} and {Vallenari}, A. and {Brown}, A.~G.~A. and {Prusti}, T. and {de Bruijne}, J.~H.~J. and {Arenou}, F. and {Babusiaux}, C. and {Biermann}, M. and {Creevey}, O.~L. and {Ducourant}, C. and {Evans}, D.~W. and {Eyer}, L. and {Guerra}, R. and {Hutton}, A. and {Jordi}, C. and {Klioner}, S.~A. and {Lammers}, U.~L. and {Lindegren}, L. and {Luri}, X. and {Mignard}, F. and {Panem}, C. and {Pourbaix}, D. and {Randich}, S. and {Sartoretti}, P. and {Soubiran}, C. and {Tanga}, P. and {Walton}, N.~A. and {Bailer-Jones}, C.~A.~L. and {Bastian}, U. and {Drimmel}, R. and {Jansen}, F. and {Katz}, D. and {Lattanzi}, M.~G. and {van Leeuwen}, F. and {Bakker}, J. and {Cacciari}, C. and {Casta{\~n}eda}, J. and {De Angeli}, F. and {Fabricius}, C. and {Fouesneau}, M. and {Fr{\'e}mat}, Y. and {Galluccio}, L. and {Guerrier}, A. and {Heiter}, U. and {Masana}, E. and {Messineo}, R. and {Mowlavi}, N. and {Nicolas}, C. and {Nienartowicz}, K. and {Pailler}, F. and {Panuzzo}, P. and {Riclet}, F. and {Roux}, W. and {Seabroke}, G.~M. and {Sordo}, R. and {Th{\'e}venin}, F. and {Gracia-Abril}, G. and {Portell}, J. and {Teyssier}, D. and {Altmann}, M. and {Andrae}, R. and {Audard}, M. and {Bellas-Velidis}, I. and {Benson}, K. and {Berthier}, J. and {Blomme}, R. and {Burgess}, P.~W. and {Busonero}, D. and {Busso}, G. and {C{\'a}novas}, H. and {Carry}, B. and {Cellino}, A. and {Cheek}, N. and {Clementini}, G. and {Damerdji}, Y. and {Davidson}, M. and {de Teodoro}, P. and {Nu{\~n}ez Campos}, M. and {Delchambre}, L. and {Dell'Oro}, A. and {Esquej}, P. and {Fern{\'a}ndez-Hern{\'a}ndez}, J. and {Fraile}, E. and {Garabato}, D. and {Garc{\'\i}a-Lario}, P. and {Gosset}, E. and {Haigron}, R. and {Halbwachs}, J.-L. and {Hambly}, N.~C. and {Harrison}, D.~L. and {Hern{\'a}ndez}, J. and {Hestroffer}, D. and {Hodgkin}, S.~T. and {Holl}, B. and {Jan{\ss}en}, K. and {Jevardat de Fombelle}, G. and {Jordan}, S. and {Krone-Martins}, A. and {Lanzafame}, A.~C. and {L{\"o}ffler}, W. and {Marchal}, O. and {Marrese}, P.~M. and {Moitinho}, A. and {Muinonen}, K. and {Osborne}, P. and {Pancino}, E. and {Pauwels}, T. and {Recio-Blanco}, A. and {Reyl{\'e}}, C. and {Riello}, M. and {Rimoldini}, L. and {Roegiers}, T. and {Rybizki}, J. and {Sarro}, L.~M. and {Siopis}, C. and {Smith}, M. and {Sozzetti}, A. and {Utrilla}, E. and {van Leeuwen}, M. and {Abbas}, U. and {{\'A}brah{\'a}m}, P. and {Abreu Aramburu}, A. and {Aerts}, C. and {Aguado}, J.~J. and {Ajaj}, M. and {Aldea-Montero}, F. and {Altavilla}, G. and {{\'A}lvarez}, M.~A. and {Alves}, J. and {Anders}, F. and {Anderson}, R.~I. and {Anglada Varela}, E. and {Antoja}, T. and {Baines}, D. and {Baker}, S.~G. and {Balaguer-N{\'u}{\~n}ez}, L. and {Balbinot}, E. and {Balog}, Z. and {Barache}, C. and {Barbato}, D. and {Barros}, M. and {Barstow}, M.~A. and {Bartolom{\'e}}, S. and {Bassilana}, J.-L. and {Bauchet}, N. and {Becciani}, U. and {Bellazzini}, M. and {Berihuete}, A. and {Bernet}, M. and {Bertone}, S. and {Bianchi}, L. and {Binnenfeld}, A. and {Blanco-Cuaresma}, S. and {Blazere}, A. and {Boch}, T. and {Bombrun}, A. and {Bossini}, D. and {Bouquillon}, S. and {Bragaglia}, A. and {Bramante}, L. and {Breedt}, E. and {Bressan}, A. and {Brouillet}, N. and {Brugaletta}, E. and {Bucciarelli}, B. and {Burlacu}, A. and {Butkevich}, A.~G. and {Buzzi}, R. and {Caffau}, E. and {Cancelliere}, R. and {Cantat-Gaudin}, T. and {Carballo}, R. and {Carlucci}, T. and {Carnerero}, M.~I. and {Carrasco}, J.~M. and {Casamiquela}, L. and {Castellani}, M. and {Castro-Ginard}, A. and {Chaoul}, L. and {Charlot}, P. and {Chemin}, L. and {Chiaramida}, V. and {Chiavassa}, A. and {Chornay}, N. and {Comoretto}, G. and {Contursi}, G. and {Cooper}, W.~J. and {Cornez}, T. and {Cowell}, S. and {Crifo}, F. and {Cropper}, M. and {Crosta}, M. and {Crowley}, C. and {Dafonte}, C. and {Dapergolas}, A. and {David}, M. and {David}, P. and {de Laverny}, P. and {De Luise}, F. and {De March}, R.},
        title = "{Gaia Data Release 3. Summary of the content and survey properties}",
      journal = {\aap},
         year = 2023,
        month = jun,
       volume = {674},
          eid = {A1},
        pages = {A1},
          doi = {10.1051/0004-6361/202243940},
archivePrefix = {arXiv},
       eprint = {2208.00211},
 primaryClass = {astro-ph.GA},
       adsurl = {https://ui.adsabs.harvard.edu/abs/2023A&A...674A...1G}
}

@ARTICLE{2024PASA...41...61O,
       author = {{Onken}, Christopher A. and {Wolf}, Christian and {Bessell}, Michael S. and {Chang}, Seo-Won and {Luvaul}, Lance C. and {Tonry}, John L. and {White}, Marc C. and {Da Costa}, Gary S.},
        title = "{SkyMapper Southern Survey: Data release 4}",
      journal = {\pasa},
         year = 2024,
        month = oct,
       volume = {41},
          eid = {e061},
        pages = {e061},
          doi = {10.1017/pasa.2024.53},
archivePrefix = {arXiv},
       eprint = {2402.02015},
 primaryClass = {astro-ph.CO},
       adsurl = {https://ui.adsabs.harvard.edu/abs/2024PASA...41...61O}
}

@software{2021ascl.soft12006K,
       author = {{Karpov}, Sergey},
        title = "{STDPipe: Simple Transient Detection Pipeline}",
 howpublished = {Astrophysics Source Code Library, record ascl:2112.006},
         year = 2021,
        month = dec,
          eid = {ascl:2112.006},
archivePrefix = {ascl},
       eprint = {2112.006},
       adsurl = {https://ui.adsabs.harvard.edu/abs/2021ascl.soft12006K}
}

@INPROCEEDINGS{GRANDMA,
       author = {{Agayeva}, S. and {Alishov}, S. and {Antier}, S. and {Ayvazian}, V.~R. and {Bai}, J.~M. and {Baransky}, A. and {Barynova}, K. and {Basa}, S. and {Beradze}, S. and {Bertin}, E. and {Berthier}, J. and {Bla{\v{z}}ek}, M. and {Bo{\"e}r}, M. and {Burkhonov}, O. and {Burrell}, A. and {Cailleau}, A. and {Chabert}, B. and {Chen}, J.~C. and {Christensen}, N. and {Coleiro}, A. and {Corre}, D. and {Coughlin}, M.~W. and {Coward}, D. and {Crisp}, H. and {Delattre}, C. and {Dietrich}, T. and {Ducoin}, J.~G. and {Duverne}, P.~A. and {Eymar}, L. and {Fock-Hang}, P. and {Gendre}, B. and {Hello}, P. and {Howell}, E.~J. and {Inasaridze}, R. Ya. and {Ismailov}, N. and {Kann}, D.~A. and {Kapanadze}, G.~V. and {Karpov}, S. and {Klotz}, A. and {Kochiashvili}, N. and {Lachaud}, C. and {Leroy}, N. and {Le Van Su}, A. and {Li}, W.~X. and {Lin}, W.~L. and {Lognone}, P. and {Marchal-Duval}, G. and {Marron}, R. and {Ma{\v{s}}ek}, M. and {Mo}, J. and {Moore}, J. and {Morris}, D. and {Natsvlishvili}, R. and {Noysena}, K. and {Orange}, N.~B. and {Perrigault}, S. and {Peyrot}, A. and {Prouza}, M. and {Sadibekova}, T. and {Samadov}, D. and {Simon}, A. and {Stachie}, C. and {Teng}, J.~P. and {Thierry}, P. and {Th{\"o}ne}, C.~C. and {Tillayev}, Y. and {Turpin}, D. and {de Ugarte Postigo}, A. and {Vachier}, F. and {Vardosanidze}, M. and {Vasylenko}, V. and {Vidadi}, Z. and {Wang}, C.~J. and {Wang}, X.~F. and {Yan}, S.~Y. and {Zhang}, J.~C. and {Zhang}, J.~J. and {Zhang}, X.~H.},
        title = "{Grandma: A Network to Coordinate Them All}",
    booktitle = {Revista Mexicana de Astronomia y Astrofisica Conference Series},
         year = 2021,
       series = {Revista Mexicana de Astronomia y Astrofisica Conference Series},
       volume = {53},
        month = sep,
        pages = {198-205},
          doi = {10.22201/ia.14052059p.2021.53.39},
archivePrefix = {arXiv},
       eprint = {2008.03962},
 primaryClass = {astro-ph.HE},
       adsurl = {https://ui.adsabs.harvard.edu/abs/2021RMxAC..53..198A}
}

@ARTICLE{KW_GCN,
       author = {{Ridnaia}, A. and {Frederiks}, D. and {Lysenko}, A. and {Svinkin}, D. and {Tsvetkova}, A. and {Ulanov}, M. and {Cline}, T. and {Konus-Wind Team}},
        title = "{Konus-Wind detection of GRB 250424A}",
      journal = {GRB Coordinates Network},
         year = 2025,
        month = apr,
       volume = {40243},
        pages = {1},
       adsurl = {https://ui.adsabs.harvard.edu/abs/2025GCN.40243....1R}
}

@ARTICLE{AstroSat_GCN,
       author = {{Harsha}, K.~H. and {Tembhurnikar}, M. and {Waratkar}, G. and {Vibhute}, A. and {Bhalerao}, V. and {Bhattacharya}, D. and {Rao}, A.~R. and {Vadawale}, S. and {AstroSat CZTI Collaboration}},
        title = "{GRB 250424A: AstroSat CZTI detection of a bright long burst}",
      journal = {GRB Coordinates Network},
         year = 2025,
        month = apr,
       volume = {40231},
        pages = {1},
       adsurl = {https://ui.adsabs.harvard.edu/abs/2025GCN.40231....1H}
}

@ARTICLE{EIRSAT_GCN,
       author = {{McKenna}, C. and {McDermott}, P. and {Murphy}, D. and {de Barra}, C. and {Ulyanov}, A. and {Finneran}, G. and {Corcoran}, G. and {Cotter}, L. and {Empey}, A. and {Fisher}, J. and {Gibson Kiely}, F. and {Thompson}, J. and {McKeown}, D. and {Martin-Carrillo}, A. and {Hanlon}, L. and {McBreen}, S. and {Eirsat-1 Team}},
        title = "{GRB 250424A: EIRSAT-1 GMOD Detection}",
      journal = {GRB Coordinates Network},
         year = 2025,
        month = apr,
       volume = {40249},
        pages = {1},
       adsurl = {https://ui.adsabs.harvard.edu/abs/2025GCN.40249....1M}
}

@ARTICLE{CALET_GCN,
       author = {{Nakahira}, S. and {Yoshida}, A. and {Sakamoto}, T. and {Sugita}, S. and {Kawakubo}, Y. and {Yamaoka}, K. and {Asaoka}, Y. and {Torii}, S. and {Akaike}, Y. and {Kobayashi}, K. and {Shimizu}, Y. and {Tamura}, T. and {Cannady}, N. and {Cherry}, M.~L. and {Ricciarini}, S. and {Marrocchesi}, P.~S. and {Calet Collaboration}},
        title = "{GRB 250424A: CALET Gamma-Ray Burst Monitor detection}",
      journal = {GRB Coordinates Network},
         year = 2025,
        month = may,
       volume = {40298},
        pages = {1},
       adsurl = {https://ui.adsabs.harvard.edu/abs/2025GCN.40298....1N}
}

@article{Piran2005,
   title={The physics of gamma-ray bursts},
   volume={76},
   ISSN={1539-0756},
   url={http://dx.doi.org/10.1103/RevModPhys.76.1143},
   DOI={10.1103/revmodphys.76.1143},
   number={4},
   journal={Reviews of Modern Physics},
   publisher={American Physical Society (APS)},
   author={Piran, Tsvi},
   year={2005},
   month=jan, pages={1143–1210} }

@ARTICLE{Piran1999,
       author = {{Piran}, T.},
        title = "{Gamma-ray bursts and the fireball model}",
      journal = {\physrep},
         year = 1999,
        month = jun,
       volume = {314},
       number = {6},
        pages = {575-667},
          doi = {10.1016/S0370-1573(98)00127-6},
archivePrefix = {arXiv},
       eprint = {astro-ph/9810256},
 primaryClass = {astro-ph},
       adsurl = {https://ui.adsabs.harvard.edu/abs/1999PhR...314..575P}
}

@ARTICLE{Mészáros1997,
       author = {{M{\'e}sz{\'a}ros}, P. and {Rees}, M.~J.},
        title = "{Optical and Long-Wavelength Afterglow from Gamma-Ray Bursts}",
      journal = {\apj},
         year = 1997,
        month = feb,
       volume = {476},
       number = {1},
        pages = {232-237},
          doi = {10.1086/303625},
archivePrefix = {arXiv},
       eprint = {astro-ph/9606043},
 primaryClass = {astro-ph},
       adsurl = {https://ui.adsabs.harvard.edu/abs/1997ApJ...476..232M}
}

@ARTICLE{Panaitescu2002,
       author = {{Panaitescu}, A. and {Kumar}, P.},
        title = "{Properties of Relativistic Jets in Gamma-Ray Burst Afterglows}",
      journal = {\apj},
         year = 2002,
        month = jun,
       volume = {571},
       number = {2},
        pages = {779-789},
          doi = {10.1086/340094},
       adsurl = {https://ui.adsabs.harvard.edu/abs/2002ApJ...571..779P}
}

@ARTICLE{zhang&Liang&zhang2007,
       author = {{Zhang}, Bin-Bin and {Liang}, En-Wei and {Zhang}, Bing},
        title = "{A Comprehensive Analysis of Swift XRT Data. I. Apparent Spectral Evolution of Gamma-Ray Burst X-Ray Tails}",
      journal = {\apj},
         year = 2007,
        month = sep,
       volume = {666},
       number = {2},
        pages = {1002-1011},
          doi = {10.1086/519548},
archivePrefix = {arXiv},
       eprint = {astro-ph/0612246},
 primaryClass = {astro-ph},
       adsurl = {https://ui.adsabs.harvard.edu/abs/2007ApJ...666.1002Z}
}

@ARTICLE{Panaitescu2011,
       author = {{Panaitescu}, A. and {Vestrand}, W.~T.},
        title = "{Optical afterglows of gamma-ray bursts: peaks, plateaus and possibilities}",
      journal = {\mnras},
         year = 2011,
        month = jul,
       volume = {414},
       number = {4},
        pages = {3537-3546},
          doi = {10.1111/j.1365-2966.2011.18653.x},
archivePrefix = {arXiv},
       eprint = {1009.3947},
 primaryClass = {astro-ph.HE},
       adsurl = {https://ui.adsabs.harvard.edu/abs/2011MNRAS.414.3537P}
}

@ARTICLE{Wuxf2013,
       author = {{Wu}, Xue-Feng and {Hou}, Shu-Jin and {Lei}, Wei-Hua},
        title = "{Giant X-Ray Bump in GRB 121027A: Evidence for Fall-back Disk Accretion}",
      journal = {\apjl},
         year = 2013,
        month = apr,
       volume = {767},
       number = {2},
          eid = {L36},
        pages = {L36},
          doi = {10.1088/2041-8205/767/2/L36},
archivePrefix = {arXiv},
       eprint = {1302.4878},
 primaryClass = {astro-ph.HE},
       adsurl = {https://ui.adsabs.harvard.edu/abs/2013ApJ...767L..36W}
}

@ARTICLE{Ree1998,
       author = {{Rees}, M.~J. and {M{\'e}sz{\'a}ros}, P.},
        title = "{Refreshed Shocks and Afterglow Longevity in Gamma-Ray Bursts}",
      journal = {\apjl},
         year = 1998,
        month = mar,
       volume = {496},
       number = {1},
        pages = {L1-L4},
          doi = {10.1086/311244},
archivePrefix = {arXiv},
       eprint = {astro-ph/9712252},
 primaryClass = {astro-ph},
       adsurl = {https://ui.adsabs.harvard.edu/abs/1998ApJ...496L...1R}
}

@ARTICLE{Sari2000,
       author = {{Sari}, Re'em and {M{\'e}sz{\'a}ros}, Peter},
        title = "{Impulsive and Varying Injection in Gamma-Ray Burst Afterglows}",
      journal = {\apjl},
         year = 2000,
        month = may,
       volume = {535},
       number = {1},
        pages = {L33-L37},
          doi = {10.1086/312689},
archivePrefix = {arXiv},
       eprint = {astro-ph/0003406},
 primaryClass = {astro-ph},
       adsurl = {https://ui.adsabs.harvard.edu/abs/2000ApJ...535L..33S}
}

@ARTICLE{Fynbo2006Nat,
       author = {{Fynbo}, Johan P.~U. and {Watson}, Darach and {Th{\"o}ne}, Christina C. and {Sollerman}, Jesper and {Bloom}, Joshua S. and {Davis}, Tamara M. and {Hjorth}, Jens and {Jakobsson}, P{\'a}ll and {J{\o}rgensen}, Uffe G. and {Graham}, John F. and {Fruchter}, Andrew S. and {Bersier}, David and {Kewley}, Lisa and {Cassan}, Arnaud and {Castro Cer{\'o}n}, Jos{\'e} Mar{\'\i}a and {Foley}, Suzanne and {Gorosabel}, Javier and {Hinse}, Tobias C. and {Horne}, Keith D. and {Jensen}, Brian L. and {Klose}, Sylvio and {Kocevski}, Daniel and {Marquette}, Jean-Baptiste and {Perley}, Daniel and {Ramirez-Ruiz}, Enrico and {Stritzinger}, Maximilian D. and {Vreeswijk}, Paul M. and {Wijers}, Ralph A.~M. and {Woller}, Kristian G. and {Xu}, Dong and {Zub}, Marta},
        title = "{No supernovae associated with two long-duration {\ensuremath{\gamma}}-ray bursts}",
      journal = {\nat},
         year = 2006,
        month = dec,
       volume = {444},
       number = {7122},
        pages = {1047-1049},
          doi = {10.1038/nature05375},
archivePrefix = {arXiv},
       eprint = {astro-ph/0608313},
 primaryClass = {astro-ph},
       adsurl = {https://ui.adsabs.harvard.edu/abs/2006Natur.444.1047F}
}

@ARTICLE{Della2006,
       author = {{Della Valle}, M. and {Malesani}, D. and {Benetti}, S. and {Chincarini}, G. and {Stella}, L. and {Tagliaferri}, G.},
        title = "{Supernova 2005nc and GRB 050525A}",
      journal = {\iaucirc},
         year = 2006,
        month = mar,
       volume = {8696},
        pages = {1},
       adsurl = {https://ui.adsabs.harvard.edu/abs/2006IAUC.8696....1D}
}

@ARTICLE{Galama1998Nat,
       author = {{Galama}, T.~J. and {Vreeswijk}, P.~M. and {van Paradijs}, J. and {Kouveliotou}, C. and {Augusteijn}, T. and {B{\"o}hnhardt}, H. and {Brewer}, J.~P. and {Doublier}, V. and {Gonzalez}, J.-F. and {Leibundgut}, B. and {Lidman}, C. and {Hainaut}, O.~R. and {Patat}, F. and {Heise}, J. and {in't Zand}, J. and {Hurley}, K. and {Groot}, P.~J. and {Strom}, R.~G. and {Mazzali}, P.~A. and {Iwamoto}, K. and {Nomoto}, K. and {Umeda}, H. and {Nakamura}, T. and {Young}, T.~R. and {Suzuki}, T. and {Shigeyama}, T. and {Koshut}, T. and {Kippen}, M. and {Robinson}, C. and {de Wildt}, P. and {Wijers}, R.~A.~M.~J. and {Tanvir}, N. and {Greiner}, J. and {Pian}, E. and {Palazzi}, E. and {Frontera}, F. and {Masetti}, N. and {Nicastro}, L. and {Feroci}, M. and {Costa}, E. and {Piro}, L. and {Peterson}, B.~A. and {Tinney}, C. and {Boyle}, B. and {Cannon}, R. and {Stathakis}, R. and {Sadler}, E. and {Begam}, M.~C. and {Ianna}, P.},
        title = "{An unusual supernova in the error box of the {\ensuremath{\gamma}}-ray burst of 25 April 1998}",
      journal = {\nat},
         year = 1998,
        month = oct,
       volume = {395},
       number = {6703},
        pages = {670-672},
          doi = {10.1038/27150},
archivePrefix = {arXiv},
       eprint = {astro-ph/9806175},
 primaryClass = {astro-ph},
       adsurl = {https://ui.adsabs.harvard.edu/abs/1998Natur.395..670G}
}

@ARTICLE{Stanek2003,
       author = {{Stanek}, K.~Z. and {Matheson}, T. and {Garnavich}, P.~M. and {Martini}, P. and {Berlind}, P. and {Caldwell}, N. and {Challis}, P. and {Brown}, W.~R. and {Schild}, R. and {Krisciunas}, K. and {Calkins}, M.~L. and {Lee}, J.~C. and {Hathi}, N. and {Jansen}, R.~A. and {Windhorst}, R. and {Echevarria}, L. and {Eisenstein}, D.~J. and {Pindor}, B. and {Olszewski}, E.~W. and {Harding}, P. and {Holland}, S.~T. and {Bersier}, D.},
        title = "{Spectroscopic Discovery of the Supernova 2003dh Associated with GRB 030329}",
      journal = {\apjl},
         year = 2003,
        month = jul,
       volume = {591},
       number = {1},
        pages = {L17-L20},
          doi = {10.1086/376976},
archivePrefix = {arXiv},
       eprint = {astro-ph/0304173},
 primaryClass = {astro-ph},
       adsurl = {https://ui.adsabs.harvard.edu/abs/2003ApJ...591L..17S}
}

@INPROCEEDINGS{xspec1996,
       author = {{Arnaud}, K.~A.},
        title = "{XSPEC: The First Ten Years}",
    booktitle = {Astronomical Data Analysis Software and Systems V},
         year = 1996,
       editor = {{Jacoby}, George H. and {Barnes}, Jeannette},
       series = {Astronomical Society of the Pacific Conference Series},
       volume = {101},
        month = jan,
        pages = {17},
       adsurl = {https://ui.adsabs.harvard.edu/abs/1996ASPC..101...17A}
}

@ARTICLE{hostgalaxy2025,
       author = {{P{\'e}rez-Fournon}, I. and {Poidevin}, F. and {Cano-Morales}, D. and {Hern{\'a}ndez-D{\'\i}az}, A.~E. and {Correa-Plasencia}, I. and {L{\'o}pez-Oramas}, A.},
        title = "{GRB 250424A: likely host galaxy}",
      journal = {GRB Coordinates Network},
         year = 2025,
        month = apr,
       volume = {40227},
        pages = {1},
       adsurl = {https://ui.adsabs.harvard.edu/abs/2025GCN.40227....1P}
}

@ARTICLE{gcn_redshift,
       author = {{Saccardi}, A. and {Malesani}, D.~B. and {Corcoran}, G. and {Covino}, S. and {Habeeb}, N. and {Izzo}, L. and {Levan}, A.~J. and {Martin-Carrillo}, A. and {Palmerio}, J.~T. and {Pugliese}, G. and {Schneider}, B. and {Tanvir}, N.~R. and {Vergani}, S.~D. and {Wiersema}, K. and {Stargate Collaboration}},
        title = "{GRB 250424A: VLT/X-shooter spectroscopic redshift z = 0.310}",
      journal = {GRB Coordinates Network},
         year = 2025,
        month = apr,
       volume = {40228},
        pages = {1},
       adsurl = {https://ui.adsabs.harvard.edu/abs/2025GCN.40228....1S}
}

@ARTICLE{GRM_trigger,
       author = {{He}, Jiang and {Sun}, Jian-Chao and {Dong}, Yong-Wei and {Wu}, Bo-Bing and {Zheng}, Shi-Jie and {Li}, Lu and {Liu}, Jiang-Tao and {Liu}, Xin and {Shi}, Hao-Li and {Song}, Li-Ming and {Wang}, Rui-Jie and {Zhang}, Juan and {Zhang}, Li and {Zhang}, Shuang-Nan and {Zhao}, Xiao-Yun and {Liu}, Xing-Guang},
        title = "{SVOM-GRM trigger performance study and verification}",
      journal = {Experimental Astronomy},
         year = 2025,
        month = feb,
       volume = {59},
       number = {1},
          eid = {15},
        pages = {15},
          doi = {10.1007/s10686-025-09983-x},
       adsurl = {https://ui.adsabs.harvard.edu/abs/2025ExA....59...15H}
}

@ARTICLE{GRM_analysis,
      title={BREAKFAST: A Framework for general joint BA duty and follow-up guidance of multiple $\gamma$-ray monitors}, 
      author={Chen-Wei Wang and Peng Zhang and Shao-Lin Xiong and Yue Huang and Wen-Jun Tan and Zheng-Hang Yu and Yue Wang and Wang-Chen Xue and Chao Zheng and Hao-Xuan Guo and Ce Cai and Yong-Wei Dong and Jiang He and Cheng-Kui Li and Xiao-Bo Li and Jia-Cong Liu and Xing-Hao Luo and Xiang Ma and Yang-Zhao Ren and Li-Ming Song and Ping Wang and Jin Wang and Bo-Bing Wu and Shuo Xiao and Sheng-Lun Xie and Shu-Xu Yi and Xue-Yuan Zao and Xiao-Yun Zhao and Li Zhang and Shuang-Nan Zhang and Yan-Qiu Zhang and Shi-Jie Zheng},
      year={2025},
      eprint={2510.15816},
      archivePrefix={arXiv},
      primaryClass={astro-ph.IM},
      url={https://arxiv.org/abs/2510.15816}, 
}

@ARTICLE{2000Kumar,
       author = {{Kumar}, Pawan and {Panaitescu}, Alin},
        title = "{Steepening of Afterglow Decay for Jets Interacting with Stratified Media}",
      journal = {\apjl},
         year = 2000,
        month = sep,
       volume = {541},
       number = {1},
        pages = {L9-L12},
          doi = {10.1086/312888},
archivePrefix = {arXiv},
       eprint = {astro-ph/0003264},
 primaryClass = {astro-ph},
       adsurl = {https://ui.adsabs.harvard.edu/abs/2000ApJ...541L...9K}
}

@ARTICLE{Minaev2020,
       author = {{Minaev}, P.~Y. and {Pozanenko}, A.~S.},
        title = "{The E$_{p,I}$-E$_{iso}$ correlation: type I gamma-ray bursts and the new classification method}",
      journal = {\mnras},
         year = 2020,
        month = feb,
       volume = {492},
       number = {2},
        pages = {1919-1936},
          doi = {10.1093/mnras/stz3611},
archivePrefix = {arXiv},
       eprint = {1912.09810},
 primaryClass = {astro-ph.HE},
       adsurl = {https://ui.adsabs.harvard.edu/abs/2020MNRAS.492.1919M}
}

@misc{minaev2020b,
      title={GRB 200415A: magnetar giant flare or short gamma-ray burst?}, 
      author={Pavel Minaev and Alexei Pozanenko},
      year={2020},
      eprint={2008.12752},
      archivePrefix={arXiv},
      primaryClass={astro-ph.HE},
      url={https://arxiv.org/abs/2008.12752}, 
}

@ARTICLE{2009ZhangB,
       author = {{Zhang}, Bing and {Zhang}, Bin-Bin and {Virgili}, Francisco J. and {Liang}, En-Wei and {Kann}, D. Alexander and {Wu}, Xue-Feng and {Proga}, Daniel and {Lv}, Hou-Jun and {Toma}, Kenji and {M{\'e}sz{\'a}ros}, Peter and {Burrows}, David N. and {Roming}, Peter W.~A. and {Gehrels}, Neil},
        title = "{Discerning the Physical Origins of Cosmological Gamma-ray Bursts Based on Multiple Observational Criteria: The Cases of z = 6.7 GRB 080913, z = 8.2 GRB 090423, and Some Short/Hard GRBs}",
      journal = {\apj},
         year = 2009,
        month = oct,
       volume = {703},
       number = {2},
        pages = {1696-1724},
          doi = {10.1088/0004-637X/703/2/1696},
archivePrefix = {arXiv},
       eprint = {0902.2419},
 primaryClass = {astro-ph.HE},
       adsurl = {https://ui.adsabs.harvard.edu/abs/2009ApJ...703.1696Z}
}

@ARTICLE{2016HI4PI,
       author = {{HI4PI Collaboration} and {Ben Bekhti}, N. and {Fl{\"o}er}, L. and {Keller}, R. and {Kerp}, J. and {Lenz}, D. and {Winkel}, B. and {Bailin}, J. and {Calabretta}, M.~R. and {Dedes}, L. and {Ford}, H.~A. and {Gibson}, B.~K. and {Haud}, U. and {Janowiecki}, S. and {Kalberla}, P.~M.~W. and {Lockman}, F.~J. and {McClure-Griffiths}, N.~M. and {Murphy}, T. and {Nakanishi}, H. and {Pisano}, D.~J. and {Staveley-Smith}, L.},
        title = "{HI4PI: A full-sky H I survey based on EBHIS and GASS}",
      journal = {\aap},
         year = 2016,
        month = oct,
       volume = {594},
          eid = {A116},
        pages = {A116},
          doi = {10.1051/0004-6361/201629178},
archivePrefix = {arXiv},
       eprint = {1610.06175},
 primaryClass = {astro-ph.GA},
       adsurl = {https://ui.adsabs.harvard.edu/abs/2016A&A...594A.116H}
}

@ARTICLE{Zheng09,
       author = {{Zheng}, Wei-Kang and {Deng}, Jin-Song and {Wang}, Jing},
        title = "{Statistical studies of optically dark gamma-ray bursts in the Swift era}",
      journal = {Research in Astronomy and Astrophysics},
         year = 2009,
        month = oct,
       volume = {9},
       number = {10},
        pages = {1103-1118},
          doi = {10.1088/1674-4527/9/10/003},
archivePrefix = {arXiv},
       eprint = {0906.2244},
 primaryClass = {astro-ph.HE},
       adsurl = {https://ui.adsabs.harvard.edu/abs/2009RAA.....9.1103Z}
}

@ARTICLE{Piran2004,
       author = {{Piran}, Tsvi},
        title = "{The physics of gamma-ray bursts}",
      journal = {Reviews of Modern Physics},
         year = 2004,
        month = oct,
       volume = {76},
       number = {4},
        pages = {1143-1210},
          doi = {10.1103/RevModPhys.76.1143},
archivePrefix = {arXiv},
       eprint = {astro-ph/0405503},
 primaryClass = {astro-ph},
       adsurl = {https://ui.adsabs.harvard.edu/abs/2004RvMP...76.1143P}
}

@ARTICLE{Zhang&Meszaros2004,
       author = {{Zhang}, Bing and {M{\'e}sz{\'a}ros}, Peter},
        title = "{Gamma-Ray Bursts: progress, problems \& prospects}",
      journal = {International Journal of Modern Physics A},
         year = 2004,
        month = jan,
       volume = {19},
       number = {15},
        pages = {2385-2472},
          doi = {10.1142/S0217751X0401746X},
archivePrefix = {arXiv},
       eprint = {astro-ph/0311321},
 primaryClass = {astro-ph},
       adsurl = {https://ui.adsabs.harvard.edu/abs/2004IJMPA..19.2385Z}
}

@ARTICLE{Veilleux&Osterbrock1987,
       author = {{Veilleux}, Sylvain and {Osterbrock}, Donald E.},
        title = "{Spectral Classification of Emission-Line Galaxies}",
      journal = {\apjs},
         year = 1987,
        month = feb,
       volume = {63},
        pages = {295},
          doi = {10.1086/191166},
       adsurl = {https://ui.adsabs.harvard.edu/abs/1987ApJS...63..295V}
}

@ARTICLE{fermi,
       author = {{Meegan}, Charles and {Lichti}, Giselher and {Bhat}, P.~N. and {Bissaldi}, Elisabetta and {Briggs}, Michael S. and {Connaughton}, Valerie and {Diehl}, Roland and {Fishman}, Gerald and {Greiner}, Jochen and {Hoover}, Andrew S. and {van der Horst}, Alexander J. and {von Kienlin}, Andreas and {Kippen}, R. Marc and {Kouveliotou}, Chryssa and {McBreen}, Sheila and {Paciesas}, W.~S. and {Preece}, Robert and {Steinle}, Helmut and {Wallace}, Mark S. and {Wilson}, Robert B. and {Wilson-Hodge}, Colleen},
        title = "{The Fermi Gamma-ray Burst Monitor}",
      journal = {\apj},
         year = 2009,
        month = sep,
       volume = {702},
       number = {1},
        pages = {791-804},
          doi = {10.1088/0004-637X/702/1/791},
archivePrefix = {arXiv},
       eprint = {0908.0450},
 primaryClass = {astro-ph.IM},
       adsurl = {https://ui.adsabs.harvard.edu/abs/2009ApJ...702..791M}
}

@ARTICLE{2026Dengc,
       author = {{Deng}, Chen and {Huang}, Yong-Feng and {Kurban}, Abdusattar and {Geng}, Jin-Jun and {Xu}, Fan and {Dong}, Xiao-Fei and {Gao}, Hao-Xuan and {Liang}, En-Wei and {Li}, Liang},
        title = "{Modeling the Multiwavelength Afterglow of Short Gamma-Ray Bursts with a Plateau Phase}",
      journal = {\apj},
         year = 2026,
        month = mar,
       volume = {1000},
       number = {1},
          eid = {97},
        pages = {97},
          doi = {10.3847/1538-4357/ae486b},
archivePrefix = {arXiv},
       eprint = {2511.11396},
 primaryClass = {astro-ph.HE},
       adsurl = {https://ui.adsabs.harvard.edu/abs/2026ApJ..1000...97D}
}

@ARTICLE{Dainotti2020,
       author = {{Dainotti}, M.~G. and {Livermore}, S. and {Kann}, D.~A. and {Li}, L. and {Oates}, S. and {Yi}, S. and {Zhang}, B. and {Gendre}, B. and {Cenko}, B. and {Fraija}, N.},
        title = "{The Optical Luminosity-Time Correlation for More than 100 Gamma-Ray Burst Afterglows}",
      journal = {\apjl},
         year = 2020,
        month = dec,
       volume = {905},
       number = {2},
          eid = {L26},
        pages = {L26},
          doi = {10.3847/2041-8213/abcda9},
archivePrefix = {arXiv},
       eprint = {2011.14493},
 primaryClass = {astro-ph.HE},
       adsurl = {https://ui.adsabs.harvard.edu/abs/2020ApJ...905L..26D}
}

@ARTICLE{Tangch2019,
       author = {{Tang}, Chen-Han and {Huang}, Yong-Feng and {Geng}, Jin-Jun and {Zhang}, Zhi-Bin},
        title = "{Statistical Study of Gamma-Ray Bursts with a Plateau Phase in the X-Ray Afterglow}",
      journal = {\apjs},
         year = 2019,
        month = nov,
       volume = {245},
       number = {1},
          eid = {1},
        pages = {1},
          doi = {10.3847/1538-4365/ab4711},
archivePrefix = {arXiv},
       eprint = {1905.07929},
 primaryClass = {astro-ph.HE},
       adsurl = {https://ui.adsabs.harvard.edu/abs/2019ApJS..245....1T}
}

@ARTICLE{Haol2005,
       author = {{Hao}, Lei and {Strauss}, Michael A. and {Tremonti}, Christy A. and {Schlegel}, David J. and {Heckman}, Timothy M. and {Kauffmann}, Guinevere and {Blanton}, Michael R. and {Fan}, Xiaohui and {Gunn}, James E. and {Hall}, Patrick B. and {Ivezi{\'c}}, {\v{Z}}eljko and {Knapp}, Gillian R. and {Krolik}, Julian H. and {Lupton}, Robert H. and {Richards}, Gordon T. and {Schneider}, Donald P. and {Strateva}, Iskra V. and {Zakamska}, Nadia L. and {Brinkmann}, J. and {Brunner}, Robert J. and {Szokoly}, Gyula P.},
        title = "{Active Galactic Nuclei in the Sloan Digital Sky Survey. I. Sample Selection}",
      journal = {\aj},
         year = 2005,
        month = apr,
       volume = {129},
       number = {4},
        pages = {1783-1794},
          doi = {10.1086/428485},
archivePrefix = {arXiv},
       eprint = {astro-ph/0501059},
 primaryClass = {astro-ph},
       adsurl = {https://ui.adsabs.harvard.edu/abs/2005AJ....129.1783H}
}

@ARTICLE{Wang&wei2008,
       author = {{Wang}, J. and {Wei}, J.~Y.},
        title = "{Understanding the AGN-Host Connection in Partially Obscured Active Galactic Nuclei. I. The Nature of AGN+H II Composites}",
      journal = {\apj},
         year = 2008,
        month = may,
       volume = {679},
       number = {1},
        pages = {86-100},
          doi = {10.1086/587048},
archivePrefix = {arXiv},
       eprint = {0802.0548},
 primaryClass = {astro-ph},
       adsurl = {https://ui.adsabs.harvard.edu/abs/2008ApJ...679...86W}
}

@ARTICLE{Francis1992,
       author = {{Francis}, Paul J. and {Hewett}, Paul C. and {Foltz}, Craig B. and {Chaffee}, Frederic H.},
        title = "{An Objective Classification Scheme for QSO Spectra}",
      journal = {\apj},
         year = 1992,
        month = oct,
       volume = {398},
        pages = {476},
          doi = {10.1086/171870},
       adsurl = {https://ui.adsabs.harvard.edu/abs/1992ApJ...398..476F}
}

@ARTICLE{Bruzual2003,
       author = {{Bruzual}, G. and {Charlot}, S.},
        title = "{Stellar population synthesis at the resolution of 2003}",
      journal = {\mnras},
         year = 2003,
        month = oct,
       volume = {344},
       number = {4},
        pages = {1000-1028},
          doi = {10.1046/j.1365-8711.2003.06897.x},
archivePrefix = {arXiv},
       eprint = {astro-ph/0309134},
 primaryClass = {astro-ph},
       adsurl = {https://ui.adsabs.harvard.edu/abs/2003MNRAS.344.1000B}
}

@ARTICLE{Cardelli1989,
       author = {{Cardelli}, Jason A. and {Clayton}, Geoffrey C. and {Mathis}, John S.},
        title = "{The Relationship between Infrared, Optical, and Ultraviolet Extinction}",
      journal = {\apj},
         year = 1989,
        month = oct,
       volume = {345},
        pages = {245},
          doi = {10.1086/167900},
       adsurl = {https://ui.adsabs.harvard.edu/abs/1989ApJ...345..245C}
}

@INPROCEEDINGS{Kriss1994,
       author = {{Kriss}, G.},
        title = "{Fitting Models to UV and Optical Spectral Data}",
    booktitle = {Astronomical Data Analysis Software and Systems III},
         year = 1994,
       editor = {{Crabtree}, D.~R. and {Hanisch}, R.~J. and {Barnes}, J.},
       series = {Astronomical Society of the Pacific Conference Series},
       volume = {61},
        month = jan,
        pages = {437},
       adsurl = {https://ui.adsabs.harvard.edu/abs/1994ASPC...61..437K}
}

@ARTICLE{Maiolino2008,
       author = {{Maiolino}, R. and {Nagao}, T. and {Grazian}, A. and {Cocchia}, F. and {Marconi}, A. and {Mannucci}, F. and {Cimatti}, A. and {Pipino}, A. and {Ballero}, S. and {Calura}, F. and {Chiappini}, C. and {Fontana}, A. and {Granato}, G.~L. and {Matteucci}, F. and {Pastorini}, G. and {Pentericci}, L. and {Risaliti}, G. and {Salvati}, M. and {Silva}, L.},
        title = "{AMAZE. I. The evolution of the mass-metallicity relation at z > 3}",
      journal = {\aap},
         year = 2008,
        month = sep,
       volume = {488},
       number = {2},
        pages = {463-479},
          doi = {10.1051/0004-6361:200809678},
archivePrefix = {arXiv},
       eprint = {0806.2410},
 primaryClass = {astro-ph},
       adsurl = {https://ui.adsabs.harvard.edu/abs/2008A&A...488..463M}
}

@ARTICLE{Kewley2004,
       author = {{Kewley}, Lisa J. and {Geller}, Margaret J. and {Jansen}, Rolf A.},
        title = "{[O II] as a Star Formation Rate Indicator}",
      journal = {\aj},
         year = 2004,
        month = apr,
       volume = {127},
       number = {4},
        pages = {2002-2030},
          doi = {10.1086/382723},
archivePrefix = {arXiv},
       eprint = {astro-ph/0401172},
 primaryClass = {astro-ph},
       adsurl = {https://ui.adsabs.harvard.edu/abs/2004AJ....127.2002K}
}

@ARTICLE{Heckman2004,
       author = {{Heckman}, Timothy M. and {Kauffmann}, Guinevere and {Brinchmann}, Jarle and {Charlot}, St{\'e}phane and {Tremonti}, Christy and {White}, Simon D.~M.},
        title = "{Present-Day Growth of Black Holes and Bulges: The Sloan Digital Sky Survey Perspective}",
      journal = {\apj},
         year = 2004,
        month = sep,
       volume = {613},
       number = {1},
        pages = {109-118},
          doi = {10.1086/422872},
archivePrefix = {arXiv},
       eprint = {astro-ph/0406218},
 primaryClass = {astro-ph},
       adsurl = {https://ui.adsabs.harvard.edu/abs/2004ApJ...613..109H}
}

@ARTICLE{Kauffmann2003,
       author = {{Kauffmann}, Guinevere and {Heckman}, Timothy M. and {Tremonti}, Christy and {Brinchmann}, Jarle and {Charlot}, St{\'e}phane and {White}, Simon D.~M. and {Ridgway}, Susan E. and {Brinkmann}, Jon and {Fukugita}, Masataka and {Hall}, Patrick B. and {Ivezi{\'c}}, {\v{Z}}eljko and {Richards}, Gordon T. and {Schneider}, Donald P.},
        title = "{The host galaxies of active galactic nuclei}",
      journal = {\mnras},
         year = 2003,
        month = dec,
       volume = {346},
       number = {4},
        pages = {1055-1077},
          doi = {10.1111/j.1365-2966.2003.07154.x},
archivePrefix = {arXiv},
       eprint = {astro-ph/0304239},
 primaryClass = {astro-ph},
       adsurl = {https://ui.adsabs.harvard.edu/abs/2003MNRAS.346.1055K}
}

@ARTICLE{Kewley2001,
       author = {{Kewley}, L.~J. and {Dopita}, M.~A. and {Sutherland}, R.~S. and {Heisler}, C.~A. and {Trevena}, J.},
        title = "{Theoretical Modeling of Starburst Galaxies}",
      journal = {\apj},
         year = 2001,
        month = jul,
       volume = {556},
       number = {1},
        pages = {121-140},
          doi = {10.1086/321545},
archivePrefix = {arXiv},
       eprint = {astro-ph/0106324},
 primaryClass = {astro-ph},
       adsurl = {https://ui.adsabs.harvard.edu/abs/2001ApJ...556..121K}
}

@ARTICLE{Kewley2006,
       author = {{Kewley}, Lisa J. and {Groves}, Brent and {Kauffmann}, Guinevere and {Heckman}, Tim},
        title = "{The host galaxies and classification of active galactic nuclei}",
      journal = {\mnras},
         year = 2006,
        month = nov,
       volume = {372},
       number = {3},
        pages = {961-976},
          doi = {10.1111/j.1365-2966.2006.10859.x},
archivePrefix = {arXiv},
       eprint = {astro-ph/0605681},
 primaryClass = {astro-ph},
       adsurl = {https://ui.adsabs.harvard.edu/abs/2006MNRAS.372..961K}
}

@INPROCEEDINGS{Tody1986,
       author = {{Tody}, Doug},
        title = "{The IRAF Data Reduction and Analysis System}",
    booktitle = {Instrumentation in astronomy VI},
         year = 1986,
       editor = {{Crawford}, David L.},
       series = {Society of Photo-Optical Instrumentation Engineers (SPIE) Conference Series},
       volume = {627},
        month = jan,
        pages = {733},
          doi = {10.1117/12.968154},
       adsurl = {https://ui.adsabs.harvard.edu/abs/1986SPIE..627..733T}
}

@INPROCEEDINGS{Tody1993,
       author = {{Tody}, Doug},
        title = "{IRAF in the Nineties}",
    booktitle = {Astronomical Data Analysis Software and Systems II},
         year = 1993,
       editor = {{Hanisch}, R.~J. and {Brissenden}, R.~J.~V. and {Barnes}, J.},
       series = {Astronomical Society of the Pacific Conference Series},
       volume = {52},
        month = jan,
        pages = {173},
       adsurl = {https://ui.adsabs.harvard.edu/abs/1993ASPC...52..173T}
}

@ARTICLE{Gehrels2006,
       author = {{Gehrels}, N. and {Norris}, J.~P. and {Barthelmy}, S.~D. and {Granot}, J. and {Kaneko}, Y. and {Kouveliotou}, C. and {Markwardt}, C.~B. and {M{\'e}sz{\'a}ros}, P. and {Nakar}, E. and {Nousek}, J.~A. and {O'Brien}, P.~T. and {Page}, M. and {Palmer}, D.~M. and {Parsons}, A.~M. and {Roming}, P.~W.~A. and {Sakamoto}, T. and {Sarazin}, C.~L. and {Schady}, P. and {Stamatikos}, M. and {Woosley}, S.~E.},
        title = "{A new {\ensuremath{\gamma}}-ray burst classification scheme from GRB060614}",
      journal = {\nat},
         year = 2006,
        month = dec,
       volume = {444},
       number = {7122},
        pages = {1044-1046},
          doi = {10.1038/nature05376},
archivePrefix = {arXiv},
       eprint = {astro-ph/0610635},
 primaryClass = {astro-ph},
       adsurl = {https://ui.adsabs.harvard.edu/abs/2006Natur.444.1044G}
}

@ARTICLE{Troja2016,
       author = {{Troja}, E. and {Sakamoto}, T. and {Cenko}, S.~B. and {Lien}, A. and {Gehrels}, N. and {Castro-Tirado}, A.~J. and {Ricci}, R. and {Capone}, J. and {Toy}, V. and {Kutyrev}, A. and {Kawai}, N. and {Cucchiara}, A. and {Fruchter}, A. and {Gorosabel}, J. and {Jeong}, S. and {Levan}, A. and {Perley}, D. and {Sanchez-Ramirez}, R. and {Tanvir}, N. and {Veilleux}, S.},
        title = "{An Achromatic Break in the Afterglow of the Short GRB 140903A: Evidence for a Narrow Jet}",
      journal = {\apj},
         year = 2016,
        month = aug,
       volume = {827},
       number = {2},
          eid = {102},
        pages = {102},
          doi = {10.3847/0004-637X/827/2/102},
archivePrefix = {arXiv},
       eprint = {1605.03573},
 primaryClass = {astro-ph.HE},
       adsurl = {https://ui.adsabs.harvard.edu/abs/2016ApJ...827..102T}
}

@ARTICLE{Lesage2023,
       author = {{Lesage}, S. and {Veres}, P. and {Briggs}, M.~S. and {Goldstein}, A. and {Kocevski}, D. and {Burns}, E. and {Wilson-Hodge}, C.~A. and {Bhat}, P.~N. and {Huppenkothen}, D. and {Fryer}, C.~L. and {Hamburg}, R. and {Racusin}, J. and {Bissaldi}, E. and {Cleveland}, W.~H. and {Dalessi}, S. and {Fletcher}, C. and {Giles}, M.~M. and {Hristov}, B.~A. and {Hui}, C.~M. and {Mailyan}, B. and {Malacaria}, C. and {Poolakkil}, S. and {Roberts}, O.~J. and {von Kienlin}, A. and {Wood}, J. and {Ajello}, M. and {Arimoto}, M. and {Baldini}, L. and {Ballet}, J. and {Baring}, M.~G. and {Bastieri}, D. and {Gonzalez}, J. Becerra and {Bellazzini}, R. and {Bissaldi}, E. and {Blandford}, R.~D. and {Bonino}, R. and {Bruel}, P. and {Buson}, S. and {Cameron}, R.~A. and {Caputo}, R. and {Caraveo}, P.~A. and {Cavazzuti}, E. and {Chiaro}, G. and {Cibrario}, N. and {Ciprini}, S. and {Orestano}, P. Cristarella and {Crnogorcevic}, M. and {Cuoco}, A. and {Cutini}, S. and {D'Ammando}, F. and {De Gaetano}, S. and {Di Lalla}, N. and {Di Venere}, L. and {Dom{\'\i}nguez}, A. and {Fegan}, S.~J. and {Ferrara}, E.~C. and {Fleischhack}, H. and {Fukazawa}, Y. and {Funk}, S. and {Fusco}, P. and {Galanti}, G. and {Gammaldi}, V. and {Gargano}, F. and {Gasbarra}, C. and {Gasparrini}, D. and {Germani}, S. and {Giacchino}, F. and {Giglietto}, N. and {Gill}, R. and {Giroletti}, M. and {Granot}, J. and {Green}, D. and {Grenier}, I.~A. and {Guiriec}, S. and {Gustafsson}, M. and {Hays}, E. and {Hewitt}, J.~W. and {Horan}, D. and {Hou}, X. and {Kuss}, M. and {Latronico}, L. and {Laviron}, A. and {Lemoine-Goumard}, M. and {Li}, J. and {Liodakis}, I. and {Longo}, F. and {Loparco}, F. and {Lorusso}, L. and {Lovellette}, M.~N. and {Lubrano}, P. and {Maldera}, S. and {Manfreda}, A. and {Mart{\'\i}-Devesa}, G. and {Mazziotta}, M.~N. and {McEnery}, J.~E. and {Mereu}, I. and {Meyer}, M. and {Michelson}, P.~F. and {Mizuno}, T. and {Monzani}, M.~E. and {Morselli}, A. and {Moskalenko}, I.~V. and {Negro}, M. and {Nuss}, E. and {Omodei}, N. and {Orlando}, E. and {Ormes}, J.~F. and {Paneque}, D. and {Panzarini}, G. and {Persic}, M. and {Pesce-Rollins}, M. and {Pillera}, R. and {Piron}, F. and {Poon}, H. and {Porter}, T.~A. and {Principe}, G. and {Rain{\`o}}, S. and {Rando}, R. and {Rani}, B. and {Razzano}, M. and {Razzaque}, S. and {Reimer}, A. and {Reimer}, O. and {Ryde}, F. and {S{\'a}nchez-Conde}, M. and {Parkinson}, P.~M. Saz and {Scotton}, L. and {Serini}, D. and {Sgr{\`o}}, C. and {Sharma}, V. and {Siskind}, E.~J. and {Spandre}, G. and {Spinelli}, P. and {Tajima}, H. and {Torres}, D.~F. and {Valverde}, J. and {Venters}, T. and {Wadiasingh}, Z. and {Wood}, K. and {Zaharijas}, G.},
        title = "{Fermi-GBM Discovery of GRB 221009A: An Extraordinarily Bright GRB from Onset to Afterglow}",
      journal = {\apjl},
         year = 2023,
        month = aug,
       volume = {952},
       number = {2},
          eid = {L42},
        pages = {L42},
          doi = {10.3847/2041-8213/ace5b4},
archivePrefix = {arXiv},
       eprint = {2303.14172},
 primaryClass = {astro-ph.HE},
       adsurl = {https://ui.adsabs.harvard.edu/abs/2023ApJ...952L..42L}
}

@software{2014heasoft,
       author = {{Nasa High Energy Astrophysics Science Archive Research Center (Heasarc)}},
        title = "{HEAsoft: Unified Release of FTOOLS and XANADU}",
 howpublished = {Astrophysics Source Code Library, record ascl:1408.004},
         year = 2014,
        month = aug,
          eid = {ascl:1408.004},
archivePrefix = {ascl},
       eprint = {1408.004},
       adsurl = {https://ui.adsabs.harvard.edu/abs/2014ascl.soft08004N}
}

@ARTICLE{Evans2007,
       author = {{Evans}, P.~A. and {Beardmore}, A.~P. and {Page}, K.~L. and {Tyler}, L.~G. and {Osborne}, J.~P. and {Goad}, M.~R. and {O'Brien}, P.~T. and {Vetere}, L. and {Racusin}, J. and {Morris}, D. and {Burrows}, D.~N. and {Capalbi}, M. and {Perri}, M. and {Gehrels}, N. and {Romano}, P.},
        title = "{An online repository of Swift/XRT light curves of {\ensuremath{\gamma}}-ray bursts}",
      journal = {\aap},
         year = 2007,
        month = jul,
       volume = {469},
       number = {1},
        pages = {379-385},
          doi = {10.1051/0004-6361:20077530},
archivePrefix = {arXiv},
       eprint = {0704.0128},
 primaryClass = {astro-ph},
       adsurl = {https://ui.adsabs.harvard.edu/abs/2007A&A...469..379E}
}
